\documentclass[a4paper,12pt,final]{article}
\usepackage{etex}

\usepackage[sc]{mathpazo}
\usepackage[table]{xcolor}

\usepackage{threeparttable}
\usepackage{standalone}
\usepackage{booktabs, dcolumn,multirow}
\usepackage{booktabs}
\usepackage{threeparttable}

\usepackage{siunitx}

\usepackage{afterpage}

\usepackage{rotating}
\usepackage{tikz}

\usepackage{latexsym}
\usepackage{amsfonts}
\usepackage{amssymb}
\usepackage{amsthm}
\usepackage{amsmath}
\usepackage{enumerate}
\usepackage[mathscr]{eucal}
\usepackage{multicol}
\usepackage{multirow}
\usepackage{setspace}
\usepackage{graphicx}
\usepackage{pdfpages}
\usepackage{pdflscape}

\usepackage{algorithm}
\usepackage{algpseudocode}

\usepackage{anyfontsize}
\usepackage{t1enc}

\usepackage{caption}
\usepackage{subcaption}
\DeclareCaptionFont{captiofont}{\fontsize{9pt}{10.8pt}\selectfont}
\usepackage{verbatim}
\definecolor{myred}{rgb}{0.89412, 0.10196, 0.10980}
\definecolor{myblue}{rgb}{0.21569, 0.49412, 0.72157}
\definecolor{mygreen}{rgb}{0.30196, 0.68627, 0.29020}
\definecolor{mygray}{rgb}{0.90, 0.90, 0.90}

\usepackage[plainpages=false,
breaklinks=true,
colorlinks=true,
urlcolor=gray,
citecolor=myblue, %blue , RoyalBlue
linkcolor=myblue, %red, %black BrickRed
bookmarks=true,
bookmarksopen=true,
bookmarksopenlevel=1,
pdfstartview=FitH,
pdfview=FitH]{hyperref}

\usepackage{color,hyperref}

\usepackage[round]{natbib}
\usepackage[lmargin=1in, rmargin=1in, tmargin=1in, bmargin=1in, paperwidth=220mm, paperheight=290mm]{geometry}

\usepackage{booktabs}
\usepackage{tabulary}
\usepackage{tabularx}
\usepackage{multirow}
\usepackage{array}

\usepackage{enumerate}

\usepackage{sectsty}
\usepackage{lipsum}
\sectionfont{\centering}

\usepackage{tocloft}
\usepackage{titlesec}
\titleformat{\section}
{\large\bfseries}{\thesection}{0.5em}{}

\renewcommand\thesection{\arabic{section}}

\renewcommand\thesubsection{\thesection.\arabic{subsection}}
\titleformat{\subsection}
{\normalfont\itshape}{\thesubsection}{0.5em}{}

\renewcommand\thesubsubsection{\thesection.\arabic{subsection}.\arabic{subsubsection}}
\titleformat{\subsubsection}{\normalfont\itshape}{\thesubsubsection}{0.5em}{}

\usepackage{setspace}

\titlespacing\section{0pt}{10pt plus 4pt minus 2pt}{5pt plus 2pt minus 2pt}
\titlespacing\subsection{0pt}{10pt plus 4pt minus 2pt}{0pt plus 2pt minus 2pt}
\titlespacing\subsubsection{0pt}{10pt plus 4pt minus 2pt}{0pt plus 2pt minus 2pt}

\providecommand{\keywords}[1]{\textbf{Keywords:}  #1}
\providecommand{\JEL}[1]{\textbf{JEL:}  #1}

\makeatletter
\newcommand*{\myfnsymbolsingle}[1]{%
\ensuremath{%
\ifcase#1% 0
\or % 1
*%
\or % 2
\dagger
\or % 3
\ddagger
\or % 4
\mathsection
\or % 5
\mathparagraph
\else % >= 6
\@ctrerr
\fi
}%
}
\makeatother

\usepackage{alphalph}
\newalphalph{\myfnsymbolmult}[mult]{\myfnsymbolsingle}{}

\renewcommand*{\thefootnote}{%
\myfnsymbolmult{\value{footnote}}%
}

\makeatletter
\def\@xfootnote[#1]{%
\protected@xdef\@thefnmark{#1}%
\@footnotemark\@footnotetext}
\makeatother

\usepackage{bbding}         % \FiveStarOpen

\usepackage[symbol]{footmisc}

\usepackage{chngcntr}

\usepackage{scrwfile}
\TOCclone[\contentsname]{toc}{atoc}

\AfterTOCHead[toc]{%
}
\AfterTOCHead[atoc]{%
\edef\maintocdepth{\the\value{tocdepth}}%
\value{tocdepth}=-10000\relax%
}

\usepackage{blindtext}

\makeatletter
\@ifundefined{theorem}{\newtheorem{theorem}{Theorem}[section]}{}
\@ifundefined{lemma}{}{}
\@ifundefined{corollary}{}{}
\@ifundefined{assumption}{}{}
\@ifundefined{proposition}{\newtheorem{proposition}[theorem]{Proposition}}{} % <-- add this
\makeatother

\ifdefined\proof\else
\newenvironment{proof}[1][Proof]{\par\noindent\textbf{#1.}\ }{\hfill$\square$\par}
\fi

\begin{document}

\setcounter{footnote}{0}

\newpage
\title{\Large \setcounter{footnote}{2}Managing Portfolios Across the Return Distribution\thanks{We are grateful to Zoran Ivkovich, Jonathan Lewellen, Antonio Galvao, Bruno Feunou, Jeroen Rombouts, Mykola Babiak, Matej Nevrla, Michael Ellington, Li Xia, Balazs Szorenyi from Yahoo Research, and Alexander Remorov from BlackRock's Systematic Active Equities for valuable discussions and comments. We appreciate the insightful comments from numerous seminar presentations, FinEML 2025, CIML 2025, ``Frontiers of Causal Inference and Machine Learning'', Hungarian Machine Learning Days 2025,. The support of the Czech Science Foundation within the project 24-11555S as well as Charles University GAUK 394825 and Research Centre program No. 24/SSH/020 is gratefully acknowledged. The replication code in \textsf{Python} and setup for the paper is available at \url{https://github.com/Attilasarkany/Q-A2C-Replication}. Note the paper previously circulated with the title ``Tailoring Portfolio Choice via Quantile-Targeted Policies''.}
\vspace{20pt}}

\author{\setcounter{footnote}{11}Jozef Barun\'{i}k\thanks{Institute of Economic Studies, Charles University, Opletalova 26, 110 00, Prague, CR and Institute of Information Theory and Automation, Czech Academy of Sciences , Pod Vodarenskou Vezi 4, 18200, Prague, Czech Republic. E-mail: \url{barunik@fsv.cuni.cz}\,\, Web: \href{https://barunik.github.io/}{barunik.github.io}} \\ {\small\textit{Charles University and}} \\
{\small\textit{Czech Academy of Sciences}}
\and
\setcounter{footnote}{0}Attila S\'ark\'any\thanks{Institute of Economic Studies, Charles University, Opletalova 26, 110 00, Prague, CR and Institute of Information Theory and Automation, Academy of Sciences of the Czech Republic, Pod Vodarenskou Vezi 4, 18200, Prague, Czech Republic. E-mail: \url{95attila.sarkany@gmail.com}} \\
{\small\textit{Charles University and}} \\
{\small\textit{Czech Academy of Sciences}}
\and
\setcounter{footnote}{6}Luk\'{a}\v{s} Jan\'{a}sek\thanks{Institute of Economic Studies, Charles University, Opletalova 26, 110 00, Prague, CR and Institute of Information Theory and Automation, Academy of Sciences of the Czech Republic, Pod Vodarenskou Vezi 4, 18200, Prague, Czech Republic. E-mail: \url{lukas.janasek@fsv.cuni.cz}} \\
{\small\textit{Charles University and}} \\
{\small\textit{Czech Academy of Sciences}}}

%\date{15.6.2026 version}
\date{\today}
% \date{\hspace{2em}}

\maketitle

\begin{abstract}

We develop a dynamic portfolio-choice framework in which investors target the region of the payoff distribution that the portfolio is designed to improve. Out of sample, the estimated policies form an ordered frontier: the policy focused on the downside delivers the strongest left-tail protection and the highest Sharpe ratio, while the policy focused on the upper quantile earns the highest mean return. The gains over volatility-managed portfolios are concentrated in periods when downside-tail dispersion is high. Evidence from fund flows in income, growth and downside protection products supports the interpretation of the quantile index as a reduced-form mandate measure.\vspace{10pt}

\keywords{Tail risk; Dynamic portfolio choice; Quantile preferences; Volatility management; Multifactor investing; Distributional reinforcement learning.}

\JEL{G11, G12, C61, C63}
\end{abstract}

\renewcommand{\thefootnote}{\arabic{footnote}}
\setcounter{footnote}{0}

% \linespread{1.0}
\onehalfspacing
%\doublespacing

%\tableofcontents

% \newpage

% \begin{center}
% \textbf{Conflict-of-interest disclosure statement}
% \end{center}
% \hspace{1cm}\\\textbf{}
%  \noindent Jozef Barun\'{i}k \\
%  \noindent I have nothing to disclose \\

%  \noindent Luk\'{a}\v{s} Jan\'{a}sek \\
%  \noindent I have nothing to disclose \\

%  \noindent Attila S\'ark\'any \\
%   \noindent I have nothing to disclose \\

\section{Introduction}

Volatility-managed portfolios reduce exposure to risk when volatility is high \citep{MoreiraMuir2017,demiguel2024multifactor}. Although this rule is counterintuitive in terms of the risk-return trade-off, it is effective in a Gaussian or location-scale world where conditional means and variances summarise the relevant movement in payoff quantiles \citep{Chamberlain1983}. However, asset returns generally deviate from these assumptions due to heavy tails, asymmetry, conditional skewness, downside risk, and regime dependence \citep{Mandelbrot1963,Fama1965,Cont2001,HarveySiddique2000,AngChenXing2006,GuidolinTimmermann2007}. Consequently, the returns of two different assets may exhibit similar conditional volatility yet differ significantly in terms of their downside tails. The most volatile asset may not necessarily have the worst left tail. An investor who needs to identify potential losses or gains in the relevant payoff region then requires more than a puzzling rule for scaling aggregate exposures. They require an objective that specifies the payoff region that the portfolio is designed to improve.

In our work, the quantile index, defined as $\tau$, will determine the relevant payoff region for the portfolio objective.  Low values of $\tau$ define objectives focused on the downside that protect lower-tail payoffs. Conversely, high values of $\tau$ define objectives that are oriented towards the upside and preserve exposure to favourable outcomes. Intermediate values represent a balance between downside protection and upside participation. The quantile index can be connected to investment mandates that are naturally distributional. Income, capital preservation, buffer, low-volatility and growth products differ in the payoff region they emphasise. Quantile targeting provides a simplified method of mapping such payoff-region objectives onto dynamic factor weights. We later test this interpretation using fund-flow data and find that investors in income, growth and downside-protection products respond differently to various regions of recent fund performance.

\begin{figure}[t]
\begin{center}
\includegraphics[width=0.5\textwidth]{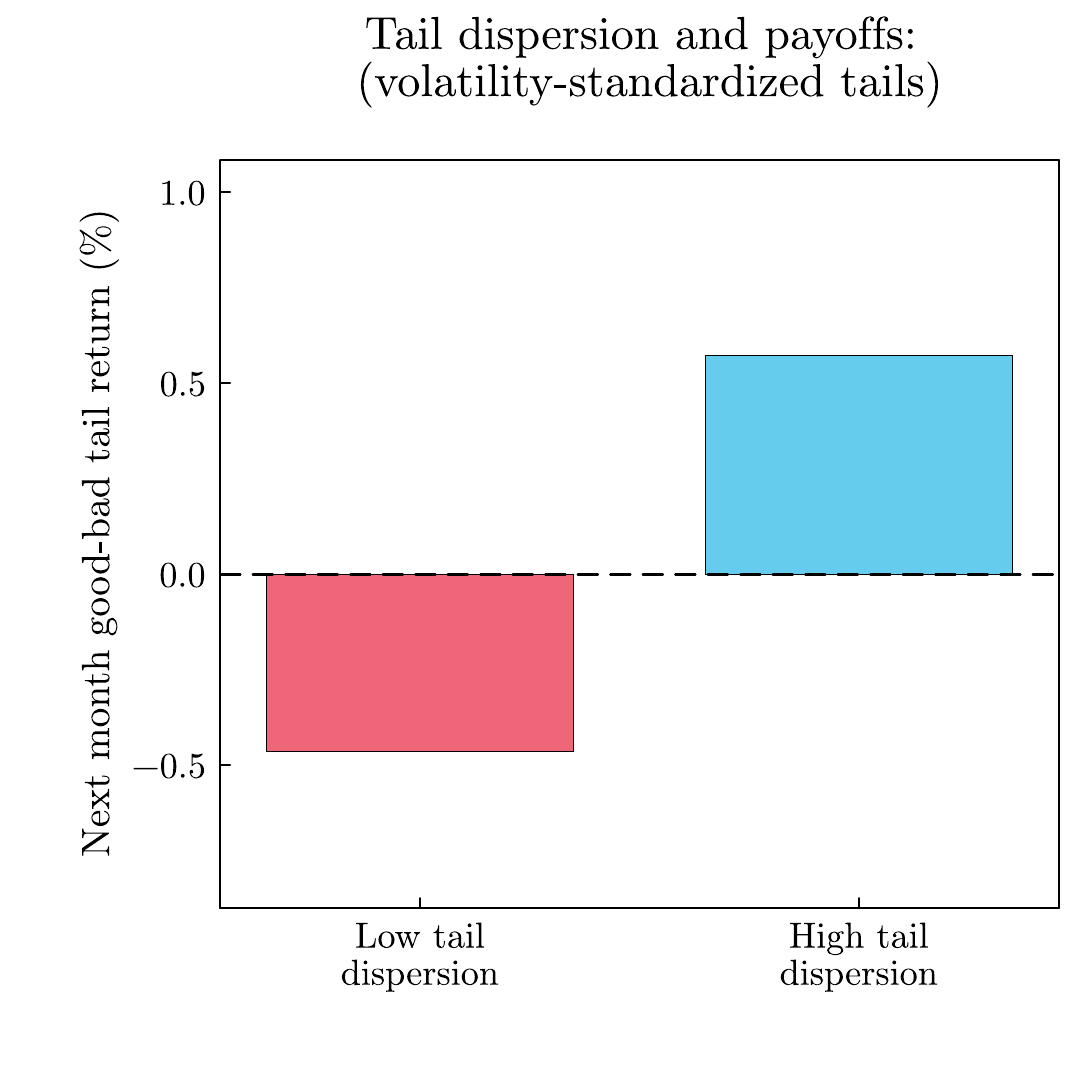}
\end{center}
\caption{\textbf{Tail dispersion and factor-selection payoffs.} This figure reports the next-month return of a good-tail-minus-bad-tail factor portfolio across low and high tail-dispersion states. Each month, factors are sorted by their standardized downside quantile, $Q_{0.10}[r_{i,t}]/\sigma_{i,t}$, where $Q_{0.10}[r_{i,t}]$ is the current-month empirical 10th percentile of daily factor returns and $\sigma_{i,t}$ is current-month factor volatility. The good-tail portfolio equally weights factors in the top tercile of this signal, corresponding to less severe standardized downside tails, while the bad-tail portfolio equally weights factors in the bottom tercile, corresponding to more severe standardized downside tails. The plotted return is the next-month return difference between the good-tail and bad-tail portfolios. Low and high tail-dispersion states are defined by the bottom and top terciles of the cross-sectional dispersion of $Q_{0.10}[r_{i,t}]/\sigma_{i,t}$. The payoff to selecting factors with better standardized downside tails is concentrated in high tail-dispersion states: the high-minus-low difference is 1.04 percentage points. This pattern is consistent with the view that tail dispersion creates the cross-sectional opportunity for selective de-risking.}
\label{fig:tail_selection_payoffs}
\end{figure}

Figure \ref{fig:tail_selection_payoffs} illustrates the distinction. Each month, we rank the returns used in the subsequent empirical analysis according to the severity of their volatility-standardised downside tail, $Q_{0.10}[r_{i,t}]/\sigma_{i,t}$, where $Q_{0.10}[r_{i,t}]$ is the current-month empirical 10th percentile of the $i$th daily factor return, $r_{i,t}$, and $\sigma_{i,t}$ is the current-month factor volatility. We then form a next-month good-tail-minus-bad-tail portfolio based on the severity of the standardised downside tails. As the signal is standardised by factor volatility, this process isolates variation in tail shape rather than only variation in scale. This selection rule results in a negative payoff when cross-sectional tail dispersion is low, but a positive payoff when tail dispersion is high. The difference between high- and low-dispersion states is 1.04 percentage points. Therefore, tail dispersion identifies states in which factor selection based on conditional downside tails is economically relevant beyond volatility timing alone. While volatility signals that the conditional distribution has changed, cross-sectional tail heterogeneity identifies where the change is most severe.

We formalize the ideas using recursive quantile preferences. The recursion evaluates the conditional quantile of current portfolio payoff plus continuation value. This makes the objective dynamic and time consistent while preserving the direct interpretation of $\tau$ as the payoff region targeted by the portfolio. The model is related to Roy’s safety-first logic because lower quantiles summarize shortfall-relevant outcomes \citep{Roy1952}. At the same time, it differs from expected-utility and mean–variance formulations because quantile preferences violate independence, and the targeted payoff region is explicit rather than generated indirectly by curvature.

We argue that when factor payoff distributions are changed asymmetrically, managed portfolios should become quantile specific. Downside-focused policies should reduce exposure most strongly to factors whose conditional left tails deteriorate. Upside-oriented policies should de-risk less aggressively and preserve more exposure to factors with attractive upper-tail payoffs. Quantile targeting then changes both the scale and the composition of risk. It turns volatility management from a scale-based allocation rule into a family of managed factor allocations indexed by the payoff region the portfolio is designed to improve.

We test these predictions using daily returns on eleven investable factors spanning the market, standard style factors, momentum, and industry portfolios. All portfolio rules are estimated in an out-of-sample exercise, where state variables are standardized using only past information. We compare the quantile-targeted portfolios with an unmanaged benchmark, a market-volatility-managed benchmark, and a factor-level volatility-scaled benchmark, where each factor exposure is scaled by its own volatility, testing whether quantile targeting adds information beyond conditional scale.

The empirical results are strongest for downside-focused mandates. The $\tau$ = 0.1 policy has the lowest volatility, the least negative CVaR, the shallowest drawdown profile, and the highest Sharpe ratio among the quantile-targeted portfolios. The $\tau$ = 0.9 policy earns the highest average return but provides weaker downside protection, while the median policy lies between the two. This ordering shows that changing the targeted payoff region changes the dynamic allocation, not only its average risk exposure.

Second, the quantile-targeted portfolios carry additional information in comparison to volatility management. Relative to the market-volatility-managed benchmark, all three quantile-targeted portfolios generate positive annualized alphas. Their betas with respect to the benchmark are below one and increase monotonically with $\tau$. The low-$\tau$ portfolio is a defensive, lower-beta version of volatility management, while the high-$\tau$ portfolio stays closer to the benchmark and preserves more upside exposure.

Third, quantile targeting is most useful when market volatility is high and the cross-sectional dispersion of factor downside tails is large. In such states, market-level volatility timing is too coarse because it provides only one aggregate risk signal. Factor-level volatility scaling adds a cross-sectional rule, but it ranks exposures by conditional scale rather than by deterioration in the targeted payoff region. Quantile targeting uses the cross section of conditional tail risks to determine which factor exposures should be reduced first. The low-$\tau$ portfolio delivers the clearest improvement in its targeted lower tail in these states, and the median portfolio also improves. The high-$\tau$ portfolio preserves more upside exposure, but does not uniformly improve its targeted upper quantile relative to standard volatility timing. The empirical value of the approach is strongest for downside-focused and balanced payoff objectives.

We also provide direct evidence on the mechanism, and relate active factor weights to factor-level downside deterioration, measured using trailing expected shortfall. The estimated slopes are negative and monotone in $\tau$. The downside-focused portfolio cuts exposure most aggressively to factors whose left tails deteriorate the most, while the upside-oriented policy attenuates this response. The gains come from selective cross-sectional reweighting linked to the targeted payoff region.

Next, factor-level scaling absorbs part of the defensive behavior of the low-$\tau$ policy, but does not fully span the quantile-targeted portfolios. We document gains in states in which volatility changes not only the amount of risk, but also the shape and cross-sectional ordering of factor tails. In high-volatility states, the quantile-targeted portfolios trace a downside–upside frontier, while market-level and factor-level volatility-managed benchmarks occupy particular points on the same trade-off map.

The paper contributes to three literatures. First, it identifies a boundary condition for volatility management \citep{MoreiraMuir2017,CederburgODohertyWangYan2020,demiguel2024multifactor}. Market- and factor-level volatility are useful state variables, but they select exposures by conditional scale. This is sufficient in location-scale environments and incomplete when volatility changes the shape and cross-sectional ordering of payoff tails. Second, the paper introduces recursive quantile preferences into dynamic multifactor allocation. The framework connects safety-first portfolio choice to state-dependent factor timing while keeping the targeted payoff region explicit \citep{Roy1952,ArzacBawa1977,Basak1995,BasakShapiro2001,AngChenSundaresan2013,Cochrane2022}. Third, the paper provides out-of-sample evidence that quantile-targeted policies form an ordered downside–upside frontier. Active weights move away from factors whose left tails deteriorate, and the gains are largest in high-volatility states with large dispersion in downside-tail risk connecting to the results of \cite{BarberisHuang2001,RoutledgeZin2010,DahlquistFaragoTedongap2017}.

\section{Why and When Tail Objectives Matter for Dynamic Portfolio Choice}
\label{sec:theory}

We begin by formalizing the portfolio objective that rationalizes the discussion. Standard portfolio models encode risk through utility curvature and evaluate policies through expected utility or mean-variance trade-offs \citep{e5a1bb8f-41b7-35c6-95cd-8b366d3e99bc}. We instead model the portfolio as targeting a chosen quantile of discounted payoffs. The quantile index is a modeling primitive that specifies which part of the payoff distribution the portfolio rule is designed to improve.

Quantiles are useful in portfolio problems that are naturally expressed in terms of floors, shortfall or value-at-risk constraints, solvency thresholds, funding ratios, or liability-relative outcomes rather than the mean of terminal wealth \citep{Roy1952,ArzacBawa1977,Basak1995,BasakShapiro2001,AngChenSundaresan2013,Cochrane2022}. In such settings, the economically relevant question is not only how much risk an investor takes on average, but which part of the payoff distribution the investor wants to protect or exploit.

Once the object of interest is a tail of the distribution rather than symmetric variance alone, portfolio choice becomes naturally quantile-specific. Quantiles encode tail priorities without imposing a particular parametric utility function and, in our application, generate directly testable implications for when and where investors should de-risk. We use this objective as a reduced-form representation of tail-focused mandates. It is related to safety-first portfolio choice, but it is not an expected-utility representation: quantile preferences violate independence.

\subsection{Quantile Preferences as a Reduced-Form Model of Tail-Focused Risk Attitudes}

While quantile preferences were introduced by \citet{manski1988ordinal} and subsequently axiomatized and studied by \citet{chambers2009axiomatization}, \citet{rostek2010quantile}, and \citet{de2019dynamic}, we use them as a convenient tool for mapping tail-focused objectives into dynamic portfolio rules. For a random variable $X$ with cumulative distribution function $F_X$, the $\tau$-quantile is
\begin{equation}
Q_{\tau}[X] = \inf \{x \in \mathbb{R}: F_X(x) \ge \tau\}, \qquad \tau \in (0,1).
\end{equation}
A preference relation $\succeq_{\tau}$ is then a $\tau$-quantile preference if, for some fixed $\tau \in (0,1)$,
\begin{equation}
X \succeq_{\tau} Y \iff Q_{\tau}[u(X)] \ge Q_{\tau}[u(Y)],
\end{equation}
where $u(\cdot)$ is strictly increasing. Because quantiles are invariant under strictly monotone transformations, this is equivalent to
\begin{equation}
X \succeq_{\tau} Y \iff Q_{\tau}[X] \ge Q_{\tau}[Y].
\end{equation}

The parameter $\tau$ has a direct economic interpretation. Lower values of $\tau$ place more weight on downside outcomes and correspond to greater downside focus; higher values of $\tau$ emphasize upside potential. In this sense, $\tau$ is a reduced-form representation of which part of the payoff distribution the investor wants the portfolio to perform well in. Recursive quantile preferences translate heterogeneous tail priorities into heterogeneous timing and allocation rules. They isolate the part of the distribution that matters for the investor's objective and generate clear predictions for when and where de-risking should occur.

\subsection{Relation to Expected Utility: What Is and Is Not New}

Recursive quantile preferences should be viewed as complementary approach to expected-utility and behavioral approaches. Traditional expected utility can rationalize de-risking when volatility rises through concave utility: an investor who dislikes dispersion scales down risk when conditional variance increases. On the other hand, behavioral models can also emphasize asymmetry, reference dependence, disappointment, and downside sensitivity \citep{tversky1992advances,BarberisHuang2001,RoutledgeZin2010,DahlquistFaragoTedongap2017}. In contrast to traditional approaches, we seek a parsimonious, dynamically tractable criterion that makes the targeted part of the payoff distribution explicit. 

The recursion is also related in spirit, but not in substance, to Epstein-Zin recursive utility \citep{EpsteinZin1989,EpsteinZin1991}. Both frameworks make current choices depend on a recursively evaluated continuation object. Epstein-Zin preferences aggregate current consumption and a certainty equivalent of future utility and are used to separate risk aversion from intertemporal substitution. Our recursion instead evaluates the conditional quantile of current portfolio payoff plus continuation value. The parameter $\tau$ indexes the investor's target region of the payoff distribution. In this sense, Epstein-Zin asks how agents trade-off current consumption and uncertain continuation utility, while recursive quantile preferences ask which conditional payoff percentile a portfolio mandate targets.

The following proposition formalizes the von Neumann--Morgenstern non-representation theorem showing why quantile preferences are not a special case of expected utility. The proof (provided by \autoref{app:proofprop1}) is simple, but it is useful because it clarifies why the preference object is genuinely different.

\begin{proposition}[Quantile preferences violate independence]\label{prop:quantile_non_vnm}
Fix $\tau \in (0,1)$ and define preferences over monetary lotteries by $X \succeq_{\tau} Y \quad \Longleftrightarrow \quad Q_{\tau}[X] \geq Q_{\tau}[Y]$. On any rich lottery domain containing finite-support monetary lotteries, $\succeq_{\tau}$ violates the independence axiom. Hence it admits no von Neumann--Morgenstern expected-utility representation on that domain.
\end{proposition}

In contrast to conventional curvature, the economic content carried by recursive quantile preferences is direct selection of the payoff region that drives the decision. The framework yields preference-indexed predictions for (i) how much downside protection is purchased, (ii) how much upside is sacrificed, and (iii) which factors are cut first when high volatility thickens left tails disproportionately. Lower $\tau$ emphasizes downside protection, while higher $\tau$ emphasizes upside potential. This can be seen in the simple one-period problem, in which an investor with initial wealth $W_0$ decides on the proportion of risky asset to allocate based on
\begin{equation}
\max_{\alpha \in [0,1]} Q_\tau(W_1),
\end{equation}
with wealth dynamics
\[
W_1 = W_0\left[\alpha R_1 + (1-\alpha)R_f\right], \qquad \alpha \in [0,1],
\] and the gross return on the risky asset $R_1$, and certain rate $R_f$.

A useful way to interpret this problem is through Roy's safety-first principle \citep{Roy1952}. In Roy's formulation, the investor chooses a portfolio to reduce the probability that terminal wealth falls below a disaster level or minimum acceptable payoff. Assuming continuous return distributions, for any fixed floor $\underline{W}$ and probability level $\tau \in (0,1)$,
\[
\Pr(W_1 \leq \underline{W}) \leq \tau
\quad \Longleftrightarrow \quad
Q_\tau(W_1) \geq \underline{W}.
\]
Thus, maximizing the $\tau$-quantile of wealth can be read as a reduced-form safety-first problem: rather than specifying the disaster threshold directly, the investor chooses the part of the payoff distribution that must be protected \citep{Roy1952}. This interpretation connects our framework to classical shortfall-based portfolio choice while remaining tractable in environments where volatility changes the conditional shape of payoffs asymmetrically across assets. Our framework is also closely related to the generalized safety-first model of \citet{ArzacBawa1977}. Their investor controls the probability that terminal wealth falls below a critical level and, conditional on satisfying that safety requirement, maximizes expected wealth.

We take this lower-tail logic out of a static setting into a dynamic multifactor environment in which volatility or other state variables deform conditional tails asymmetrically across assets. In such a setting, safety-first considerations  determine which factor exposures should be reduced first when the lower tail of continuation wealth deteriorates.

Note that when returns are symmetric and the investor targets the median, quantile-based and mean-based choices can coincide. In general, however, they do not. Quantile choice is useful when the investor cares about downside or upside outcomes that are not well summarized by mean and variance alone.\footnote{The example in \autoref{app:relationclassic} provides a more detailed illustration for the interested reader.} This is also the environment in which the classical justification for mean-variance methods becomes restrictive: with a riskless asset, mean and variance summarize portfolio choice only under strong distributional structure \citep{Chamberlain1983}. Outside Gaussian or location-scale benchmarks, mean and variance do not identify downside tail exposure. Appendix~\ref{app:mv_notails} gives a simple example with the same mean and variance as a Gaussian benchmark but a substantially worse lower quantile, clarifying why a tail-targeted objective can add economic content in our setting, where volatility changes the shape of conditional payoff distributions asymmetrically across portfolios.

Quantile preferences are useful because they isolate which tail deteriorates, by how much, and which portfolios should be cut first. In the static two-asset case, this logic already delivers an useful implication: the investor holds the risky asset if its targeted quantile exceeds the risk-free rate and otherwise retreats to the safe asset. The main novelty of the paper begins once this static comparison is embedded in a dynamic, state-dependent environment. In our setting, high-volatility states worsen left tails much more for some portfolios than for others. Quantile preferences add economic content because they make explicit how tail priorities map into selective, state-contingent portfolio timing. The model should be interpreted as a tail-focused portfolio objective suited to environments with state-dependent asymmetry, not as a universal empirical claim about investor behavior.

\subsection{State-Dependent Interiors}\label{sec:state_dependent_interiors}

The static safety-first problem is useful but incomplete. With one remaining payoff date and a risk-free asset, quantile choice has a simple corner structure. For wealth dynamics
\begin{equation}
W_{t+1}=W_t\left[\alpha_t R_{t+1}+(1-\alpha_t)R_f\right], \qquad \alpha_t\in[0,1],
\end{equation}
maximizing the targeted quantile of next-period wealth is equivalent to comparing the targeted quantile of the risky return with the risk-free rate:
\begin{equation}
\alpha_t^*=\begin{cases}
1, & \text{if } Q_\tau[R_{t+1}\mid s_t]>R_f,\\
0, & \text{if } Q_\tau[R_{t+1}\mid s_t]<R_f,\\
\text{any } \alpha_t\in[0,1], & \text{if } Q_\tau[R_{t+1}\mid s_t]=R_f.
\end{cases}
\end{equation}
Thus, without economically relevant state dependence in continuation values, the quantile objective gives an all-or-nothing comparison between the targeted payoff quantile and the safe payoff.

The dynamic problem becomes different once current choices affect the distribution of continuation payoffs. Let the state be $s_t=(W_t,z_t)$, where $z_t$ is a volatility regime, and let the risky return distribution and transition probabilities depend on $z_t$. The recursive quantile problem is then
\begin{equation}\label{eq:recursive_quantile_state_dependent}
V_\tau(W_t,z_t)=\max_{\alpha_t\in[0,1]}
Q_\tau\left[
W_t\left\{\alpha_t R_{t+1}(z_{t+1})+(1-\alpha_t)R_f\right\}
+\beta V_\tau(W_{t+1},z_{t+1})
\,\middle|\, z_t
\right].
\end{equation}
The important object is the random continuation value $V_\tau(W_{t+1},z_{t+1})$. Because the next regime is uncertain, the continuation payoff cannot generally be taken outside the quantile. The investor is no longer comparing only the one-period risky-return quantile with the risk-free rate. She is choosing how today's exposure changes the conditional distribution of current wealth plus continuation value.

\begin{figure}[t!]
    \begin{center}
        \includegraphics[width=0.4\textwidth]{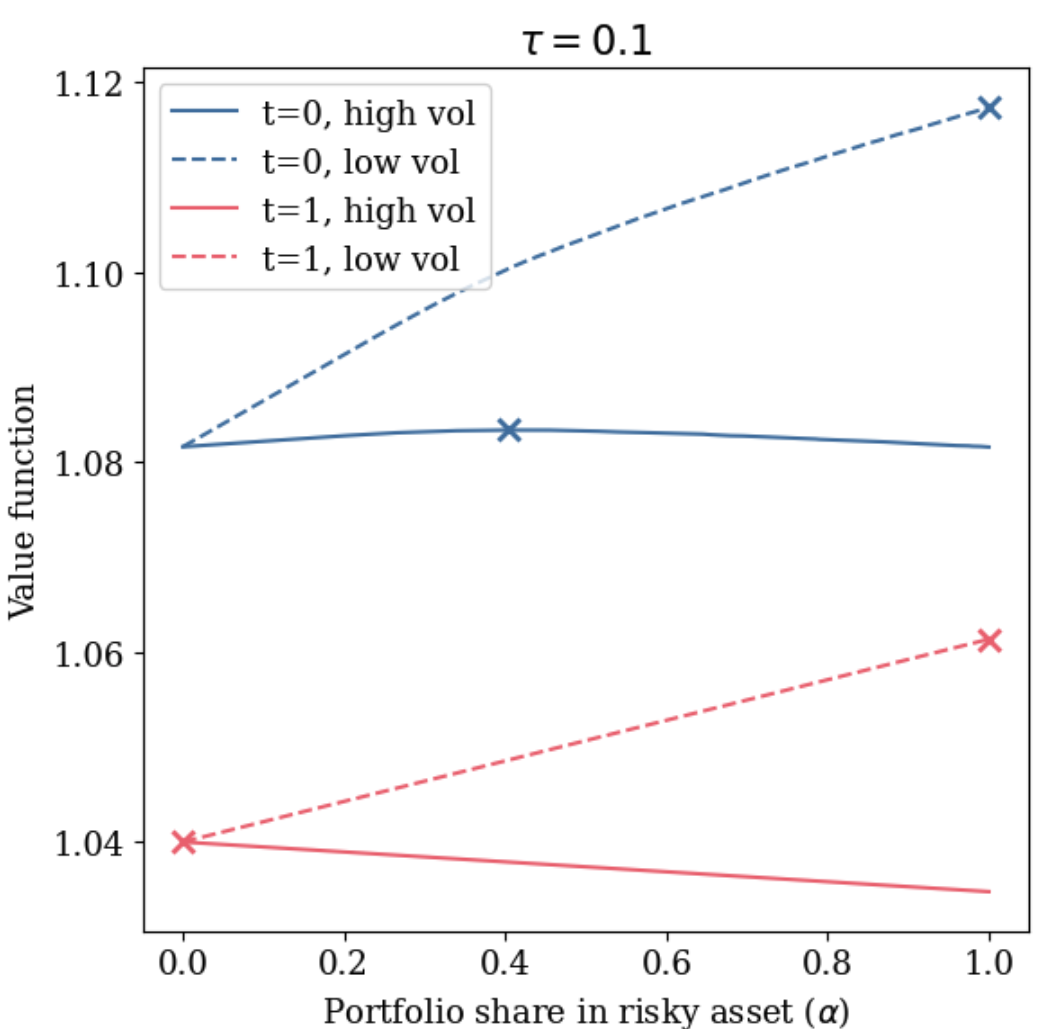}
    \end{center}
    \caption{\textbf{Value functions under volatility regimes and quantile preferences.}
This figure plots the value function as a function of the risky portfolio share in a two-period regime-switching example. Terminal-period value functions are shown in red and initial-period value functions in blue; solid lines correspond to high volatility and dashed lines to low volatility. Crosses denote optimal allocations. The figure illustrates the key dynamic mechanism: once future volatility regimes are uncertain, a downside-focused investor can optimally choose an interior allocation even in a currently high-volatility state.}
\label{fig:valuefunctions_example}
\end{figure}

This is where interior policies arise. A downside-focused investor in a high-volatility state may still hold a positive risky share if a modest exposure preserves the possibility of moving into a better future regime and shifts the targeted lower tail of continuation wealth to the right. At the same time, too much risky exposure widens the current left tail and lowers the targeted quantile. The optimum can therefore be interior: the investor balances the tail cost of current volatility against the continuation value of retaining exposure to favorable future states. The original two-period calculation in Appendix~\ref{app:multi-period-example} is a transparent illustration of this mechanism; the expression with a squared discount factor is specific to that two-period example and is not the general dynamic value function.

Figure~\ref{fig:valuefunctions_example} illustrates the mechanism. In the terminal period, the downside-focused investor follows the static corner logic and avoids the risky asset in the high-volatility state. At the initial date, however, the same investor chooses an interior allocation in the high-volatility state because regime transitions make the continuation value state dependent. A small risky position improves the targeted lower quantile by keeping access to better future regimes, whereas a large risky position again makes the left tail too severe.

This example clarifies why recursive quantiles are an interesting object for portfolio choice. In a static problem, tail-focused investors often retreat fully from risky assets in bad states. In a dynamic problem, they may optimally retain some exposure because state transitions change the payoff distribution they care about. Volatility management then emerges from the investor's tail objective rather than being imposed as an external scaling rule. In a multifactor setting, the same logic applies cross-sectionally: the policy must decide not only how much total risk to take, but which factor exposures preserve or damage the targeted continuation tail.

\section{The Dynamic Quantile Portfolio Problem}

We formulate the investor's problem as a Markov decision process. At time $t$, the investor observes a state $s_t \in \mathcal{S}$ containing exogenous market conditions and endogenous portfolio characteristics, chooses portfolio weights $\alpha_t \in \mathcal{A}(s_t)$, receives a one-period reward $r(\alpha_t,s_t)$, and transitions to a new state $s_{t+1}$.

A naive dynamic quantile objective would maximize the quantile of discounted cumulative rewards,
\begin{equation}
v_{\tau}^{\pi}(s_0)=Q_{\tau}\left[\sum_{t=0}^{\infty}\beta^t r(\alpha_t,s_t)\right],
\end{equation}
but this formulation is generally time-inconsistent because the quantile operator is nonlinear. Following \cite{de2019dynamic,de2025dynamic}, we instead use the dynamically consistent recursive formulation
\begin{equation}
v_{\tau}^{\pi}(s_t)
=
Q_{\tau}\left[r(\alpha_t,s_t)+\beta v_{\tau}^{\pi}(s_{t+1}) \mid s_t\right],
\end{equation}
with optimal value function
\begin{equation}
v_{\tau}^{*}(s_t)
=
\max_{\alpha_t \in \mathcal{A}(s_t)}
Q_{\tau}\left[r(\alpha_t,s_t)+\beta v_{\tau}^{*}(s_{t+1}) \mid s_t\right].
\end{equation}

This recursion is the dynamic counterpart of the reduced-form tail objective above. It says that the investor chooses current portfolio weights to improve the targeted quantile of the sum of current payoff and continuation value. In our application, that continuation value is where state dependence enters: current allocations matter because they reshape future tail risk through the evolution of the state.

The recursive problem is computationally hard. The nonlinearity of the quantile operator generates nested conditional quantiles, so even when the one-period problem is simple, the dynamic problem quickly becomes analytically intractable in realistic multifactor environments with many portfolios, endogenous states, and trading frictions. The next section introduces the tractable method we use to approximate the optimal quantile-targeted policy in high-dimensional settings.

\subsection{Solving the Dynamic Quantile Portfolio Problem}

The previous section shows that the investor's problem is a recursive dynamic quantile problem in which current portfolio choices reshape future tail risk through state transitions. In realistic multifactor environments, this problem has no tractable closed-form solution. The difficulty comes from nested conditional quantiles, high-dimensional state variables, and trading frictions. We use a computational method to approximate the optimal quantile-targeted policy.

Our algorithm adapts an actor-critic reinforcement learning to a dynamic portfolio problem with recursive quantile objectives, portfolio weights constrained to the simplex, endogenous state variables, and trading frictions. The role of the method is tractability: it allows us to study the economic implications of quantile-targeted portfolio choice in environments that are analytically intractable but empirically relevant. The computational ingredients of our method build on existing actor--critic and distributional reinforcement-learning tools, as well as the dynamic quantile-learning framework in \cite{JANASEK2026105355}.

\subsection{A Quantile Actor--Critic Approximation}

For a targeted quantile level $0<\tau<1$, the critic approximates the continuation value of the portfolio problem, while the actor chooses a distribution over feasible portfolio weights. Let $V_{\omega}^{\tau}(s)$ denote the critic approximation to the recursive value function at quantile $\tau$, and let $\pi_{\theta}(\alpha \mid s)$ denote the actor's stochastic policy over admissible portfolio weights $\alpha \in \mathcal{A}(s)$.

Given a state transition $(s_t,\alpha_t,r_{t+1},s_{t+1})$, the critic forms the one-step target
\begin{equation}
y_{t+1}^{\tau}=r_{t+1}+\beta V_{\bar{\omega}}^{\tau}(s_{t+1}),
\end{equation}
where $\bar{\omega}$ denotes a slowly updated target critic. The critic is trained by quantile regression: it updates $V_{\omega}^{\tau}(s_t)$ so that the temporal-difference error
\begin{equation}
\delta_{t}^{\tau}=y_{t+1}^{\tau}-V_{\omega}^{\tau}(s_t)
\end{equation}
tracks the appropriate recursive quantile.

The actor is updated using a quantile-based policy-gradient signal. Actions that improve the investor's targeted quantile are reinforced, while actions that worsen it are discouraged. Because the learning signal is tied directly to the relevant quantile of continuation value, the resulting policies are interpretable: low-$\tau$ investors learn rules that protect downside outcomes, whereas high-$\tau$ investors learn rules that preserve or amplify upside exposure.

First, the objective is a recursively defined, time-consistent quantile value function rather than an expected-return or mean--variance criterion. Second, the action is a portfolio weight vector that must satisfy economically meaningful constraints, including non-negativity and a full-investment budget constraint. Third, the state includes both exogenous market conditions and endogenous portfolio variables, so the policy affects future tail risk through state evolution. Finally, the environment includes transaction costs and portfolio frictions, which are essential for realistic portfolio dynamics.

These features matter for interpretation. The method approximates the solution to a dynamic portfolio problem whose objective is explicitly tied to a chosen part of the payoff distribution. This is what allows the estimated policies to be read economically as downside-focused, median-focused, or upside-focused portfolio rules.

\subsection{Approximation Properties and Implementation Summary}
The recursive quantile formulation in Section 2 defines quantile Bellman operators for both a fixed policy and the optimal control problem. Under standard regularity conditions, these operators are contractions, so the associated recursive quantile value functions are well defined. The critic targets a fixed point of the policy-specific quantile Bellman operator, while the actor updates the policy in the direction that improves the investor's targeted quantile.

In the exact tabular case, these arguments yield standard fixed-point existence and convergence results. Under function approximation, the critic converges to a projected fixed point of the quantile Bellman operator, and the actor converges to a stationary point of the corresponding quantile objective under a standard two-timescale argument.

In the empirical analysis, we use an on-policy actor--critic implementation with a stochastic portfolio policy defined on the simplex. The critic estimates the continuation value, and the actor maps the current state into a distribution over feasible portfolio weights. To stabilize learning, we use a target critic and rolling normalization of state variables. In addition, we train the critic on a grid of quantile levels $0<\tau_1<\cdots<\tau_p<1$ rather than a single quantile $\tau$ to exploit the information from full return distribution. We include monotonicity regularization across the estimated quantiles.

The implementation is deliberately kept parsimonious. We maintain a common set of global hyperparameters across $\tau$, use multiple random seeds for robustness, and evaluate performance out of sample. Details on network architecture, feature scaling, training schedules, pseudocode, and hyperparameter choices are reported in \autoref{app:qac_details}.

This computational approach makes it possible to solve the recursive dynamic quantile portfolio problem in realistic environments with many portfolios and frictions. \autoref{app:simdynexample} studies its properties in a stylized regime-switching setting to illustrate the mechanism before turning to the empirical portfolio applications.

\section{Out-of-Sample Evidence}

This section assesses the performance of tail-managed policies when applied to new data. We expect the estimated policies to constitute a family of managed portfolios indexed by preference, whereby low-$\tau$ allocations behave more defensively and high-$\tau$ allocations preserve more upside. These portfolios  should also identify new information in relation to volatility management.

\subsection{Data}

Our universe comprises a cross-section of eleven investable factors spanning the market, style, and industry. We use a richer opportunity set than \cite{demiguel2024multifactor} because the main mechanism is inherently about selective reallocation across factors rather than uniform scaling of aggregate risk. Specifically, we use the daily data of excess returns for the market (MKT),  small-minus-big (SMB), high-minus-low (HML), robust-minus-weak (RMW), conservative- minus-aggressive (CMA), and momentum (MOM) factors of \cite{fama2018choosing}, and mix them with cyclical and defensive industries where dispersion and downside potential differ: Manufacturing (Manuf), Business Equipment (HiTec), Consumer Durables (Cnsmr),  Health (Health) and Other\footnote{containing Mines, Constr, BldMt, Trans, Hotels, Bus Serv, Entertainment, Finance} (Other). Every factor except MKT is the return of a long-short portfolio of stocks with one dollar in the long leg and one dollar in the short leg. The MKT factor is also a zero-cost portfolio because the long market position is financed by a short position in the risk-free asset.

\begin{figure}[t]
    \centering
    \includegraphics[width=\textwidth]{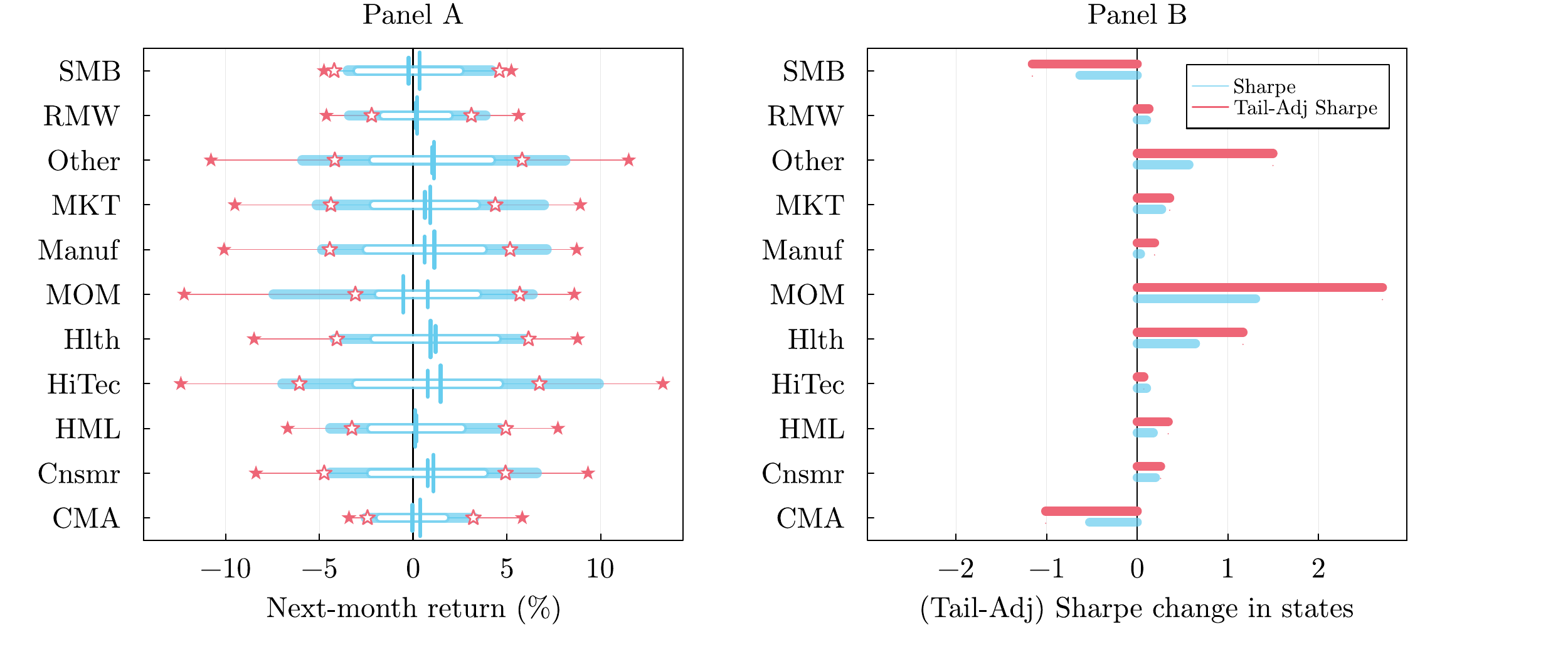}
    \caption{\textbf{Factor tails and performance across market-volatility states.}
    Panel A compares the next-month return distribution of each factor following low- and high-market-volatility months. Blue intervals show mean plus or minus one standard deviation, while red intervals show the 5th-to-95th percentile range. Solid intervals correspond to high-volatility states and hollow intervals correspond to low-volatility states. Panel B reports, for each factor, the change from low- to high-volatility states in the conventional Sharpe ratio and in a tail-adjusted Sharpe ratio. The larger changes in the tail-adjusted Sharpe ratio indicate that high-volatility states affect not only dispersion around the mean but also the tails of factor payoff distributions. The figure motivates the paper's focus on cross-sectional tail heterogeneity: volatility identifies turbulent states, but factor-level tail deterioration determines which exposures are most costly for a tail-focused investor.}
    \label{fig:vol_tail_data}
\end{figure}

Figure~\ref{fig:vol_tail_data} provides descriptive evidence that tail risk is not a simple rescaling of volatility in the factor cross section. Panel A compares the next-month return distributions of each factor following low- and high-market-volatility months. High-volatility states generally widen return distributions, but the widening is not uniform across factors and is often more pronounced in the tails than around the center of the distribution. Panel B shows that the deterioration in performance from low- to high-volatility states is larger when performance is measured using a tail-sensitive Sharpe ratio than when it is measured using the conventional Sharpe ratio. The figure illustrates the data feature that motivates the empirical design: market volatility identifies turbulent states, but the cross-sectional tail response determines which factor exposures become most costly for a tail-focused investor.

Two features are important for the analysis below. First, the distance between low- and high-volatility distributions varies substantially across factors, so a common market-volatility signal does not imply a common change in factor-level risk. Second, the tail-adjusted performance deterioration is often larger than the deterioration in the conventional Sharpe ratio. This pattern suggests that the relevant empirical object is not only conditional volatility, but the way volatility changes factor-level payoff tails. The tail-managed portfolios below are designed to exploit  this cross-sectional heterogeneity.

We use daily returns because the object of interest is a conditional quantile. Monthly returns provide too few observations to estimate lower-tail behavior reliably in expanding windows. The use of daily data is less natural for expected-return prediction, where signals are low-frequency, but it is appropriate for measuring changes in conditional tail shape. The same qualitative conclusions hold for a standard six-factor universe and for expanded universes that include market-neutral factors such as betting against beta and quality, as well as tail-shape strategies such as PUT and PPUT. To preserve focus and space, we make these results available in \autoref{ref:add_11alt_oos} as the results are materially similar.

\subsection{Empirical design and benchmarks}
\label{sec:empirical_design}

Reported portfolio returns are generated from an expanding-window exercise. First, we estimate the portfolio rule for a given $\tau$-quantile target over an initial five-year estimation window. Then, we evaluate the resulting policy over the subsequent two-year out-of-sample period. We then expand the estimation window to include the newly observed data, re-estimate the model, and evaluate performance on the next two-year out-of-sample block. Repeating this procedure sequentially yields a concatenated out-of-sample return series for each managed portfolio. This design eliminates look-ahead bias and enables the estimated policy to adapt gradually as the conditional distribution of returns evolves across market environments.\footnote{Note that we also experimented with a ten-year initial window and one-year out-of-sample blocks. The results do not change materially.}

To limit the scope for implicit data mining, we maintain a common global architecture across target quantiles, standardise state variables using rolling statistics computed only from past data, and assess robustness across multiple random seeds.

We compare three tail-managed portfolios, corresponding to $\tau \in \{0.1,0.5,0.9\}$,
with two benchmarks. The first, UC, is the unmanaged portfolio: a static, long-only, normalized fully invested Markowitz portfolio estimated on the training sample. The second, VM, is the
market-volatility-managed benchmark. In the spirit of \cite{MoreiraMuir2017,demiguel2024multifactor}, VM links risky exposure to lagged realised market volatility, reducing exposure in turbulent states and restoring it when volatility is low. Specifically, raw benchmark factor exposures are parameterised as affine functions of inverse lagged market volatility,
\[
\theta_{k,t}=a_k+b_k\frac{1}{\sigma^M_t},
\]
where $\sigma^M_t$ denotes lagged realized market volatility.
This is the economically relevant
benchmark for the main text because it isolates the paper's main question: whether making
the investor's objective tail-specific changes the form of volatility management itself, rather
than merely its average aggressiveness.

To keep the comparison economically meaningful, we implement VM under the same
investability discipline as the tail-managed portfolios. Benchmark parameters are
estimated on the training sample only, the implied exposures are mapped into tradable
long-only fully invested weights, and portfolio performance is evaluated net of transaction
costs computed from turnover in drifted pre-trade weights. The comparison is between two implementable managed portfolios that differ in the information they use and the objective they optimize.

Section \ref{sec:factorVM} considers a harder benchmark that replaces the common market-volatility signal with factor-level volatility scaling. Because that benchmark already permits cross-sectional
de-risking, it is deliberately demanding and is best interpreted as a robustness comparison
rather than as the main economic benchmark. Detailed construction of both volatility-management
benchmarks is reported in Appendix \ref{app:vm_implementation}.

\subsection{A Preference-Indexed Family of Managed Portfolios}

\begin{figure}[ht]
    \begin{center}
        \includegraphics[width=0.8\textwidth]{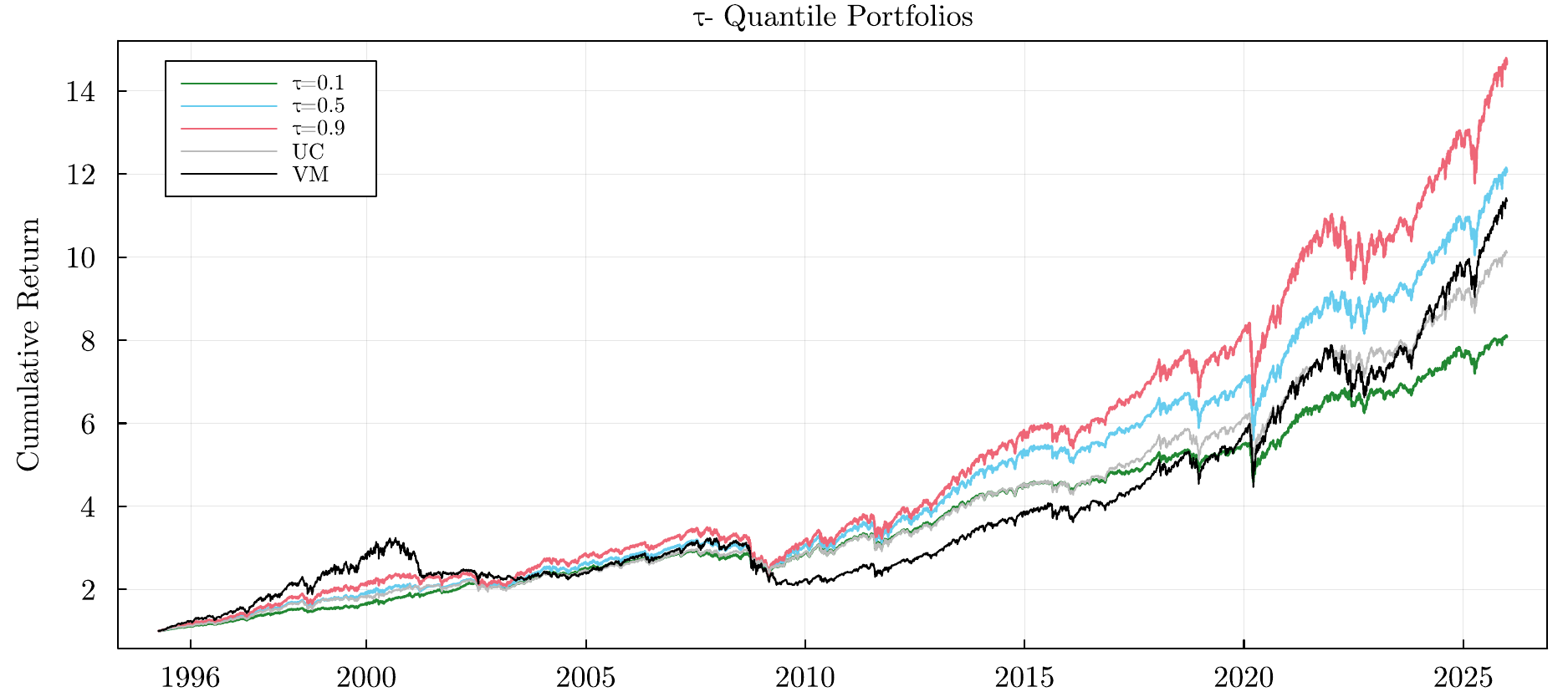}
    \end{center}
    \caption{\textbf{Out-of-sample cumulative returns of tail-managed portfolios.}
This figure plots cumulative out-of-sample returns for the unmanaged benchmark (UC), the volatility-managed benchmark (VM), and the tail-managed portfolios with $\tau \in \{0.1,0.5,0.9\}$. The three tail-managed portfolios form a preference-indexed family: the low-$\tau$ portfolio is the most defensive, the high-$\tau$ portfolio preserves the most upside, and the intermediate portfolio lies between them.}
\label{fig:q_portfolios}
\end{figure}

Figure~\ref{fig:q_portfolios} provides the compact overview of the out-of-sample behavior of the estimated policies. The three tail-managed portfolios move together through time, but they do so in an ordered manner. The low-$\tau$ portfolio is the most defensive, the high-$\tau$ portfolio captures the most upside, and the intermediate portfolio lies between them. The estimated policies should be interpreted as a family indexed by investor tail priorities rather than as isolated timing rules.

\begin{table}[ht]
\centering
\small
\caption{\textbf{Out-of-sample performance of tail-managed portfolios.}}
\label{tab:oos_summary}  
\begin{tabular}{llllll}
    \toprule
     & UC & VM & $\tau = 0.1$ & $\tau = 0.5$ & $\tau = 0.9$ \\
    \cmidrule{2-6}
    Annualized mean (\%)  & 7.96 & 8.81 & 7.10 & 8.63 & 9.46 \\
    Annualized standard deviation (\%) & 9.21 & 13.32 & 7.59 & 10.11 & 11.92 \\
    CVaR at 5\% (\%)   & -1.36 & -2.02 & -1.13 & -1.51 & -1.79 \\
    Maximum drawdown (\%)  & 21.39 & 35.26 & 21.28 & 27.02 & 33.34 \\
    Sharpe ratio (annualized)  & 0.87 & if y & 0.94 & 0.85 & 0.79 \\
    Sortino ratio (annualized)  & 0.82 & 0.63 & 0.89 & 0.81 & 0.75 \\
    Tail-adjusted Sharpe ratio & 9.13 & 7.02 & 9.98 & 9.14 & 8.42 \\\bottomrule
  \end{tabular}
  \begin{tablenotes}
\footnotesize
\item \textit{Notes:} This table reports out-of-sample performance for the unmanaged benchmark (UC), the volatility-managed benchmark (VM), and the tail-managed portfolios with $\tau \in \{0.1,0.5,0.9\}$. Annualized mean returns and annualized standard deviations are reported in percent. CVaR 5\% is the average return in the worst 5\% of monthly outcomes. Max DD and Avg DD denote maximum and average drawdown, respectively, computed from the cumulative return series. Sharpe ratios are annualized. Tail-Adj Sharpe (CVaR5) is the annualized mean return divided by the absolute value of CVaR 5\%. Lower-$\tau$ portfolios are designed to protect downside outcomes, whereas higher-$\tau$ portfolios preserve more upside. Bold entries indicate the best value in each row.
\end{tablenotes}
  \label{tab:oos_summary}
\end{table}

Table~\ref{tab:oos_summary} confirms that interpretation. The $\tau = 0.1$ portfolio is the defensive allocation that delivers the highest annualized Sharpe ratio, the strongest left-tail protection, and the shallowest drawdown profile. In addition to the strongest Sharpe ratio, it has the least negative CVaR, and the lowest maximum drawdown among the managed portfolios. By contrast, the $\tau = 0.9$ portfolio is the upside-oriented allocation earning the highest annualized mean return, but it does so with materially weaker downside outcomes, including a more negative CVaR and substantially deeper drawdowns. The $\tau = 0.5$ portfolio lies naturally between these two extremes. The ordering is coherent: lower $\tau$ buys downside protection, higher $\tau$ preserves upside, and the middle policy reflects an intermediate trade-off. It is important to note that the relatively low Sharpe ratio of VM reflects its high and noisy volatility measurement in daily data. This is also consistent with the literature showing that volatility-managed portfolios do not systematically outperform unmanaged portfolios in direct Sharpe comparisons \citep{CederburgODohertyWangYan2020}, especially in multifactor setting \citep{demiguel2024multifactor} with transaction costs.

The results show that recursive quantile preferences generate a preference-indexed menu of managed portfolios with distinct and interpretable tail profiles. Once the investor objective is tied to a particular part of the payoff distribution, the empirical output should be a family of allocations rather than a single universally optimal rule.

\subsection{Relation to Standard Volatility Management}

The next question is how these policies are genuinely related to standard volatility management. Table~\ref{tab:horse_vm} shows that relative to the volatility-managed benchmark, all three tail-managed portfolios earn positive annualized alpha. At the same time, all three have beta below one, and those betas increase monotonically with $\tau$. The low-$\tau$ portfolio has the smallest beta and the largest defensive tilt; the high-$\tau$ portfolio has the largest beta and lies closest to the volatility-managed benchmark. The $R^2$ values are also high throughout.

The tail-managed portfolios remain linked to volatility management in return space, but investor preferences systematically tilt that benchmark. Low-$\tau$ portfolios behave as lower-beta, downside-focused variants of volatility management, whereas high-$\tau$ portfolios preserve more of the benchmark's upside exposure. Tail targeting can be interpreted as a preference-specific distortion of volatility management.

The gains relative to VM show that the strategy earns alpha while loading less on the volatility-managed benchmark, especially at low $\tau$. This is expected if the investor objective alters not only the scale of risk, but also its composition across factors.

% Auto-generated by Julia
\begin{table}[!htbp]
\centering
\caption{\textbf{Comparison against the volatility-managed benchmark.}}
\label{tab:horse_vm}
\small
\setlength{\tabcolsep}{5pt}
\begin{tabular}{lrrrrrrrrrrrrrr}
\toprule
Strategy & Alpha & SE($\alpha$) & $t(\alpha)$ & Beta & SE($\beta$) & $t(\beta)$ & $R^2$ & Mean & TE & IR & $\Delta$Sharpe & $\Delta$CVaR5 \\
\midrule
$\tau$ = 0.1 & 2.77 & 0.72 & 3.83 & 0.49 & 0.01 & 43.22 & 0.743 & -1.71 & 7.79 & -0.22 & 0.27 & 0.89 \\
$\tau$ = 0.5 & 2.47 & 0.74 & 3.34 & 0.70 & 0.01 & 61.77 & 0.848 & -0.18 & 5.62 & -0.03 & 0.19 & 0.51 \\
$\tau$ = 0.9 & 2.16 & 0.85 & 2.54 & 0.83 & 0.01 & 68.45 & 0.858 & 0.66 & 5.03 & 0.13 & 0.13 & 0.23 \\
\bottomrule
\end{tabular}
\begin{tablenotes}
\footnotesize
\item \textit{Notes:} This table compares each tail-managed portfolio with the volatility-managed benchmark (VM). Alpha and beta are estimated from the daily time-series regression
\[
r_{p,t}=\alpha+\beta r_{VM,t}+\varepsilon_t,
\]
where reported alpha is annualized and expressed in percent. Standard errors are Newey--West with lag length $L=5$. $R^2$ measures the fraction of return variation explained by VM. Active Mean is the average daily return difference relative to VM, TE is annualized tracking error, and IR is the corresponding information ratio. $\Delta$Sharpe and $\Delta$CVaR5 report the difference in annualized Sharpe ratio and CVaR 5\% relative to VM. A beta below one combined with positive alpha indicates that the tail-managed portfolios should be interpreted as preference-specific tilts of volatility management rather than unrelated strategies.
\end{tablenotes}
\end{table}

\subsection{State Dependence: Where the Gains Occur}

Through the text we argue that the value of tail targeting should be state dependent. Tail targeting should matter most when volatility changes the shape of the conditional return distribution asymmetrically across factors. Table~\ref{tab:state_perf} supports that prediction.

\begin{table}[htbp]
\centering
\caption{\textbf{Conditional out-of-sample performance by volatility and market-tail states.}}
\label{tab:state_perf}
\footnotesize
\setlength{\tabcolsep}{6pt}

\begin{tabular}{lccccc}
\toprule
Metric & UC & VM & $\tau=0.1$ & $\tau=0.5$ & $\tau=0.9$ \\
\midrule

\multicolumn{6}{l}{\textit{Panel A: Volatility states}} \\
\addlinespace[0.25em]
\multicolumn{6}{l}{\quad \textit{Low volatility state}} \\
Ann. Mean (\%) & 3.31 & \textbf{6.29} & 2.32 & 3.07 & 3.57 \\
Sharpe (ann.) & 0.55 & \textbf{0.78} & 0.47 & 0.48 & 0.49 \\
Sortino (ann.) & 0.69 & \textbf{0.97} & 0.60 & 0.60 & 0.60 \\
CVaR 5\% (\%) & -0.96 & -1.26 & \textbf{-0.78} & -1.03 & -1.19 \\
Tail-Adj Sharpe (CVaR5) & 3.46 & \textbf{4.99} & 2.95 & 2.99 & 3.00 \\

\addlinespace[0.4em]
\multicolumn{6}{l}{\quad \textit{High volatility state}} \\
Ann. Mean (\%) & 8.61 & 8.17 & 6.92 & 10.05 & \textbf{11.49} \\
Sharpe (ann.) & 0.59 & 0.39 & 0.57 & \textbf{0.62} & 0.60 \\
Sortino (ann.) & 0.80 & 0.53 & 0.75 & \textbf{0.83} & 0.82 \\
CVaR 5\% (\%) & -2.07 & -3.03 & \textbf{-1.81} & -2.38 & -2.79 \\
Tail-Adj Sharpe (CVaR5) & 4.15 & 2.70 & 3.83 & \textbf{4.22} & 4.12 \\

\addlinespace[0.4em]
\midrule
\multicolumn{6}{l}{\textit{Panel B: Tail states}} \\
\addlinespace[0.25em]
\multicolumn{6}{l}{\quad \textit{Downside market state}} \\
Ann. Mean (\%) & 11.98 & \textbf{18.43} & 10.11 & 13.92 & 14.70 \\
Sharpe (ann.) & 0.84 & \textbf{0.89} & 0.87 & 0.88 & 0.79 \\
Sortino (ann.) & 1.13 & \textbf{1.21} & 1.12 & 1.18 & 1.07 \\
CVaR 5\% (\%) & -2.05 & -2.99 & \textbf{-1.72} & -2.30 & -2.73 \\
Tail-Adj Sharpe (CVaR5) & 5.86 & \textbf{6.16} & 5.86 & 6.06 & 5.39 \\

\addlinespace[0.4em]
\multicolumn{6}{l}{\quad \textit{Upside market state}} \\
Ann. Mean (\%) & 5.22 & 1.44 & 5.38 & 5.64 & \textbf{6.83} \\
Sharpe (ann.) & 0.70 & 0.14 & \textbf{0.87} & 0.69 & 0.70 \\
Sortino (ann.) & 0.93 & 0.18 & \textbf{1.17} & 0.90 & 0.93 \\
CVaR 5\% (\%) & -1.14 & -1.65 & \textbf{-0.94} & -1.26 & -1.49 \\
Tail-Adj Sharpe (CVaR5) & 4.58 & 0.87 & \textbf{5.74} & 4.47 & 4.58 \\
\bottomrule
\end{tabular}

\vspace{0.35em}

\noindent
\begin{minipage}{\textwidth}
\footnotesize
\textit{Notes:} This table reports conditional out-of-sample performance for UC, VM, and the tail-managed portfolios with $\tau \in \{0.1,0.5,0.9\}$. In Panel A, volatility states are defined using lagged 21-day realized market volatility; low- and high-volatility states correspond to the bottom and top quintiles of the volatility distribution. In Panel B, market-tail states are defined using lagged 21-day compounded market returns; downside and upside states correspond to the bottom and top quintiles of the return distribution. Annualized mean returns, Sharpe ratios, and Tail-Adj Sharpe (CVaR5) are reported within each state. CVaR 5\% is the average return in the worst 5\% of observations within the corresponding state. The table shows where the gains from tail targeting arise: VM is strongest in calm states, whereas tail-managed portfolios become more competitive when volatility is high or when market states make tail considerations more relevant. Bold entries indicate the best value in each row.
\end{minipage}

\end{table}

In low-volatility states, the volatility-managed benchmark performs well on conventional metrics. Its mean return and Sharpe ratio exceed those of the tail-managed portfolios, and the case for additional tail-sensitive timing is correspondingly weak. This is an important result rather than an inconvenience. In calm states, when conditional tails are less distorted, the theory does not predict large gains from a tail-specific overlay. At the same time, $\tau = 0.1$ portfolio delivers best protection to losses even in this state.

% Auto-generated by Julia
\begin{table}[!htb]
\centering
\caption{\textbf{Comparison against VM across states}}
\label{tab:vm_horserace_states}
\footnotesize
\begin{threeparttable}
\begin{tabular}{llrrrrrrrrrr}
\toprule
Strategy & State & Alpha & SE($\alpha$) & $t(\alpha)$ & Beta & SE($\beta$) & $t(\beta)$ & $R^2$ & Mean & IR & $\Delta$CVaR5 \\
\midrule
& \multicolumn{11}{l}{\textit{Panel A: Volatility states}} \\
\cmidrule{2-12}
\addlinespace[0.25em]
$\tau$ = 0.1 & Low  & -1.22 & 0.86 & -1.41 & 0.56 & 0.01 & 41.91 & 0.820 & -3.97 & -0.97 & 0.47 \\
$\tau$ = 0.5 & Low  & -1.65 & 0.83 & -1.99 & 0.75 & 0.01 & 54.33 & 0.895 & -3.22 & -1.12 & 0.23 \\
$\tau$ = 0.9 & Low  & -1.93 & 0.85 & -2.26 & 0.87 & 0.01 & 65.88 & 0.916 & -2.72 & -1.15 & 0.07 \\
\addlinespace[0.3em]
$\tau$ = 0.1 & High &  2.86 & 2.34 &  1.22 & 0.50 & 0.02 & 25.23 & 0.753 & -1.25 & -0.10 & 1.22 \\
$\tau$ = 0.5 & High &  4.22 & 2.46 &  1.72 & 0.71 & 0.02 & 37.47 & 0.853 &  1.88 &  0.22 & 0.65 \\
$\tau$ = 0.9 & High &  4.57 & 2.89 &  1.58 & 0.85 & 0.02 & 40.80 & 0.857 &  3.32 &  0.42 & 0.24 \\
\cmidrule{2-12}
& \multicolumn{11}{l}{\textit{Panel B: Tail states}} \\
\cmidrule{2-12}
\addlinespace[0.25em]
$\tau$ = 0.1 & Downside &  0.77 & 2.10 &  0.37 & 0.51 & 0.02 & 27.73 & 0.815 & -8.32 & -0.73 & 1.27 \\
$\tau$ = 0.5 & Downside &  0.62 & 1.99 &  0.31 & 0.72 & 0.02 & 41.91 & 0.908 & -4.50 & -0.60 & 0.69 \\
$\tau$ = 0.9 & Downside & -1.15 & 2.22 & -0.52 & 0.86 & 0.02 & 48.82 & 0.917 & -3.73 & -0.61 & 0.26 \\
\addlinespace[0.3em]
$\tau$ = 0.1 & Upside   &  4.75 & 1.68 &  2.82 & 0.44 & 0.03 & 17.37 & 0.540 &  3.94 &  0.55 & 0.72 \\
$\tau$ = 0.5 & Upside   &  4.72 & 1.94 &  2.44 & 0.64 & 0.03 & 19.72 & 0.650 &  4.20 &  0.69 & 0.39 \\
$\tau$ = 0.9 & Upside   &  5.73 & 2.30 &  2.50 & 0.76 & 0.04 & 19.68 & 0.657 &  5.39 &  0.87 & 0.16 \\
\bottomrule
\end{tabular}
\begin{tablenotes}[flushleft]
\footnotesize
\item[] \textit{Notes:} Each row reports the regression
$r_{p,t}=\alpha+\beta r_{VM,t}+\varepsilon_t$
estimated within the indicated state or crisis window. Reported alpha is annualized
and expressed in percent. Standard errors are Newey-West. State definitions follow
Table~\ref{tab:state_perf}. Crisis windows correspond to the dot-com bust, the global
financial crisis, the COVID crash, and the 2022 bear market. Because these windows
are short, alpha estimates are noisier than in the full sample and should be interpreted
as supportive rather than definitive evidence.
\end{tablenotes}
\end{threeparttable}
\end{table}

In high-volatility states, the tail-managed portfolios become more competitive, and in several cases outperform the benchmark on the measures that are most relevant for the paper's mechanism. The $\tau$ portfolios improve on VM in terms of Sharpe and tail-adjusted Sharpe, and the $\tau = 0.1$ portfolio delivers the strongest downside protection, as reflected in its less negative CVaR. Thus, once the environment turns volatile, the preference-indexed trade-off becomes economically important: the investor can buy better downside protection, or preserve more upside, by moving away from a volatility-timing rule.

A similar pattern appears when we condition on lagged market returns. In downside market states, the benchmark remains competitive, but in upside market states the tail-managed portfolios dominate VM much more clearly. All three tail-managed portfolios deliver substantially stronger mean returns than VM in these states, and the low-$\tau$ allocation combines that improvement with the strongest Sharpe ratio and tail-adjusted Sharpe. This pattern shows that the model is not merely crash insurance. It also reallocates toward factors whose upper-tail payoffs remain attractive when the state becomes favorable.

Table~\ref{tab:vm_horserace_states} supports the previous discussion ( Table \ref{tab:vm_horserace_crisis} in \autoref{app:OOSresults_additional} shows additional results for crisis periods). Overall, we provide evidence that in calm states when tail-specific heterogeneity is weak, the volatility-managed benchmark remains a competitive rule, although it delivers weaker downside protection and a different portfolio composition. In contrast, when volatility is high or market conditions make tail asymmetries more relevant, the tail-managed portfolios become more competitive and outperform the benchmark on several mandate-relevant dimensions The value of tail targeting is state dependent because the underlying tail distortions are state dependent.

The results suggest that tail-managed portfolios are similar to volatility managed portfolios but carry significantly higher alpha in states of high cross-sectional heterogeneity. The preceding spanning regressions are unconditional within coarse states. To see if there are periods driving this relation, we next estimate a local time-varying version,
\begin{equation}
    r^{\tau}_{p,t} = \alpha_t + \beta_t r_{VM,t} + \varepsilon_t,
\end{equation}
where both $\alpha_t$ and $\beta_t$ are allowed to vary smoothly over the out-of-sample period. This exercise follows the conditional performance-evaluation logic that managed portfolios should be assessed relative to the information set and state variables that drive their exposures \citep{FersonSchadt1996,FersonHarvey1999}.

\begin{figure}[ht]
    \begin{center}
        \includegraphics[width=0.82\textwidth]{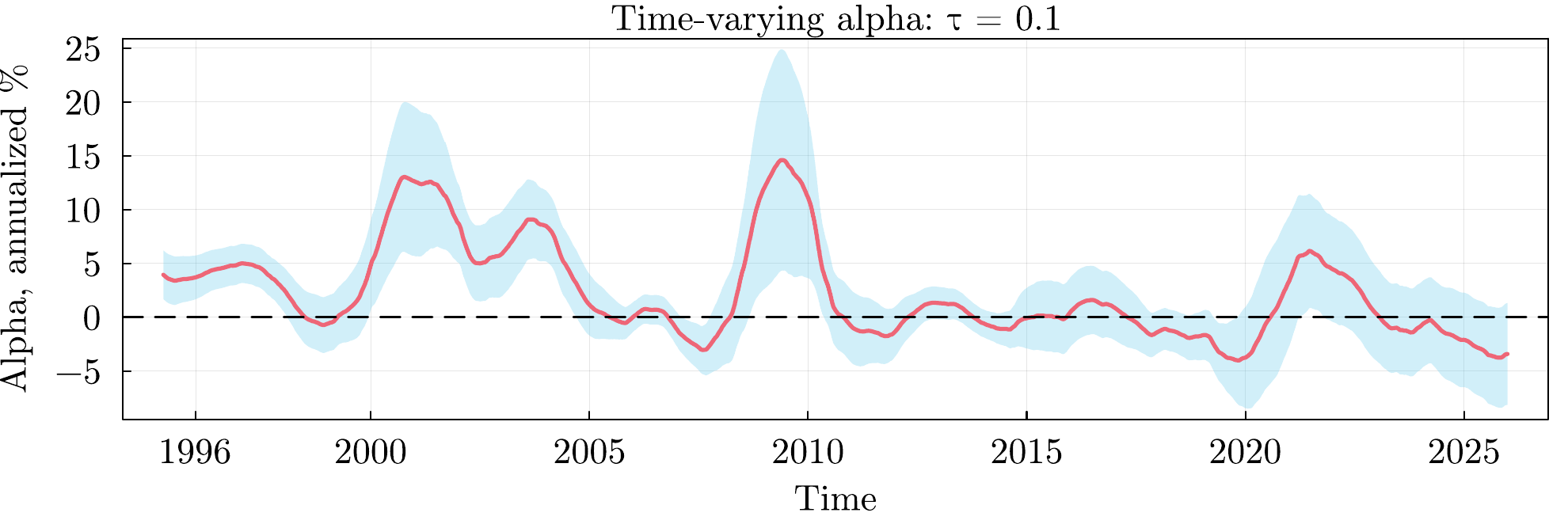}
        \includegraphics[width=0.82\textwidth]{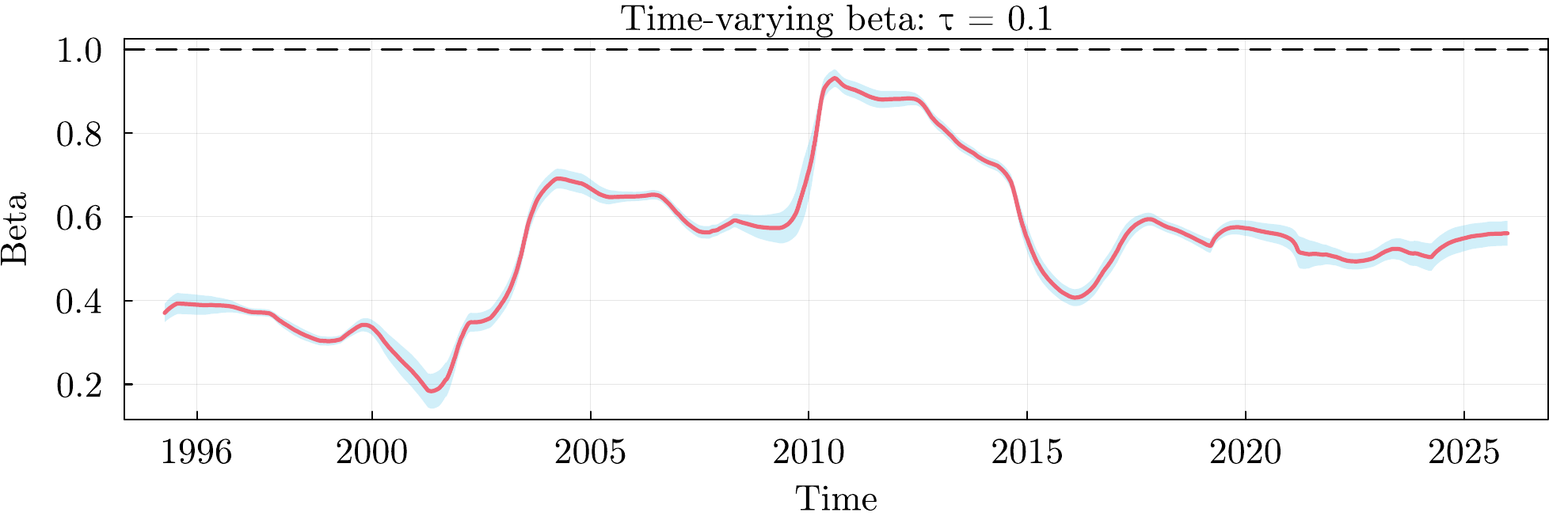}
    \end{center}
    \caption{\textbf{Conditional spanning by the volatility-managed benchmark.}
    This figure reports local time-varying estimates from the regression
    $r^{0.1}_{p,t}=\alpha_t+\beta_t r_{VM,t}+\varepsilon_t$ for the downside-focused
    tail-managed portfolio. The top panel plots the annualized alpha, expressed in
    percent, and the bottom panel plots the beta with respect to the market-volatility-managed
    benchmark. Shaded areas denote pointwise confidence bands based on local
    Newey--West standard errors. The dashed horizontal line in the top panel corresponds
    to zero alpha, and the dashed horizontal line in the bottom panel corresponds to unit
    beta. A beta persistently below one together with positive alpha in selected periods
    indicates that the low-$\tau$ portfolio is a defensive, preference-specific tilt of
    volatility management rather than a strategy fully spanned by the VM benchmark.}
    \label{fig:conditional_spanning}
\end{figure}

Figure~\ref{fig:conditional_spanning} reports the time-varying alpha and beta for the downside-focused portfolio, $\tau=0.1$. The beta is always below one and remains far from the unit benchmark, confirming that the low-$\tau$ policy behaves as a defensive, lower-beta version of volatility management whose exposure changes over time. At the same time, the alpha is episodic rather than constant. It is positive and economically large in periods when market conditions make cross-sectional tail differences especially relevant, including the early 2000s, the global financial crisis period, and the post-COVID volatility episode. These are the states in which the theory predicts that a tail-specific objective should add value: market-level volatility timing determines how much aggregate risk to take, but it does not determine which factor exposures should be cut first.

The time-varying evidence refines the interpretation of Table~\ref{tab:horse_vm} and Table~\ref{tab:vm_horserace_states}. Tail-managed portfolios are closely related to volatility management in return space, but they are not spanned by it in the economically relevant states. The low-$\tau$ portfolio loads persistently less on VM and earns its abnormal performance when the cross-section of factor tails creates an opportunity for selective de-risking. The value of tail targeting does not come from abandoning volatility management, but from turning it into a preference-specific and cross-sectionally selective rule. Figure~\ref{fig:conditional_spanning_all} in Appendix~\ref{app:OOSresults_additional} reports the corresponding time-varying alpha and beta estimates for the remaining quantile levels.

\section{Mechanism: Selective De-Risking Across Factors}

Next, we turn from performance to mechanism. Tail-managed policies differ across $\tau$ for two related reasons. First, market volatility reshapes the conditional return distribution asymmetrically across factors, worsening the lower tails of some factors much more than their conditional means. Second, downside-focused investors respond by selectively reducing exposure to those factors whose left tails deteriorate the most. We evaluate both parts of this mechanism directly.

\subsection{Volatility reshapes factor-return tails}

\begin{figure}[h]
    \begin{center}
        \includegraphics[width=0.6\textwidth]{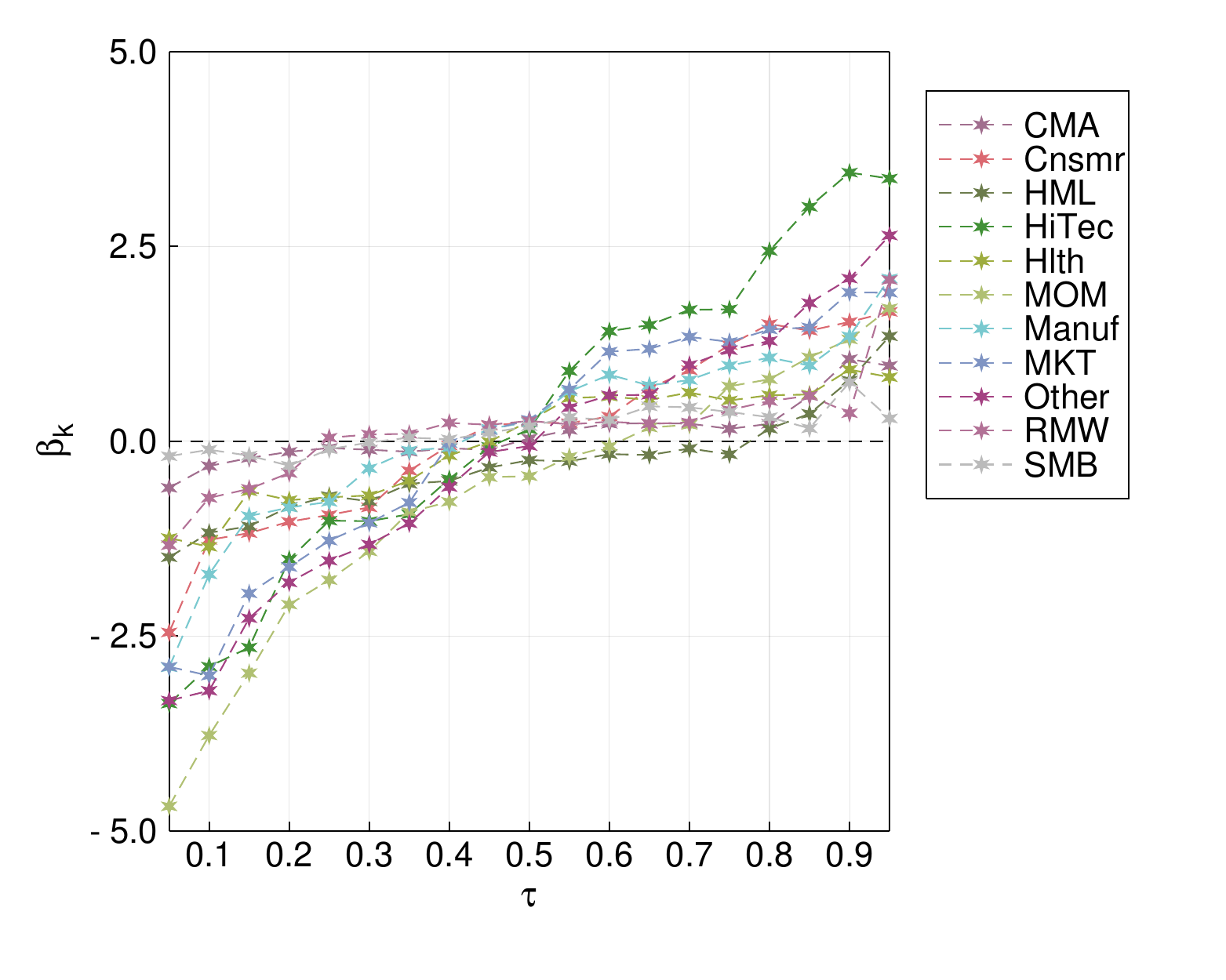}
    \end{center}
    \caption{\textbf{Quantile sensitivity of next-month factor returns to current market volatility.}
For each factor $k$, the figure plots the coefficient $\beta_k(\tau)$ from the quantile regression
\[
Q_\tau(r_{k,t+1}\mid \sigma_t)=\alpha_k(\tau)+\beta_k(\tau)\sigma_t,
\]
where $r_{k,t+1}$ is the next-month factor return and $\sigma_t$ is realized market volatility in month $t$. More negative coefficients at low quantiles imply that elevated volatility disproportionately worsens downside outcomes, whereas flatter or more positive coefficients at high quantiles indicate that upper-tail payoffs are less adversely affected. Cross-factor heterogeneity in these slopes shows that volatility reshapes conditional return distributions asymmetrically, providing the empirical motivation for tail-targeted portfolio choice.}
\label{fig:qreg_beta_by_tau}
\end{figure}

Figure~\ref{fig:qreg_beta_by_tau} shows that tail sensitivity to changes in market volatility is stronger than changes in conditional mean. For most factors, the sensitivity of returns to current market volatility is strongly negative in the lower quantiles and becomes much weaker, flat, or even positive at higher quantiles. The effect is heterogeneous across factors. High-dispersion factors such as HiTec, MOM, MKT, and Other display much steeper lower-tail sensitivity than more defensive factors.

High-volatility states imply asymmetric risk. They disproportionately damage downside outcomes in some parts of the cross-section while leaving the upper tail less affected. Once conditional payoffs shift in that way, the economically relevant object is no longer variance alone. Portfolio choice becomes quantile-specific because investors with different tail priorities face different trade-offs across factors.

The figure also shows how volatility is a state variable that may forecast how the relevant quantiles of future payoffs will move.

\subsection{The policy responds through selective cross-sectional reweighting.}

\begin{figure}[h]
    \begin{center}
        \includegraphics[width=0.5\textwidth]{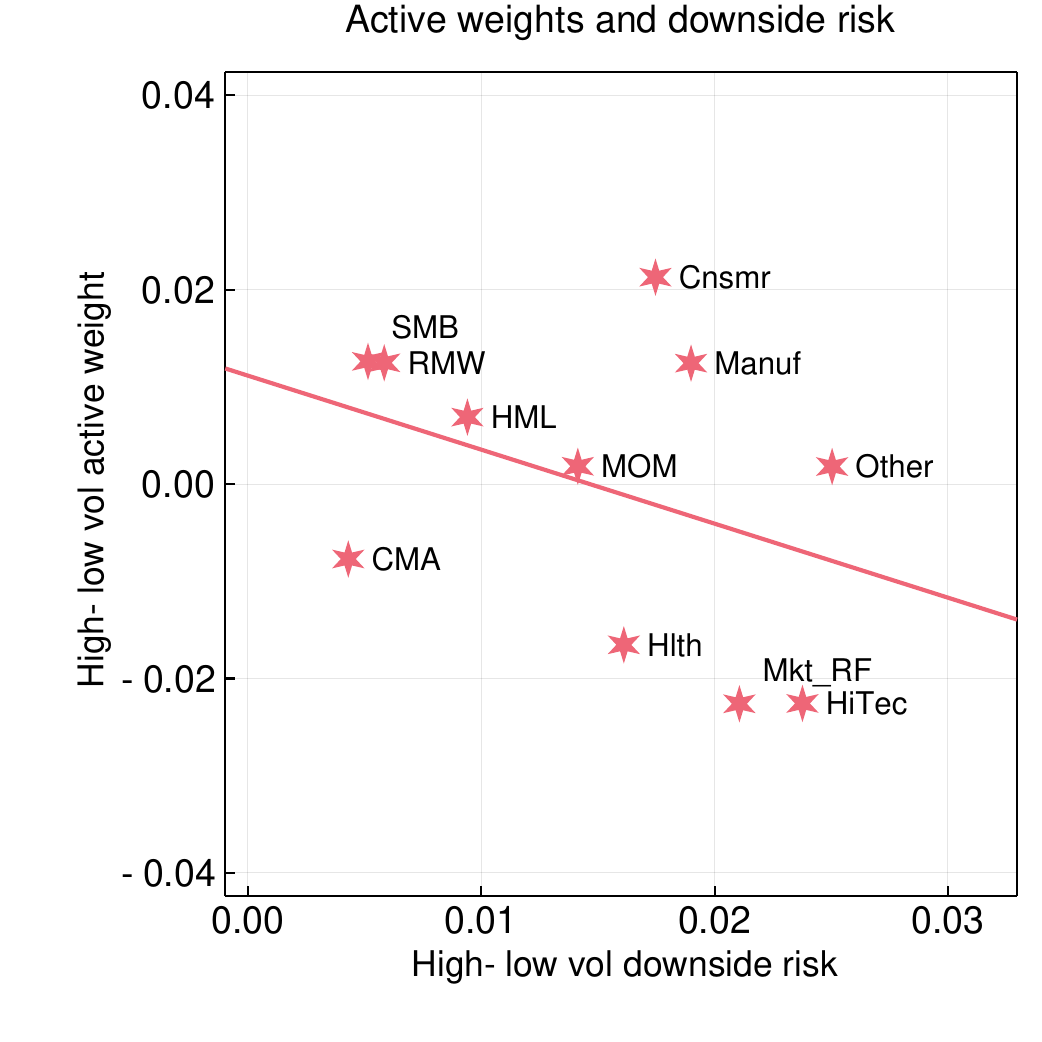}
    \end{center}
    \caption{\textbf{Selective active reallocation and factor-level downside deterioration.}
This figure plots, for each factor, the change in its active portfolio weight from low- to high-volatility states against the corresponding change in factor-level downside risk, measured using trailing expected shortfall. Active weights are measured relative to the volatility-managed benchmark (VM). A negative slope indicates that the policy reduces exposure more strongly in factors whose left tails deteriorate more when volatility rises. The downward-sloping relation for the low-$\tau$ policy shows that tail-targeted de-risking operates through selective cross-sectional reweighting rather than uniform volatility scaling.}
\label{fig:active_vs_es_scatter}
\end{figure}

Figure~\ref{fig:active_vs_es_scatter} and Table~\ref{tab:mechanism_lefttail} provide the paper's direct mechanism test. The theory implies that low-$\tau$ investors should selectively reduce exposure to the factors whose downside tails deteriorate the most. We evaluate that prediction by relating each factor's active portfolio weight to a factor-level measure of left-tail deterioration based on trailing expected shortfall. The evidence is strongly supportive. Relative to the volatility-managed benchmark, the cross-sectional slope is negative. Thus, the most downside-focused policy contracts exposure aggressively in the factors whose left tails worsen most, whereas the upside-focused policy attenuates that response substantially.

The low-$\tau$ portfolio reweights the portfolio across factors in a way that is tightly aligned with cross-sectional differences in left-tail deterioration. That is the selective de-risking logic developed in earlier sections. It is also what connects the paper to the multifactor volatility-management evidence in \cite{demiguel2024multifactor}: in a multifactor setting, the important timing question is not only how much aggregate risk to take, but where in the cross section to take it.

Taken together, Figures~\ref{fig:qreg_beta_by_tau} and~\ref{fig:active_vs_es_scatter} show both sides of the mechanism. Volatility first reshapes the conditional distribution of factor returns asymmetrically across quantiles; the portfolio rule then responds by reallocating away from the factors whose left tails worsen most. This is why tail-managed portfolios should be interpreted as preference-based, tail-specific microfoundations for volatility management. The volatility response emerges endogenously from the investor's objective once conditional tails become state dependent.

\begin{table}[!htbp]
\centering
\caption{\textbf{Active reallocation toward factors with worse downside deterioration.}}
\label{tab:mechanism_lefttail}
\small
\setlength{\tabcolsep}{7pt}

\begin{tabular}{lccc}
\toprule
& All & Vol-up & High-vol \\
\midrule
\multicolumn{4}{l}{\textit{Panel A. Level specification: active weight on left-tail badness (vs.\ VM)}}\\
\addlinespace[0.25em]

$\tau = 0.1$
& \begin{tabular}[c]{@{}c@{}}-0.0591\\[-0.15em] {\footnotesize[-165.66]}\end{tabular}
& \begin{tabular}[c]{@{}c@{}}-0.0592\\[-0.15em] {\footnotesize[-116.19]}\end{tabular}
& \begin{tabular}[c]{@{}c@{}}-0.0697\\[-0.15em] {\footnotesize[-93.27]}\end{tabular} \\

$\tau = 0.5$
& \begin{tabular}[c]{@{}c@{}}-0.0363\\[-0.15em] {\footnotesize[-106.56]}\end{tabular}
& \begin{tabular}[c]{@{}c@{}}-0.0361\\[-0.15em] {\footnotesize[-74.19]}\end{tabular}
& \begin{tabular}[c]{@{}c@{}}-0.0454\\[-0.15em] {\footnotesize[-63.20]}\end{tabular} \\

$\tau = 0.9$
& \begin{tabular}[c]{@{}c@{}}-0.0216\\[-0.15em] {\footnotesize[-62.07]}\end{tabular}
& \begin{tabular}[c]{@{}c@{}}-0.0214\\[-0.15em] {\footnotesize[-42.94]}\end{tabular}
& \begin{tabular}[c]{@{}c@{}}-0.0295\\[-0.15em] {\footnotesize[-38.35]}\end{tabular} \\
\bottomrule
\end{tabular}

\vspace{0.35em}

\noindent
\begin{minipage}{\textwidth}
\footnotesize
\textit{Notes:} This table reports date-by-date cross-sectional regressions of active factor weights on factor-level downside deterioration. The dependent variable is the active weight of a factor relative to the benchmark, and the regressor is factor-level left-tail badness measured using lagged expected shortfall. Panel A uses the volatility-managed benchmark (VM) as the reference portfolio; ``Vol-up'' denotes dates on which lagged market volatility increases, and ``High-vol'' denotes dates in the top quintile of lagged 21-day realized market volatility. Entries are time-series averages of the cross-sectional slope estimates, with $t$-statistics in brackets. A more negative slope means that the policy reduces exposure more strongly in factors whose downside tails worsen more. The monotone ordering across $\tau$ shows that selective de-risking is strongest for downside-focused portfolios.
\end{minipage}

\end{table}

Table \ref{tab:scqg_market_vm} translates the same mechanism into an investor-level performance object. The state-conditional quantile gain asks whether a given tail-managed portfolio improves the part of the return distribution it is designed to target, conditional on the state. Relative to the market-volatility-managed benchmark, the low-$\tau$ portfolio delivers the clearest gains in the targeted lower tail, and those gains are especially large when volatility is high and cross-sectional tail dispersion across factors is also high. The $\tau=0.5$ portfolio also becomes useful in those same stressed environments. By contrast, the $\tau=0.9$ portfolio does not improve its own targeted upper-tail quantile relative to market-volatility timing. Tail targeting is most useful relative to standard market-volatility timing for downside-focused and balanced mandates, when volatility creates large cross-sectional differences in factor tail risk.

\begin{table}[!htbp]
\centering
\caption{\textbf{State-conditional quantile gains relative to the market-volatility-managed benchmark}}
\label{tab:scqg_market_vm}
\small
\setlength{\tabcolsep}{8pt}

\begin{tabular}{lccc}
\toprule
State & $\tau = 0.1$ & $\tau = 0.5$ & $\tau = 0.9$ \\
\midrule
All sample                                  & 152.8 & -9.2 & -23.9 \\
High volatility                            & 227.1 & 37.1 & -15.2 \\
Downside tail dispersion                   & 272.3 & 52.3 & -13.2 \\
Tail Opportunity State                     & 236.2 & 57.9 & -24.7 \\
Upside state + Downside tail dispersion    & 263.4 & 60.4 & -10.7 \\
\addlinespace[0.25em]
Active mean (ann., \%) in Tail Opportunity State & 2.1 & 6.6 & 8.8 \\
Active Sharpe in Tail Opportunity State           & 0.08 & 0.22 & 0.28 \\
\bottomrule
\end{tabular}

\vspace{0.35em}

\noindent
\begin{minipage}{\textwidth}
\footnotesize
\textit{Notes:} This table reports the state-conditional quantile gain (SCQG), measured in basis points, of each tail-managed portfolio relative to the market-volatility-managed benchmark. For a portfolio with target quantile $\tau$, $\mathrm{SCQG}_{\tau}(s)=Q_{\tau}\!\left(R^{\tau}_{t,t+21}\mid s_t\right)- Q_{\tau}\!\left(R^{MV}_{t,t+21}\mid s_t\right)$,
where $R_{t,t+21}$ is the next 21-trading-day compounded return and $s_t$ denotes the conditioning state. Positive values indicate that the tail-managed portfolio improves its own targeted part of the return distribution relative to market-volatility management. High volatility is defined as the top quintile of lagged 21-trading-day realized market volatility. Upside market state dispersion is defined as the top quintile of the lagged cross-sectional spread in factor downside quantiles, computed as the max-minus-min of rolling 63-trading-day factor $Q_{0.10}$. The Tail Opportunity State is the intersection of high volatility and Upside market state dispersion. The final two rows report the annualized active mean return and active Sharpe ratio in that state.
\end{minipage}

\end{table}

The conditional quantile gains are concentrated in what we call tail-opportunity states: months with high market volatility and large cross-sectional dispersion in factor downside tails. In these states, the $\tau$ = 0.1 portfolio improves its targeted lower quantile by 236 basis points relative to market-volatility management, and the $\tau$ = 0.5 portfolio also improves. The $\tau$ = 0.9 portfolio does not improve its upper-tail quantile relative to the benchmark. Quantile targeting adds most for downside-focused and balanced mandates, precisely when the cross section of factor tails makes market-level volatility timing too coarse.

Taken together, Table~\ref{tab:scqg_market_vm}, Figure~\ref{fig:active_vs_es_scatter}, and Table~\ref{tab:mechanism_lefttail} all point to the same conclusion. Relative to standard market-volatility timing, tail targeting is most useful when volatility not only rises but also generates large cross-sectional differences in factor downside deterioration. In those states, investors need to manage tail shape, not just total variance.

\section{Factor-Level Volatility Scaling Benchmark}
\label{sec:factorVM}

The previous sections show that tail-managed portfolios are closely related to
volatility management, but that the relation is preference-specific and state dependent.
We now ask a more demanding question. How much of this evidence remains once the
benchmark is allowed to scale each factor by its own volatility? This benchmark is much
harder to beat because it already embeds part of the cross-sectional de-risking logic that
motivates the paper. It  is an important robustness benchmark that asks whether tail targeting adds information
beyond factor-level volatility scaling.

To answer this question, we estimate the same conditional-spanning regression used in
Figure~\ref{fig:conditional_spanning}, but replace the market-volatility-managed benchmark
by the factor-level volatility-scaled benchmark:
\[
    r^{\tau}_{p,t}
    =
    \alpha_t
    +
    \beta_t r^{FVM}_{t}
    +
    \varepsilon_t,
\]
where $r^{FVM}_{t}$ denotes the return on the factor-level volatility-scaled benchmark. The
parameters $\alpha_t$ and $\beta_t$ are estimated locally over time. This specification asks
whether the tail-managed portfolio is simply a time-varying exposure to a benchmark
that already scales each factor by its own realized volatility, or whether there remains
state-dependent abnormal performance associated with the investor's tail objective.

\begin{figure}[ht]
    \begin{center}
        \includegraphics[width=0.82\textwidth]{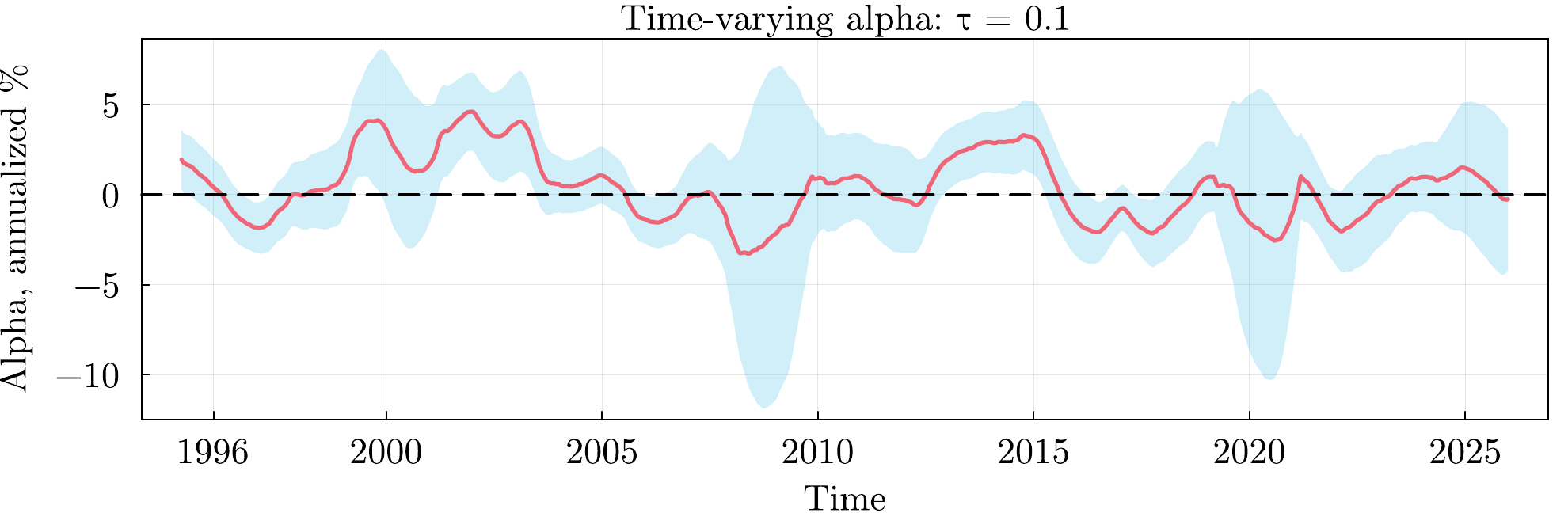}
        \includegraphics[width=0.82\textwidth]{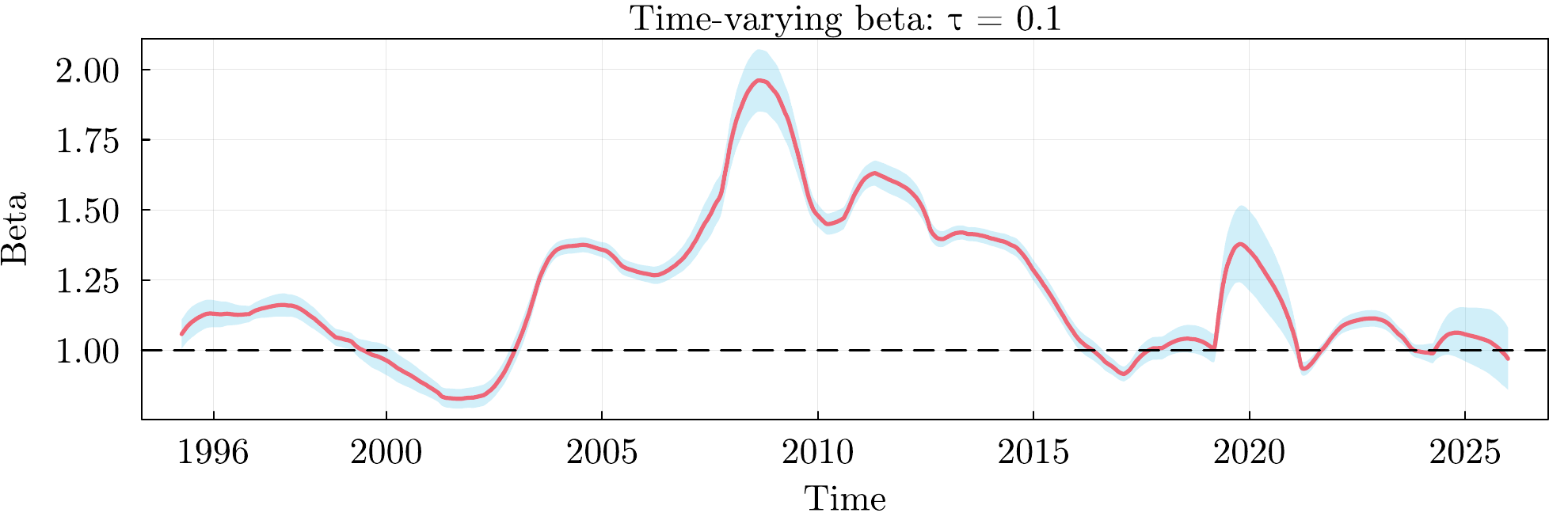}
    \end{center}
    \caption{\textbf{Conditional spanning by the factor-level volatility-scaled benchmark.}
    This figure reports local time-varying estimates from the regression
    $r^{0.1}_{p,t}=\alpha_t+\beta_t r^{FVM}_{t}+\varepsilon_t$ for the downside-focused
    tail-managed portfolio. The benchmark $r^{FVM}_{t}$ is the factor-level
    volatility-scaled portfolio. The top panel plots the annualized alpha, expressed in
    percent, and the bottom panel plots the beta with respect to the benchmark. Shaded
    areas denote pointwise confidence bands based on local Newey--West standard errors.
    The dashed horizontal line in the top panel corresponds to zero alpha, and the dashed
    horizontal line in the bottom panel corresponds to unit beta. Relative to the
    market-volatility-managed benchmark, the beta is closer to one, showing that
    factor-level volatility scaling absorbs a substantial part of the defensive timing
    component of the low-$\tau$ policy. The remaining alpha is episodic, consistent with
    tail targeting adding value mainly when factor-level volatility scaling does not
    fully capture state-dependent tail heterogeneity.}
    \label{fig:conditional_spanning_factor_vm}
\end{figure}

Figure~\ref{fig:conditional_spanning_factor_vm} reports the results for the downside-focused
portfolio, $\tau=0.1$. The contrast with Figure~\ref{fig:conditional_spanning} is informative.
Relative to the market-volatility-managed benchmark, the low-$\tau$ portfolio behaved as
a persistent lower-beta, defensive version of volatility management. Relative to the
factor-level volatility-scaled benchmark, the beta is much closer to one and in several
periods exceeds one. This shows that factor-level volatility scaling absorbs a large part of
the defensive timing component of the low-$\tau$ policy. In other words, much of what a
downside-focused quantile investor does mechanically resembles factor-wise volatility
scaling: the investor cuts exposure more strongly in factors whose own risk conditions
deteriorate.

Importantly, the time-varying alpha is episodically large. Positive alpha appears especially in periods when volatility and cross-sectional tail heterogeneity make the allocation problem more demanding 
than a simple scaling problem. 
Because factor-level volatility scaling already captures part of the variation in conditional quantiles, the incremental gains from tail targeting should be smaller against this benchmark. Its remaining value should appear when volatility reshapes factor tails in ways that require an investor-specific choice about which part of the distribution to protect. Its remaining value should appear
when volatility reshapes factor tails in ways that require an investor-specific choice about
which part of the distribution to protect. Tail targeting supplies the missing objective that explains when factor-level
de-risking is desirable and how that de-risking should differ across downside, median, and
upside mandates. In this context, the episodically large alpha relative to the factor-level volatility-scaled benchmark supports the rationale of tail targeting. The corresponding estimates for $\tau=0.5$ and $\tau=0.9$ are reported in Figure~\ref{fig:conditional_spanning_factor_vm_appendix} in Appendix~\ref{app:OOSresults_additional}.

\begin{figure}[ht]
    \begin{center}
        \includegraphics[width=0.6\textwidth]{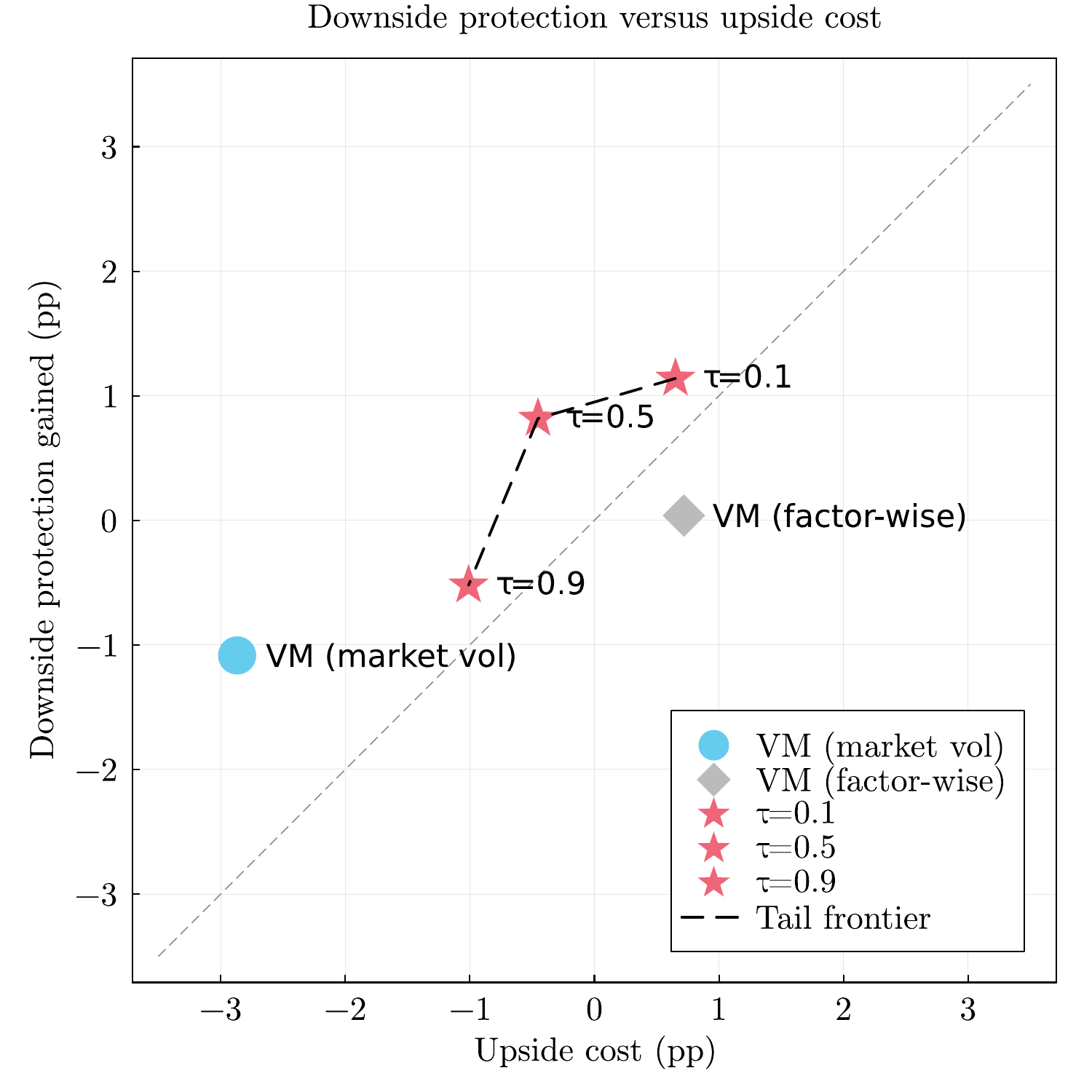}
    \end{center}
    \caption{\textbf{High-volatility downside-upside frontier for managed portfolios.}
    This figure compares the tail trade-off of the tail-managed portfolios with the
    market-volatility-managed benchmark and the factor-level volatility-scaled benchmark
    in high-volatility states. Each point represents one managed portfolio. The horizontal
    axis reports upside cost, measured as the reduction in the 90th percentile of
    next-month portfolio returns relative to the unmanaged benchmark. The vertical axis
    reports downside protection gained, measured as the improvement in the 10th
    percentile relative to the unmanaged benchmark. The dashed line connects the
    tail-managed portfolios with $\tau \in \{0.1,0.5,0.9\}$ and traces the
    preference-indexed frontier implied by investor tail priorities. The figure is not
    intended to show uniform dominance. Its purpose is to show that tail targeting
    organizes the downside-upside frontier, while the two volatility-managed benchmarks
    occupy distinct points on the same trade-off map.}
    \label{fig:tail_frontier_high_vol}
\end{figure}

Figure~\ref{fig:tail_frontier_high_vol} summarizes the same point from the perspective of
tail trade-offs. In high-volatility states, the three tail-managed portfolios trace a
preference-indexed downside-upside frontier. The low-$\tau$ portfolio purchases the most
downside protection, the high-$\tau$ portfolio gives up less upside, and the median portfolio
lies between these two cases. The two volatility-managed benchmarks occupy points on
the same trade-off map, but they do not organize the frontier by investor tail priorities. Factor-level volatility scaling can approximate a large part
of the raw de-risking behavior, especially for a downside-focused investor, but it does not
explain which point on the downside-upside frontier corresponds to a given mandate.
Tail targeting provides that interpretation.

A strong factor-level volatility-scaled benchmark captures part of the common and cross-sectional risk adjustment in the tail-managed policies. This makes the remaining evidence more informative. The low-$\tau$ portfolio is close to factor-level volatility scaling because both reduce
exposure when factor risk deteriorates, but tail targeting clarifies why such reductions
are valuable for downside-focused investors and how the same investment opportunity set
generates different allocations for median- and upside-oriented investors. The main
contribution is thus that recursive quantile preferences organize volatility management into a preference-indexed frontier of selective de-risking.

Finally, note that \autoref{ref:add_11alt_oos} confirms the results on an alternative universe of factors that include tail-shape, or quality factors.

% ======================================================================
% NEW SECTION
% Insert after the current factor-level volatility scaling section and
% before the conclusion. Renumber the conclusion accordingly.
% ======================================================================

\section{Mandate Evidence from Fund Flows}
\label{sec:mandate_flows}

The portfolio results above show that quantile-targeted policies organize dynamic allocation into a downside--upside frontier. This section provides an external check on the economic interpretation of the quantile index. We ask a directly testable question: do real-world investment mandates differ in the payoff regions to which investors allocate capital?

Fund flows reveal economically meaningful heterogeneity in distributional evaluation. Some mandates reward lower-tail stability, some place less weight on downside outcomes, and some attract capital when downside risk becomes salient. This evidence supports the use of $\tau$ as a reduced-form index of the payoff region emphasized by the mandate.

\subsection{Data and mandate classification}

Similarly to \cite{Kaniel2025,Moussawi2025}, we use SEC Form N-PORT filings for registered investment funds. The filings provide monthly fund returns, assets, and flow components. We construct a fund-month panel from June 2020 to October 2025. For each fund $i$ and month $t$, we compute net flow $Flow_{i,t+1}$, lagged fund size, lagged flow, lagged return, and rolling distributional performance measures from recent monthly returns. The resulting regression sample contains 603,026 fund-month observations and 11,818 funds.

We classify funds into mandate groups using fund names. Income-oriented mandates include terms such as income, dividend, yield, covered call, buy-write, option income, and premium income. Growth-oriented mandates include terms such as growth, aggressive growth, capital appreciation, technology, innovation, momentum, high beta, and disruptive. Downside-protection mandates are defined more narrowly and include terms such as buffer, defined outcome, downside protection, capital preservation, principal protection, hedged equity, low volatility, minimum volatility, and defensive. Funds outside these groups form the baseline category.

The classification is intentionally simple. The purpose is to test whether observed capital allocation differs across mandates in the direction predicted by the model. If mandates emphasize different payoff regions, flow-performance sensitivities should differ across mandate groups.

\subsection{Flow sensitivity to distributional performance}

For each fund-month, we compute rolling distributional performance measures $Q_{0.10,i,t}$ measuring recent lower-tail performance, $Q_{0.50,i,t}$ recent median performance, and $Q_{0.90,i,t}$ recent upper-tail performance. We then estimate
\begin{align}
Flow_{i,t+1}
=&\ b_{10} Q_{0.10,i,t}
 + b_{50} Q_{0.50,i,t}
 + b_{90} Q_{0.90,i,t} \notag \\
&+ \gamma^{Inc}_{10} Inc_i Q_{0.10,i,t}
 + \gamma^{Inc}_{50} Inc_i Q_{0.50,i,t}
 + \gamma^{Inc}_{90} Inc_i Q_{0.90,i,t} \notag \\
&+ \gamma^{Gro}_{10} Gro_i Q_{0.10,i,t}
 + \gamma^{Gro}_{50} Gro_i Q_{0.50,i,t}
 + \gamma^{Gro}_{90} Gro_i Q_{0.90,i,t} \notag \\
&+ X_{i,t}'\delta + \alpha_i + \lambda_t + \varepsilon_{i,t+1},
\label{eq:mandate_quantile_flows}
\end{align}
where $Inc_i$ and $Gro_i$ indicate income and growth mandates, $X_{i,t}$ includes lagged flow, lagged return, log assets, and return volatility, and $\alpha_i$ and $\lambda_t$ are fund and month fixed effects.

Table~\ref{tab:mandate_quantile_flows} reports the implied total sensitivities and the key mandate interactions. Baseline funds exhibit positive flow sensitivity to all three parts of the distribution. The sensitivity is 0.061 for $Q_{0.10}$, 0.080 for $Q_{0.50}$, and 0.044 for $Q_{0.90}$. Thus, ordinary fund flows respond to broad recent performance, with the strongest response around the center of the distribution.

Income-oriented funds display a much stronger response to lower-tail performance. Their total sensitivity to $Q_{0.10}$ is 0.158, compared with 0.061 for baseline funds. The income interaction with $Q_{0.10}$ is 0.097, with a $t$-statistic of 6.21. Income and payout-oriented products reward lower-tail stability, which preserves the capital base supporting income generation.

Growth-oriented funds show the opposite pattern. The growth interaction with $Q_{0.10}$ is -0.039, with a $t$-statistic of -2.67. The total lower-tail sensitivity of growth funds is only 0.023, compared with 0.061 for baseline funds. The growth interaction with $Q_{0.90}$ is not statistically significant. Thus, the distinctive feature of growth mandates is weaker sensitivity to lower-tail performance. This pattern is consistent with mandates that place less weight on downside protection and preserve more upside-oriented exposure.

\begin{table}[t]
\centering
\caption{Mandate-specific flow sensitivity to distributional performance}
\label{tab:mandate_quantile_flows}
\begin{tabular}{lccc}
\toprule
& $Q_{0.10}$ & $Q_{0.50}$ & $Q_{0.90}$ \\
\midrule
\multicolumn{4}{l}{\emph{Panel A: Implied total sensitivities}} \\
Baseline funds & 0.061*** & 0.080*** & 0.044*** \\ 
 & (5.387) & (7.590) & (3.853) \\ 
Income-oriented funds & 0.158*** & 0.092*** & 0.091*** \\ 
 & (8.792) & (4.058) & (4.781) \\ 
Growth-oriented funds & 0.023 & 0.084*** & 0.025 \\ 
 & (1.414) & (4.360) & (1.465) \\ 

\midrule
\multicolumn{4}{l}{\emph{Panel B: Incremental mandate interactions}} \\
Income $\times Q_\tau$ & 0.097*** & 0.012 & 0.047*** \\ 
 & (6.205) & (0.538) & (2.853) \\ 
Growth $\times Q_\tau$ & -0.039*** & 0.003 & -0.019 \\ 
 & (-2.671) & (0.167) & (-1.238) \\ 

\bottomrule
\end{tabular}
\begin{minipage}{0.92\textwidth}
\footnotesize
\emph{Notes:} This table reports flow sensitivities from fund-month regressions of future flows on rolling distributional performance measures, mandate interactions, controls, fund fixed effects, and month fixed effects. Panel A reports implied total sensitivities for each mandate group. Panel B reports incremental mandate interactions relative to baseline funds. $Q_{0.10}$, $Q_{0.50}$, and $Q_{0.90}$ denote recent lower-tail, median, and upper-tail fund performance. $t$-statistics are reported in parentheses. $^{***}$, $^{**}$, and $^{*}$ denote significance at the 1\%, 5\%, and 10\% levels.
\end{minipage}
\end{table}

\subsection{Downside-protection demand and risk salience}

Downside-protection products are different from ordinary performance-chasing funds. Investors may allocate to them not because recent average returns are high, but because downside risk has become salient. We then estimate a second specification that relates future flows to recent mean performance, worst recent performance, best recent performance, and return volatility:
\begin{align}
Flow_{i,t+1}
=&\ a_1 MeanPerf_{i,t}
 + a_2 WorstPerf_{i,t}
 + a_3 BestPerf_{i,t}
 + a_4 VolPerf_{i,t} \notag \\
&+ \theta_1 Down_i WorstPerf_{i,t}
 + \theta_2 Down_i BestPerf_{i,t}
 + \theta_3 Down_i VolPerf_{i,t} \notag \\
&+ X_{i,t}'\delta + \alpha_i + \lambda_t + \varepsilon_{i,t+1},
\label{eq:downside_salience}
\end{align}
where $Down_i$ identifies downside-protection funds.

Table~\ref{tab:downside_salience} shows that downside-protection funds attract capital when risk is salient. The interaction between downside-protection mandates and return volatility is 1.105, with a $t$-statistic of 5.53. The interaction with worst recent performance is 0.297, with a $t$-statistic of 4.97. The interaction with best recent performance is negative, -0.266, with a $t$-statistic of -4.86. Thus, flows into downside-protection products rise when volatility and downside-risk states become salient, and decline relative to other funds when upside performance is salient.

\begin{table}[t]
\centering
\caption{Downside-protection demand and risk salience}
\label{tab:downside_salience}
\begin{tabular}{lcc}
\toprule
Variable & Estimate & $t$-statistic \\
\midrule
Downside protection $\times$ worst recent performance & 0.297*** & 4.972 \\ 
Downside protection $\times$ return volatility & 1.105*** & 5.535 \\ 
Downside protection $\times$ best recent performance & -0.266*** & -4.862 \\ 

\bottomrule
\end{tabular}
\begin{minipage}{0.92\textwidth}
\footnotesize
\emph{Notes:} This table reports key interaction coefficients from fund-month regressions of future flows on recent mean performance, worst recent performance, best recent performance, return volatility, mandate interactions, controls, fund fixed effects, and month fixed effects. Downside-protection funds are identified using buffer, defined-outcome, downside-protection, capital-preservation, principal-protection, hedged-equity, low-volatility, minimum-volatility, defensive, and related mandate terms. $^{***}$, $^{**}$, and $^{*}$ denote significance at the 1\%, 5\%, and 10\% levels.
\end{minipage}
\end{table}

The salience evidence is important because it distinguishes downside-protection mandates from simple performance chasing. Investors allocate to these products when volatility and downside states make protection valuable. This behavior is precisely the investor-demand counterpart of the portfolio mechanism studied in the main analysis. When volatility changes payoff distributions asymmetrically, the relevant investment problem is not only how much risk to take, but which part of the payoff distribution the mandate is designed to protect.

\subsection{Implications for the quantile interpretation}

The fund-flow evidence provides an external check on the interpretation of $\tau$. The portfolio analysis shows that factor returns exhibit state-dependent tail heterogeneity. Quantile-targeted policies translate different payoff-region objectives into different dynamic allocations. The fund-flow evidence shows that real investors also evaluate products along distributional dimensions. Income-oriented investors reward lower-tail stability. Growth-oriented investors are less sensitive to lower-tail performance. Downside-protection products attract capital when volatility and downside risk become salient.

These patterns are consistent with interpreting $\tau$ as a reduced-form mandate index. We do not require $\tau$ to be directly observed. Instead, the evidence shows that observed mandates differ in the payoff regions that investors reward. This is the heterogeneity that recursive quantile preferences are designed to capture. Once investors differ in the payoff region they care about, volatility management becomes mandate-specific, and quantile targeting provides the mapping from mandate to allocation.

\section{Conclusion}

Volatility management is useful because volatility forecasts changes in conditional risk. In multifactor portfolios, however, volatility need not summarize how different payoff regions move across factors. Some factors experience a severe deterioration in their downside tails, while others retain comparatively attractive payoff distributions. This paper shows that the relevant dynamic allocation problem in these states is not only how much risky exposure to take, but which exposures should be reduced first.

We develop a dynamic portfolio framework with recursive quantile preferences to study this problem. The framework maps investor tail objectives into state-dependent portfolio rules. Downside-focused investors place greater weight on protecting the left tail of portfolio returns and therefore reduce exposure most strongly to factors whose conditional downside tails deteriorate. Investors targeting higher quantiles preserve more upside exposure and de-risk less aggressively. Quantile preferences thus transform volatility management from a scalar scaling rule into a preference-indexed theory of selective cross-sectional de-risking.

The empirical evidence supports this interpretation. Using eleven investable factors, we find that tail-managed portfolios form a coherent family of dynamic allocations. The low-$\tau$ policy is defensive: it delivers the strongest downside protection and the highest Sharpe ratio. The high-$\tau$ policy earns higher average returns but provides less protection against adverse outcomes. The median policy lies between these two cases. These patterns show that there is no single volatility-timing rule that is optimal for all investors. Different tail objectives imply different dynamic exposures.

The gains from tail targeting are concentrated in the states predicted by the model. Tail-managed portfolios add the most value when market volatility is high and cross-sectional dispersion in factor downside risk is large. In these tail opportunity states, market-level volatility timing is too coarse because it reduces portfolio risk without distinguishing among factors whose tails have changed differently. Tail targeting adds value by using the cross section of conditional tail risks to decide where risk should be reduced and where it should be preserved.

The mechanism is selective rather than mechanical. The active weights of tail-managed portfolios respond negatively to factor-level downside deterioration, and this response is strongest for downside-focused investors. This evidence distinguishes tail targeting from uniform de-risking. The portfolios do not simply cut exposure whenever volatility rises. They reallocate away from factors whose left tails worsen most, with the strength of that response determined by the investor's targeted quantile.

The comparison with factor-level volatility scaling further clarifies the contribution. A strong factor-level volatility benchmark already captures part of the cross-sectional de-risking behavior emphasized in this paper. Tail-managed portfolios therefore do not mechanically dominate volatility scaling in all states. Instead, their incremental value is episodic and state dependent. They matter most when volatility and tail heterogeneity jointly make the allocation problem richer than a variance-scaling problem. This is precisely the setting in which investor preferences over different parts of the payoff distribution become economically relevant.

The main implication is that volatility is an important state variable, but it is not always a sufficient statistic for dynamic allocation. When high-volatility states produce large cross-sectional differences in factor tail risk, investors need a rule that links their mandate to the part of the return distribution they want to protect or exploit. Recursive quantile preferences provide such a rule. They organize dynamic factor allocation into a preference-indexed frontier of selective de-risking. The mandate-flow evidence shows that this distinction is not only theoretical: actual fund investors allocate capital differently across lower-tail, median, upper-tail, and risk-salience states, depending on the stated mandate of the product.

More broadly, the results suggest that managed portfolios should be evaluated not only by whether they improve average performance or Sharpe ratios, but also by whether they improve the part of the payoff distribution that investors care about. For downside-focused mandates, the relevant success criterion is left-tail protection; for upside-oriented mandates, it is preserving exposure to favorable states. Tail-managed portfolios make these objectives explicit. They show when volatility management should be common across investors and when it should differ according to tail preferences.

The results suggest a narrower view of volatility management. Volatility is a useful state variable, but it is not an objective. Once investors care about different regions of the payoff distribution, the same volatility signal can imply different portfolio responses. Recursive quantile preferences provide one way to discipline that response. They map mandate-specific payoff regions into selective de-risking rules and explain why volatility management should be most valuable when volatility changes the cross-sectional ordering of tail risk.

\newpage
\bibliography{Bibliography}
\bibliographystyle{chicago}

\clearpage
\newpage
\appendix
\section*{Appendix}
\addcontentsline{toc}{section}{Appendices}
\counterwithin{figure}{section}
\counterwithin{table}{section}
\counterwithin{algorithm}{section}

\section{Proof of Proposition 1}
\label{app:proofprop1}

\begin{proof}
Choose $q \in (\tau,1)$, set $\lambda=\tau/q$, and choose $p\in(\tau,q)$. Let $X$ pay $1$ with probability one, let $Y$ pay $0$ with probability $p$ and $2$ otherwise, and let $Z$ pay $3$ with probability one. Then $Q_{\tau}(X)=1$ while $Q_{\tau}(Y)=0$, so $X\succ_{\tau}Y$.

Let $\lambda X \oplus (1-\lambda)Z$ denote the lottery that pays according to $X$ with probability $\lambda$ and according to $Z$ with probability $1-\lambda$, with the same notation for $Y$. In the mixture with $X$, the outcome $1$ occurs with probability $\lambda=\tau/q>\tau$, so
\[
Q_{\tau}\bigl(\lambda X \oplus (1-\lambda)Z\bigr)=1.
\]
In the mixture with $Y$, the outcome $0$ occurs with probability $\lambda p=\tau p/q<\tau$, while outcomes weakly below $2$ occur with probability $\lambda=\tau/q>\tau$. Hence
\[
Q_{\tau}\bigl(\lambda Y \oplus (1-\lambda)Z\bigr)=2.
\]
Thus
\[
\lambda Y \oplus (1-\lambda)Z \succ_{\tau} \lambda X \oplus (1-\lambda)Z,
\]
even though $X\succ_{\tau}Y$. This reverses the ranking after mixing both lotteries with the same $Z$ and violates independence. Since any von Neumann--Morgenstern expected-utility representation satisfies independence, no such representation exists.
\end{proof}

\section{Mean-variance does not identify downside tails outside Gaussian benchmarks}
\label{app:mv_notails}

A common intuition is that if two payoffs have the same mean and variance, then they have the same ``risk.'' That is true in a Gaussian location--scale benchmark, where
\[
Q_{\tau}[X]=\mu+\sigma \Phi^{-1}(\tau)
\]
is fully determined by $(\mu,\sigma)$. Outside that benchmark, mean and variance do not pin down lower-tail outcomes. The next proposition gives a simple crash-mixture construction.

\begin{proposition}[Same mean and variance, different lower tail]
\label{prop:mv_vs_tail}
Fix $\mu\in\mathbb{R}$, $\sigma^2>0$, and $\tau\in(0,1)$. Choose $p\in(0,1)$ such that $\tau\le p$, and define
\[
J:=\sigma\sqrt{\frac{1-p}{p}},
\qquad
U:=\sigma\sqrt{\frac{p}{1-p}}.
\]
Consider two excess-return distributions:
\begin{enumerate}
    \item[(i)] (\emph{Gaussian benchmark}) 
    \[
    X_A \sim \mathcal{N}(\mu,\sigma^2).
    \]
    \item[(ii)] (\emph{Crash-mixture})
    \[
    X_B=
    \begin{cases}
    \mu-J, & \text{with probability } p,\\[4pt]
    \mu+U, & \text{with probability } 1-p.
    \end{cases}
    \]
\end{enumerate}
Then
\[
\mathbb{E}[X_A]=\mathbb{E}[X_B]=\mu,
\qquad
\mathrm{Var}(X_A)=\mathrm{Var}(X_B)=\sigma^2,
\]
but for any $\tau\le p$,
\[
Q_{\tau}(X_B)=\mu-J,
\qquad
Q_{\tau}(X_A)=\mu+\sigma \Phi^{-1}(\tau).
\]
Hence the crash-mixture has a strictly worse lower tail whenever
\[
J>-\sigma \Phi^{-1}(\tau).
\]
\end{proposition}

\begin{proof}
For the Gaussian benchmark, the claims are immediate:
\[
\mathbb{E}[X_A]=\mu,
\qquad
\mathrm{Var}(X_A)=\sigma^2,
\qquad
Q_{\tau}(X_A)=\mu+\sigma\Phi^{-1}(\tau).
\]

For the crash-mixture $X_B$, the mean is
\[
\mathbb{E}[X_B]
=
p(\mu-J)+(1-p)(\mu+U)
=
\mu+\bigl[-pJ+(1-p)U\bigr].
\]
By the definitions of $J$ and $U$,
\[
pJ = p\sigma\sqrt{\frac{1-p}{p}}=\sigma\sqrt{p(1-p)},
\qquad
(1-p)U=(1-p)\sigma\sqrt{\frac{p}{1-p}}=\sigma\sqrt{p(1-p)},
\]
so $-pJ+(1-p)U=0$, and therefore
\[
\mathbb{E}[X_B]=\mu.
\]

Since $X_B$ takes only the two values $\mu-J$ and $\mu+U$, with mean $\mu$, its variance is
\[
\mathrm{Var}(X_B)
=
pJ^2+(1-p)U^2.
\]
Substituting the definitions of $J$ and $U$ gives
\[
pJ^2+(1-p)U^2
=
p\sigma^2\frac{1-p}{p}+(1-p)\sigma^2\frac{p}{1-p}
=
\sigma^2.
\]
Hence
\[
\mathrm{Var}(X_B)=\sigma^2.
\]

Finally, $X_B$ places probability mass $p$ at the crash outcome $\mu-J$. Therefore, for any $\tau\le p$,
\[
Q_{\tau}(X_B)=\mu-J.
\]
Comparing with the Gaussian benchmark yields
\[
Q_{\tau}(X_A)=\mu+\sigma\Phi^{-1}(\tau).
\]
Thus
\[
Q_{\tau}(X_B)<Q_{\tau}(X_A)
\quad\Longleftrightarrow\quad
\mu-J<\mu+\sigma\Phi^{-1}(\tau)
\quad\Longleftrightarrow\quad
J>-\sigma\Phi^{-1}(\tau),
\]
which proves the claim.
\end{proof}

\noindent
Proposition \ref{prop:mv_vs_tail} shows that once returns deviate from a Gaussian benchmark, mean and variance no longer identify downside tail exposure. Two payoffs can have the same mean and variance, yet a low-$\tau$ investor strictly prefers one over the other because its lower-tail quantile is less severe.

\subsection{Relation to the Classical Preferences and Risk}
\label{app:relationclassic}
A preference of $\tau = 0.1$ reflects pessimistic behaviour focused on downside risk and avoiding exposure to risk, whereas a $\tau = 0.9$ represents optimistic behaviour focusing on upside potential and preferring higher risk. Therefore, investors can focus directly on the downside or upside depending on their risk attitudes encoded in the value of $\tau$.
\begin{figure}[ht!]
            \begin{center}
                \includegraphics[scale=0.33]{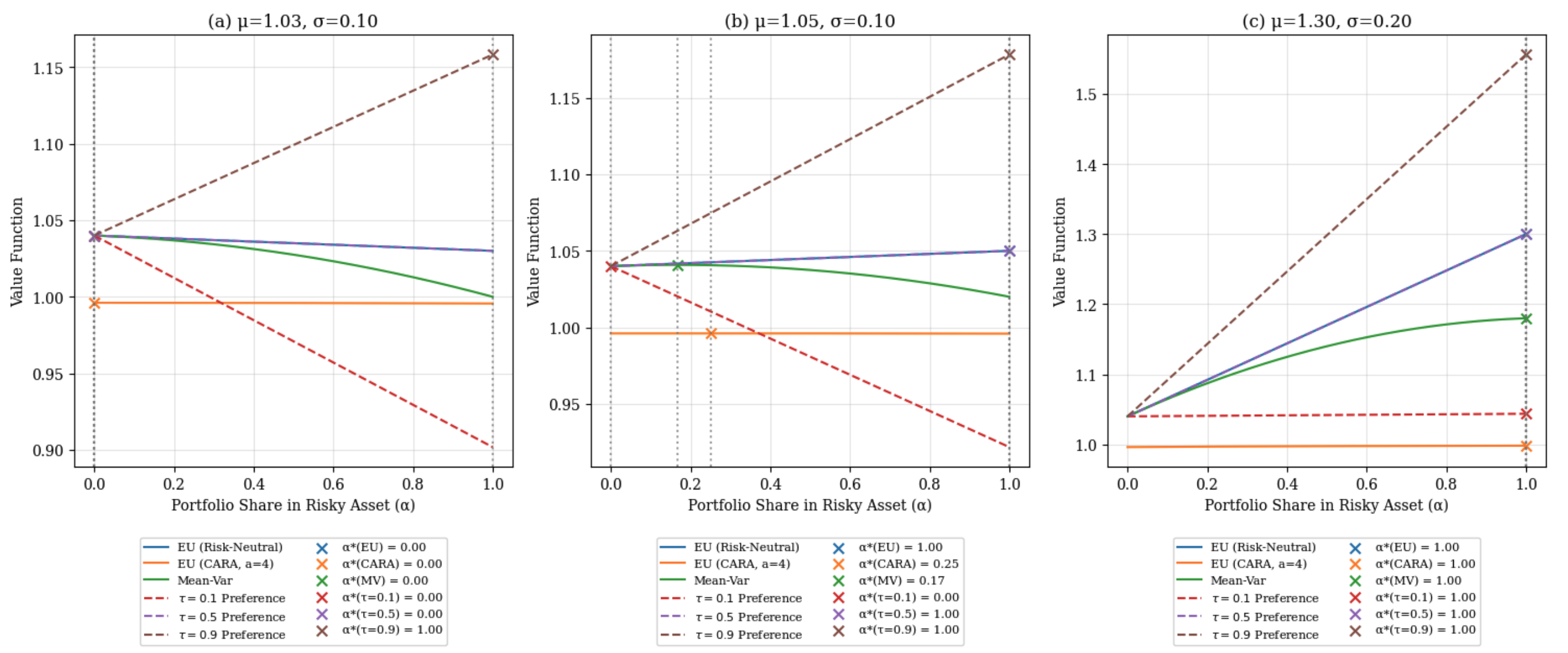}
            \end{center}
            \caption{\footnotesize{(a) $R\sim N(0.03,0.1^2)$, (b) $R\sim N(0.05,0.1^2)$, (c) $R\sim N(0.3,0.2^2)$} with $R_f=0.04$. Note: CARA utility value function is shifted by 0.75 for a better readability.}
            \label{fig:example}
\end{figure} 

\autoref{fig:example} shows three simple cases with different expected values of a risky asset, $\mu$ and variance $\sigma^2$. In \autoref{fig:example} (a), when $\mu<R_f$, risk-neutral EUs, risk-averse EUs and mean-variance preference makers choose the risk-free asset, while quantile preference makers decide according to the $\tau$. For example, when the quantile is 0.9, the risky asset is chosen since the expected value is much higher than the risk-free rate, since $\mu + \sigma \Phi^{-1}(\tau)=0.03+0.1\times 1.28=0.158$. \autoref{fig:example} (b) shows that shifting the value of $\mu$ to $\mu>R_f$ changes the decision of the EU and the mean-variance preference makers. Shifting $\mu$ to a large outcome of 0.3 also directs all decisions towards the risky asset, even with increased variance, as illustrated in \autoref{fig:example} (c). Note that the preference maker with a value of tau equal to $0.1$ also chooses the risky asset, simply because $\mu + \sigma \Phi^{-1}(\tau)=0.3+0.2\times (-1.28)=0.044$, which is larger than $R_f$. Finally, note that the risk neutral EU preference maker is defined by construction as having a preference for the value of the parameter at the $\tau=0.5$, i.e. at the value of the parameter at which the mean and median are equal as $\mu + \sigma \Phi^{-1}(0.5)=\mu$.

\section{Multiple-Period Portfolio Choice}
\label{app:multi-period-example}

\subsection{Two-Period Portfolio Choice ($T=2$)}
\label{app:secexample}
Relative to the classical Merton expected-utility \citep{merton1971optimum}, our approach applies a recursive quantile aggregator to compounded portfolio payoffs. Wealth compounds in both settings, but here the multiplicative return structure remains inside the objective because the investor targets a chosen quantile of continuation wealth rather than expected utility.
For the decision at $t=0$ we need to consider the discounted value $Y = 0 + \beta v^{\ast\tau}(W_1) | W_0$, and since $Y$ is affine and increasing in $R_1$, then
\begin{equation} 
    v^{\ast\tau}(W_0) = \beta^2 W_0 \cdot \max(Q_{\tau}[R_2], R_f) \cdot \max(Q_{\tau}[R_1], R_f)
\end{equation}
To see this, consider no shorting, $\alpha\in[0,1]$. By monotonicity and positive homogeneity of quantiles for non-negative scalings:
\begin{eqnarray}
    Y &=& 0 + \beta v^{\ast\tau}(W_1) | W_0 \\
    &=&  \beta^2 W_1 \max(Q_{\tau}\left[R_2\right],R_f) | W_0 \\
    &=&  \beta^2 W_0 \left[\alpha_0R_1+(1-\alpha_0)R_f\right] \max(Q_{\tau}\left[R_2\right],R_f)
\end{eqnarray}
and since $Y$ is affine and increasing in $R_1$, then
\begin{eqnarray} 
    v^{\ast\tau}(W_0) &=&  \max_{\alpha_{0}\in[0,1]} Q_{\tau}[\beta^2 W_0 \left[\alpha_0R_1+(1-\alpha_0)R_f\right] \max(Q_{\tau}\left[R_2\right],R_f)] \\
    &=&  \beta^2 W_0 \cdot \max(Q_{\tau}\left[R_2\right],R_f) \cdot \max_{\alpha_{0}\in[0,1]} Q_{\tau}[\left[\alpha_0R_1+(1-\alpha_0)R_f\right]] \\
    &=& \beta^2 W_0 \cdot \max(Q_{\tau}[R_2], R_f) \cdot \max(Q_{\tau}[R_1], R_f)
\end{eqnarray}

\subsection{Two-Period Portfolio Choice (T=2) with volatility regimes}

Both sources of uncertainty jointly determine the distribution whose $\tau$-quantile the investor seeks to optimise:
\begin{eqnarray}
    Y &=& \beta v^{\ast\tau}(W_1, z_1) \,\big|\, W_0, z_0 \\
    &=& \beta^2 W_1 \max\!\left(Q_{\tau}[R_2 \mid z_1],\, R_f\right) \,\big|\, W_0, z_0 \\
    &=& \beta^2 W_0 \left[\alpha_0 R_1 + (1-\alpha_0)R_f\right] 
        \max\!\left(Q_{\tau}[R_2 \mid z_1],\, R_f\right) \,\big|\, z_0 \\
    &=& 
    \begin{cases}
        \displaystyle \beta^2 W_0 \left[\alpha_0 (R_1 | z_0) + (1-\alpha_0)R_f\right] 
        \max\!\left(Q_{\tau}[R_2 \mid z_1=L],\, R_f\right) & \text{with prob. } p_{z_0L} \\[8pt]
        \displaystyle \beta^2 W_0 \left[\alpha_0 (R_1 | z_0 )+ (1-\alpha_0)R_f\right] 
        \max\!\left(Q_{\tau}[R_2 \mid z_1=H],\, R_f\right) & \text{with prob. } p_{z_0H}
    \end{cases}
\end{eqnarray}
and the investor's value function is
\begin{eqnarray} 
    v^{\ast\tau}(W_0, z_0) &=&  \max_{\alpha_{0}\in[0,1]} Q_{\tau}[\beta^2 W_0 \left[\alpha_0R_1+(1-\alpha_0)R_f\right] \max(Q_{\tau}\left[R_2 | z_1 \right],R_f) | z_0] \\
    &=& \beta^2 W_0 \max_{\alpha_{0}\in[0,1]} Q_{\tau}\left[\left[\alpha_0R_1+(1-\alpha_0)R_f\right] \max(Q_{\tau}\left[R_2 | z_1 \right],R_f) | z_0\right]
\end{eqnarray}

\subsection{Multiple-Period Portfolio Choice}

% ------------------------------
% A.0.2 Multiple-Period Portfolio Choice (terminal-wealth objective)
% ------------------------------

We keep the setting of one risky asset with gross return \(R_{t+1}\), a risk-free gross rate \(R_f\), rebalancing each period, and \(\alpha_t\in[0,1]\) the share in the risky asset. Wealth evolves as
\begin{equation}
W_{t+1}=W_t\bigl(\alpha_t R_{t+1}+(1-\alpha_t)R_f\bigr).
\end{equation}
As in the one- and two-period illustrations, only terminal wealth is rewarded. With discount factor \(\beta\in(0,1)\),
\begin{equation}
v_\tau^\ast(W_t)\;=\;\max_{\{\alpha_j\}_{j=t}^{T-1}}\;Q_\tau\!\left[\beta^{\,T-t}\,W_T\;\middle|\;W_t\right].
\end{equation}

\textbf{Proposition (Optimal policy and value)}\\
Define \(m_k:=\max\!\bigl(Q_\tau[R_k],\,R_f\bigr)\). Then, for every \(t=0,\dots,T-1\),
\begin{equation}
v_\tau^\ast(W_t)\;=\;\beta^{\,T-t}\,W_t\,\prod_{k=t+1}^{T} m_k,
\end{equation}
and the optimal allocation is the corner rule
\begin{equation}
\alpha_t^\ast
=\begin{cases}
1,& \text{if } Q_\tau[R_{t+1}]> R_f,\\[3pt]
0,& \text{if } Q_\tau[R_{t+1}]< R_f,\\[3pt]
\text{any }\alpha_t\in[0,1],& \text{if } Q_\tau[R_{t+1}]= R_f.
\end{cases}
\end{equation}

\paragraph{Proof.}
\emph{Step \(T-1\).} By monotonicity and affine equivariance of quantiles for positive scalars,
\begin{align}
v_\tau^\ast(W_{T-1})
&=\max_{\alpha_{T-1}} Q_\tau\!\Bigl[\beta\,W_{T-1}\bigl(\alpha_{T-1}R_T+(1-\alpha_{T-1})R_f\bigr)\,\Big|\,W_{T-1}\Bigr] \\
&=\beta W_{T-1}\,\max(Q_\tau[R_T],R_f)\;=\;\beta W_{T-1} m_T.
\end{align}
 Suppose \(v_\tau^\ast(W_{t+1})=\beta^{\,T-(t+1)}W_{t+1}\prod_{k=t+2}^T m_k\). Then
\begin{align}
v_\tau^\ast(W_t)
&=\max_{\alpha_t} Q_\tau\!\Bigl[\beta^{\,T-t}\,W_{t+1}\,\prod_{k=t+2}^{T} m_k\ \Big|\ W_t\Bigr]\\
&=\beta^{\,T-t}\Bigl(\prod_{k=t+2}^T m_k\Bigr)\max_{\alpha_t} Q_\tau[W_{t+1}\mid W_t]\\
&=\beta^{\,T-t}\Bigl(\prod_{k=t+2}^T m_k\Bigr)\,W_t\,\max(Q_\tau[R_{t+1}],R_f)\\
&=\beta^{\,T-t}W_t\prod_{k=t+1}^{T} m_k,
\end{align}
and the same corner rule follows. \(\square\)

Note that \emph{(i) \(T=1\)} gives \(v_\tau^\ast(W_0)=\beta\,W_0\,\max(Q_\tau[R_1],R_f)\), \emph{(ii) \(T=2\)} gives \(v_\tau^\ast(W_0)=\beta^{2}W_0\,\max(Q_\tau[R_1],R_f)\max(Q_\tau[R_2],R_f)\). If \(R_t\stackrel{d}{=}R\) i.i.d., then \(m_k\equiv m=\max(Q_\tau[R],R_f)\) and
\begin{equation}
v_\tau^\ast(W_t)=\beta^{\,T-t}\,W_t\,m^{\,T-t}.
\end{equation}

\subsection{Illustration: Multiple-Period Portfolio Choice}
\label{app:multi-period-example-optim}

A downside-focused investor ($\tau=0.1$) chooses a mixed allocation in the high-volatility initial regime due to the probability of transitioning to a more favourable low-volatility regime in the next period, which flattens the left tail of the target distribution.

More specifically, example considers the case where the risk-free asset return is $R_f = 1.04$ (corresponding to a net return of $4$\%), while the risky asset return $R$ is drawn from a normal distribution with a mean of $1.1$ (corresponding to a net return of $10$\%) and a low volatility regime standard deviation of $\sigma_L = 0.03$ and a high volatility regime standard deviation of $\sigma_H = 1.7\sigma_L$ times the low volatility regime standard deviation. We also assume that volatility is persistent, with transition probabilities of $0.7$ for both high-to-high and low-to-low transitions.Figure then plots the value function depending on the risky asset share for risk-averse investors ($\tau = 0.1$), risk-neutral investors ($\tau = 0.5$), and risk-loving investors ($\tau = 0.9$). While the risk-neutral and risk-loving investors fully invest in the risky asset regardless of the volatility regime, the risk-averse investor reflects the regime in the portfolio allocation. In the final period ($t=1$), the risk-averse agent only invests fully in the risky asset in the low volatility regime. In the high volatility regime, the agent only invests in the risk-free asset. 

\begin{figure}[ht!]
    \begin{center}
        \includegraphics[scale=0.2]{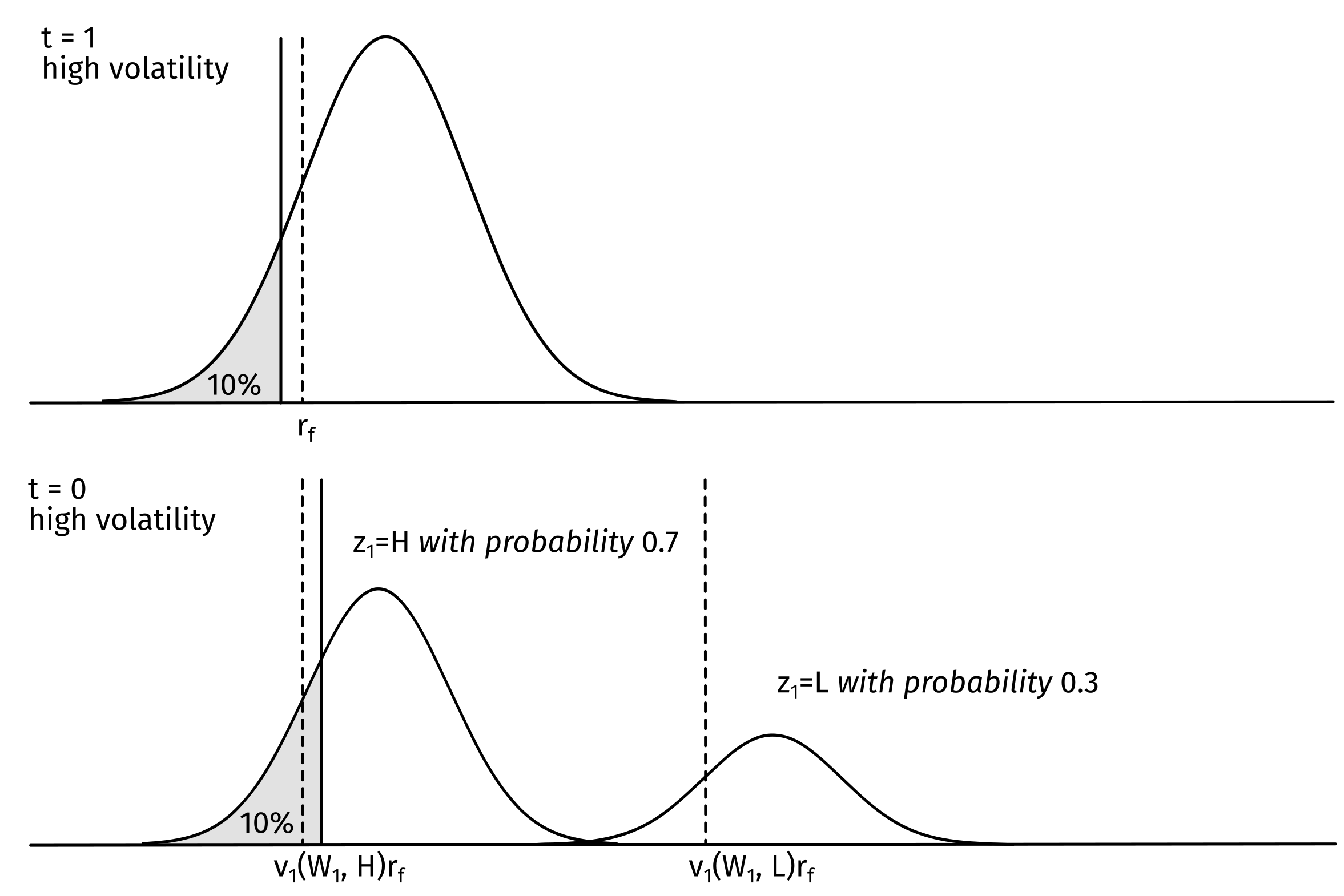}
    \end{center}
    \caption{Portfolio return distribution when mixing with risky asset in high volatility regime. Top panel compares the $0.1$ quantile of the target distribution with risk free asset return in the last period. Bottom panel shows the left tail flattening effect on the target distribution caused by the volatility regime transition in the first period.}
    \label{fig:high_regime_distribution}
\end{figure}

Crucially, unlike in the second period when the risk-averse agent faces high volatility and invests entirely in the risk-free asset, this investor selects a mixed portfolio, allocating  $46\%$ of their initial wealth to the risky asset at time $t=0$. Adding a small amount of the risky asset shifts the target quantile to the right, thinned the left tail of the next period's distribution and making this a better option, even though the current state is volatile. This is because there is a possibility of transitioning to a more plausible low-volatility regime. This investor is willing to take more risk and begins to mix with the risky asset. \autoref{fig:high_regime_distribution} shows the portfolio return distribution when mixing with the risky asset in the high volatility regime. In the second period, $t = 1$, mixing with the risky asset offers no advantage to the risk-averse investor compared to investing solely in the risk-free asset. However, the situation changes in the first period, $t = 0$. There is a 30\% probability that the next period's regime will be low volatility, which shifts part of the target distribution to the right when mixing with the risky asset (as shown by the right normal distribution in the chart). This shift thinned the left tail of the target distribution. As the risk-averse agent focuses on the left tail of the distribution, this flattening effect provides an incentive to deviate from full allocation to the risk-free asset and start mixing with the risky asset. Since the risky asset has a higher expected return than the risk-free asset, mixing with the risky asset shifts the distribution to the right. Eventually, the 0.1 quantile of the target distribution under the mixed portfolio exceeds the 0.1 quantile under full allocation to the risk-free asset. However, allocating too much to the risky asset eventually widens the target distribution, creating an overly heavy left tail. In this example, the 46\% allocation represents the optimal balance between these two effects.

The volatility regime example illustrates the impact of state dependence on investor portfolio allocation, as well as the range of possible allocations that can be created by introducing volatility regimes. The optimal allocation depends on the current state, as well as the investor's expectations regarding future states and their respective transition probabilities. This demonstrates why a downside-focused decision maker would scale down risk in high volatility states and only admit exposure when state-transition probabilities sufficiently reduce the tail risk. Volatility management is what a low-$\tau$ policy does, not an ad hoc overlay.

\section{Computational Details for the Quantile Actor--Critic}
\label{app:qac_details}

This appendix reports the computational details underlying the quantile actor--critic approximation used in the paper. The main text presents only the economic logic of the method. Here we provide the loss functions, update rules, regularization choices, implementation details, and the formal approximation results referenced in Section 3.

\subsection{Critic Approximation and Quantile Loss}

For a grid of quantile levels $0<\tau_1<\cdots<\tau_p<1$, the critic returns the vector
\[
V_{\omega}(s_t)=\left(V_{\omega}^{\tau_1}(s_t),\ldots,V_{\omega}^{\tau_p}(s_t)\right).
\]
Given a transition $(s_t,\alpha_t,r_{t+1},s_{t+1})$, the target for quantile $\tau_j$ is
\begin{equation}
y_{t+1}^{\tau_j}=r_{t+1}+\beta V_{\bar{\omega}}^{\tau_j}(s_{t+1}),
\end{equation}
and the corresponding temporal-difference error is
\begin{equation}
\delta_t^{\tau_j}=y_{t+1}^{\tau_j}-V_{\omega}^{\tau_j}(s_t).
\end{equation}
The critic is trained by minimizing the pinball loss across quantiles,
\begin{equation}
L_{\text{critic}}(\omega)
=
\sum_{j=1}^{p}
\left|\tau_j-\mathbb{1}\{\delta_t^{\tau_j}<0 \}\right|\, |\delta_t^{\tau_j}|
+
\lambda_{\omega}
\sum_{j=1}^{p-1}
\left(
V_{\omega}^{\tau_j}(s_t)-V_{\omega}^{\tau_{j+1}}(s_t)
\right)_{+},
\end{equation}
where $(x)_{+}=\max\{x,0\}$ and $\lambda_{\omega}>0$ penalizes violations of monotonicity across estimated quantiles. The critic update is
\begin{equation}
\omega \leftarrow \omega - \eta_{\omega}\nabla_{\omega} L_{\text{critic}}(\omega).
\end{equation}
The monotonicity penalty is important because the critic approximates several quantiles jointly, ensuring that lower estimated quantiles remain below higher estimated quantiles.

% TODO - consider where to place this: - we might say the main result, that for a fixed policy, there is a fixed point, since it is a contraction (give references ...)

% Importantly, note that the critic update implements a projected $\mathcal{T}_{\pi}^{\tau}$-fixed point in the pinball norm (see \autoref{app:theory} for the approximation bound)
% \begin{equation}
% V^\tau_{\omega}\ \approx\ \Pi_{\mathcal F}\,\mathcal{T}_{\pi}^{\tau}V^\tau_{\omega},
% \end{equation}
% and under assumptions \autoref{ass:critic} in \autoref{app:theory} the critic iterates converge to a stationary point. Standard stochastic-approximation and TD analyses justify these statements \citep{borkar2008stochastic}. For the specific quantile setting, the target is exactly the $\tau$-Bellman target, and distributional Reinforcement Learning provide the analogous theory and estimators for quantiles \citep{bellemare2017distributional,dabney2018distributional}. Unlike \citet{dabney2018distributional}, who use quantile regression to approximate the full action-value return distribution in a Wasserstein-consistent way and then apply mean-greedy control, our model defines a fundamentally different objective: a recursive quantile control problem in which $\tau$ indexes the investor's objective and the stochastic actor is trained to improve the corresponding $\tau$ -specific continuation value.

\subsection{Actor Update and Quantile Policy Gradient}
Let $\pi_{\theta}(\alpha_t \mid s_t)$ denote the stochastic policy over feasible portfolio weights. The actor is updated using a quantile-based policy-gradient surrogate that reinforces actions improving the investor's targeted quantile (\cite{jiang2022quantile, janasek2025gradient}). Defining the quantile-specific temporal-difference error
\begin{equation}
\delta_t^{\tau}=r_{t+1}+\beta V_{\bar{\omega}}^{\tau}(s_{t+1})-V_{\omega}^{\tau}(s_t),
\end{equation}
the actor loss is
\begin{equation}
L_{\text{actor}}(\theta)
=
\log \pi_{\theta}(\alpha_t \mid s_t) \mathbb{1}\{\delta_t^{\tau}\leq0\}
-\lambda_{\theta} H\!\left(\pi_{\theta}(\cdot \mid s_t)\right),
\end{equation}
where $H(\pi_{\theta}(\cdot \mid s_t))$ denotes policy entropy and $\lambda_{\theta}\ge 0$ controls exploration. The update is
\begin{equation}
\theta \leftarrow \theta - \eta_{\theta}\nabla_{\theta}L_{\text{actor}}(\theta).
\end{equation}
This loss has a natural interpretation. We penalize the likelihood of actions that lead to a worse outcome than the $\tau$ quantile value estimate. As a result, the algorithm reinforces actions that improve the $\tau$ quantile value function.

\subsection{Target Network, Training Loop, and Pseudocode}
We use a slowly updated target critic network,
\begin{equation}
\bar{\omega}\leftarrow \rho \omega + (1-\rho)\bar{\omega},
\end{equation}
with smoothing parameter $\rho \in (0,1)$. Training proceeds on-policy: at each episode, the actor generates portfolio weights, the environment returns realized rewards and next states, and the resulting transitions are used to update the actor and critic. Algorithm \ref{alg:quantile-rl} summarizes the full training loop.
\begin{algorithm}[H]
\footnotesize
\caption{Quantile Reinforcement Learning for Portfolio Choice}
\label{alg:quantile-rl}
\begin{algorithmic}
\State Set $\tau\in(0,1)$-quantile preference level
\State \textbf{Initialize:} 
\Statex \hspace{1em} Actor network (policy) weights $\theta$ 
\Statex \hspace{1em} Critic network (quantile value) weights $\omega$
\Statex \hspace{1em} Set Critic target weights $\bar{\omega} \gets \omega$
\Statex \hspace{1em} Quantile levels $0 < \tau_1 < \cdots < \tau_p < 1$
\Statex \hspace{1em} Experience buffer $\mathcal{D} \gets \emptyset$
\For{episode $=1$ \textbf{to} $N_\text{episodes}$}
    \State Sample initial state $s_0$
    \For{each time step $t$}
        \State \textbf{Action selection:} Sample portfolio weights $\alpha_t \sim \pi_\theta(\cdot\,|\,s_t)$ 
        \State  Apply $\alpha_t$, observe reward $r_{t+1}$ and next state $s_{t+1}$
        \State  Store transition $(s_t, \alpha_t, r_{t+1}, s_{t+1})$ in $\mathcal{D}$
        \State \textbf{Critic (Value) update:}
        \For{each quantile $\tau_j$}
            \State Compute TD target: $y_j = r_{t+1} + \beta V^{\tau_j}_{\bar{\omega}}(s_{t+1})$ 
            \State Compute TD error: $\delta^{\tau_j}_t = y_j - V^{\tau_j}_\omega(s_t)$ 
        \EndFor
        \State Update $\omega$ by descending the quantile regression loss:
        \Statex \hspace{2em} $\omega \gets \omega - \eta_\omega \nabla_\omega L_\text{critic}(\omega)$ 
        \State \textbf{Actor (Policy) update:}
        \Statex \hspace{2em} Estimate $Q_\tau[Y\,|\,\theta] \approx V_\omega(s_t)$ 
        \Statex \hspace{2em} Compute policy gradient and update $\theta$:
        \Statex \hspace{2em} $\theta \gets \theta - \eta_\theta \nabla_\theta L_\text{actor}(\theta)$ 
        \State \textbf{Target Critic soft update:} $\bar{\omega} \gets \rho \omega + (1-\rho)\bar{\omega}$ 
        \State Update state: $s_t \gets s_{t+1}$
        \If{episode terminates}
            \State break
        \EndIf
    \EndFor
    \If{episode $\bmod K = 0$}
        \State Evaluate current policy $\pi_\theta$ on validation set
        \State Update monitoring metrics
        \If{early stopping criterion satisfied}
            \State Save best $\theta, \omega$
            \State break
        \EndIf
    \EndIf
\EndFor
\State Save all relevant training metrics and model checkpoints
\State Plot training curves
\State Final evaluation: test $\pi_\theta$ on full training set
\end{algorithmic}
\end{algorithm}

\begin{table}[H]
\centering
\footnotesize
\caption{Main Parameters}
\label{tab:experiment_params}
\begin{tabular}{lcc}
\toprule
\textbf{Parameter}            &   \\
\hline
Beta                         &  0.99 \\
Episodes                      &  50 \\
Batch Size                    &  Episodic \\
Critic LR Start               &  0.01 \\
Critic LR End                 & 0.001 \\
Actor LR Start                &  0.005 \\
Actor LR End                  &  0.001 \\
Rho                           &  0.01 \\
Variance control              &  0  \\
Sigma                          &  0.5 \\
Min Epochs                    &  15 \\
Evaluation                     &  3 \\
Patience                      &  2 \\
Critic decay                  &  1.5 \\
Actor decay                   &  1.5 \\
Reward scale                  &  1221 \\
Order loss Reg                &  5 \\
Rolling window scaling              &  60 days \\
L2                            & 0.0001 \\
Number of exogenous state variables      &  7 \\
Number of  endogenous state variables      & 5 \\
Number of actions = shares      &  3 \\
Critic neurons per layer &  16 \\
Actor neurons per layer &  16 \\
TD error scale & 10
\\
Scaling gradients &  wealth,balance,shares, actor loss
\\
Transaction cost & 0.0001, 0.0005, 0.001, 0.002
\\
Interest rate on Balance & 1.0002 \\
\bottomrule
\end{tabular}
\end{table}

\clearpage

% \begin{algorithm}[h]
% \caption{Quantile Actor--Critic for Dynamic Portfolio Choice}
% \label{alg:qac}
% \begin{algorithmic}[1]
% \State Choose target quantile level $\tau$ and quantile grid $\{\tau_1,\ldots,\tau_p\}$
% \State Initialize actor parameters $\theta$, critic parameters $\omega$, and target critic parameters $\bar{\omega}\leftarrow \omega$
% \For{episode $=1,\ldots,N_{\text{episodes}}$}
%     \State Initialize state $s_0$
%     \For{$t=0,1,\ldots$ until termination}
%         \State Sample portfolio weights $\alpha_t \sim \pi_{\theta}(\cdot \mid s_t)$
%         \State Observe reward $r_{t+1}$ and next state $s_{t+1}$
%         \State Compute critic targets $y_{t+1}^{\tau_j}=r_{t+1}+\beta V_{\bar{\omega}}^{\tau_j}(s_{t+1})$
%         \State Update $\omega$ using $L_{\text{critic}}(\omega)$
%         \State Update $\theta$ using $L_{\text{actor}}(\theta)$
%         \State Soft-update target critic $\bar{\omega}\leftarrow \rho \omega + (1-\rho)\bar{\omega}$
%         \State Set $s_t \leftarrow s_{t+1}$
%     \EndFor
% \EndFor
% \end{algorithmic}
% \end{algorithm}

\subsection{Approximation Results}
The recursive quantile formulation in the main text defines policy-specific and optimal quantile Bellman operators,
\begin{equation}
(T_{\tau}^{\pi}V)(s)=Q_{\tau}\left[r(\alpha,s)+\beta V(s')\mid s\right],
\end{equation}
and
\begin{equation}
(T_{\tau}^{*}V)(s)=\max_{\alpha \in \mathcal{A}(s)}Q_{\tau}\left[r(\alpha,s)+\beta V(s')\mid s\right].
\end{equation}
Under standard boundedness and regularity conditions, these operators are contractions in the sup norm and admit unique fixed points. The critic therefore approximates the fixed point of the policy-specific operator, while the actor performs stochastic ascent on the targeted quantile objective. As a result, the critic converges to a projected fixed point and the actor converges to a stationary point of the quantile objective under a standard two-timescale argument.

The formal assumptions for the convergence, the convergence guarantees and the critic approximation error bound are summarized in \cite{janasek2025gradient}. We refer the reader also to the work of \cite{bellemare2017distributional} and \cite{dabney2018distributional} who provide a foundation for the distributional reinforcement learning and to the work of \cite{jiang2022quantile} who introduced the quantile-based policy-gradient approach.

\subsection{Architecture, Constraints, and Feature Scaling}
The actor maps the current state into a stochastic policy over feasible portfolio weights. The action is constrained to the simplex so that weights are non-negative and sum to one. In practice, we implement this by parameterizing unconstrained latent outputs and transforming them into valid portfolio weights. The critic uses the same state input but returns a vector of continuation values across quantiles. Both networks are deliberately small and use the same architecture across all target quantiles to reduce scope for overfitting and to improve comparability across investor types.

Because naive full-sample normalization would introduce look-ahead bias, all state variables are standardized using rolling-window statistics computed only from past data. We also scale reward and temporal-difference signals to improve numerical stability in low signal-to-noise financial environments.

\subsection{Hyperparameters and Practical Choices}
Table \ref{tab:hyperparameters_qac} reports the main hyperparameters. We keep the global hyperparameters fixed across target quantiles and use multiple random seeds for robustness.

\begin{table}[h]
\centering
\caption{\textbf{Main hyperparameters for the quantile actor--critic implementation.}}
\label{tab:hyperparameters_qac}
\begin{tabular}{ll}
\toprule
Parameter & Value / description \\
\midrule
Discount factor $\beta$ & 0.99 \\
Number of episodes & 50 \\
Critic learning rate & decaying schedule \\
Actor learning rate & decaying schedule \\
Target-network smoothing $\rho$ & 0.01 \\
Quantile monotonicity penalty $\lambda_{\omega}$ & 5 \\
Entropy penalty $\lambda_{\theta}$ & as specified in experiments \\
Rolling normalization window & 60 days \\
Transaction costs & experiment-specific \\
Evaluation & multiple seeds, out-of-sample \\
\bottomrule
\end{tabular}
\end{table}
The exact network architecture, training schedule, early stopping criteria, and environment-specific parameter values can be modified across applications. Its important to note that the same quantile-targeted learning principle is used throughout.

\section{Construction of the Volatility-Management Benchmarks}
\label{app:vm_implementation}

This appendix describes the construction of the volatility-management benchmarks used in
Sections 4 and 6. The purpose of these benchmarks is not to solve the investor's dynamic
quantile problem, but to provide disciplined reduced-form timing rules that can be
implemented under the same investability and trading-friction discipline as the
tail-managed portfolios. Throughout, benchmark parameters are estimated out of
sample using the same expanding-window protocol as in the main exercise.

\subsection{Out-of-sample protocol}

All benchmark returns are generated in real time. We begin with an initial five-year
estimation window, estimate the benchmark parameters using only information available in
that window, and evaluate the resulting benchmark over the subsequent two-year
out-of-sample period. The estimation sample is then expanded to include the newly
observed data, the benchmark is re-estimated, and performance is evaluated on the next
two-year block. Repeating this procedure sequentially yields a concatenated out-of-sample
return series for each benchmark.

This design ensures that benchmark performance is evaluated under the same information
set as the tail-managed portfolios. In particular, all volatility signals and any
normalization objects are constructed using lagged or past data only, so the benchmark
comparison is free of look-ahead bias by construction.

\subsection{Market-volatility-managed benchmark}

The main-text volatility-management (VM) benchmark follows the multifactor timing logic in
DeMiguel et al.\ (2024). Let $r_{k,t+1}$ denote the return on factor $k=1,\dots,K$ over the
next rebalancing period and let $\sigma^M_t$ denote lagged realized market volatility.
Benchmark factor exposures are parameterized as
\[
\theta_{k,t}=a_k+b_k\frac{1}{\sigma^M_t},
\]
where $a_k$ and $b_k$ are estimated on the training sample. The benchmark
reduces risky exposure when market volatility is high and restores exposure when market
volatility is low.

Because our empirical comparison is conducted in a long-only, fully invested portfolio
environment, the raw exposures $\theta_t=(\theta_{1,t},\dots,\theta_{K,t})'$ are converted into
tradable weights by normalization:
\[
w_t=\frac{\theta_t}{\mathbf{1}'\theta_t}.
\]
The benchmark portfolio return is then
\[
r^{VM}_{p,t+1}=w_t' r_{t+1},
\]
where $r_{t+1}=(r_{1,t+1},\dots,r_{K,t+1})'$ collects factor returns. This normalization is
economically important. It places the benchmark on the same fully invested simplex as the
tail-managed portfolios, so differences in performance reflect differences in timing and
cross-factor allocation rather than differences in admissible leverage or financing.

We estimate the timing coefficients by maximizing a net-of-cost mean--variance objective
on the training sample:
\[
\max_{a,b}\;\; \hat{\mu}_p(a,b)-\frac{\gamma}{2}\hat{\sigma}^2_p(a,b)-TC(a,b),
\]
where $\hat{\mu}_p(a,b)$ and $\hat{\sigma}^2_p(a,b)$ are the sample mean and variance of the
normalized benchmark return series implied by $(a,b)$, and $TC(a,b)$ denotes transaction
costs. In the main specification, $\gamma=3$, weights are constrained to be non-negative after
normalization so that the benchmark remains directly comparable to the long-only
tail-managed portfolios.

A limitation of this benchmark is that the optimization problem is not globally
convex in $(a,b)$. The affine timing rule is followed by a normalization step that maps raw
exposures into fully invested portfolio weights, and the transaction-cost term depends on
turnover in these normalized weights. As a result, the numerical solution can depend on the initial guess. In the empirical implementation
we therefore use randomized initializations similar to our RL algorithm and averaged the results over seeds.

\subsection{Transaction costs and turnover}

Transaction costs are computed from turnover in drifted pre-trade weights. Let
$\tilde{w}_{t-1}$ denote the portfolio weights chosen at the previous rebalancing date and let
$r_t$ denote realized factor returns between $t-1$ and $t$. The corresponding pre-trade
weights are
\[
w^{pre}_t=
\frac{\tilde{w}_{t-1}\circ(1+r_t)}
{\mathbf{1}'[\tilde{w}_{t-1}\circ(1+r_t)]},
\]
where $\circ$ denotes the element-by-element product. One-way turnover is
\[
\tau_t=\frac{1}{2}\sum_{k=1}^K |w_{k,t}-w^{pre}_{k,t}|.
\]
Given proportional trading-cost parameter $c$, average transaction costs are
\[
TC=c\cdot\frac{1}{T-1}\sum_{t=2}^T \tau_t.
\]

This turnover measure matches the trading logic of the main portfolio problem. The
benchmark is therefore not evaluated as a frictionless timing overlay, but as an
implementable managed portfolio whose trading intensity is penalized under the same
drift-based rebalancing discipline as the learned policies.

\subsection{factor-level volatility-scaling benchmark}

Secion \ref{sec:factorVM} considers a harder robustness benchmark that replaces the common market
volatility signal with factor-specific volatility signals. Rather than estimating the full
factor-level DeMiguel-style rule,
\[
\theta_{k,t}=a_k+b_k\frac{1}{\sigma_{k,t}},
\]
we use a mechanically scaled version of the in-sample unconditional mean--variance
portfolio. This choice improves numerical stability because it avoids the non-convex joint
optimization over factor-specific timing coefficients.

Specifically, we first estimate the unconditional mean--variance portfolio on the training
sample and obtain static weights \(w^{UC}\). Out of sample, we scale each factor weight by
its own lagged realized volatility,
\[
\tilde{w}_{k,t}=\frac{w^{UC}_k}{\sigma_{k,t}},
\]
and normalize the resulting exposures to lie on the fully invested simplex,
\[
w^{FVM}_{k,t}
=
\frac{\tilde{w}_{k,t}}
{\sum_{j=1}^K \tilde{w}_{j,t}}.
\]
In this benchmark, transaction costs enter the in-sample estimation only through the drift-rebalancing cost of the static unconditional portfolio. The subsequent factor-level inverse-volatility scaling is applied mechanically out of sample. We then evaluate the scaled portfolio net of transaction costs by computing turnover from drifted pre-trade weights. Thus, costs are included in performance evaluation, but the factor-level volatility-scaling rule is not itself optimized with respect to transaction costs.

The main-text benchmark asks whether a tail-specific
objective changes standard market-level volatility timing, whereas the robustness benchmark
asks whether tail targeting adds value relative to a rule that already allows
cross-sectional de-risking at the factor level.

This is why the factor-level benchmark is intentionally demanding. If tail-managed
portfolios deliver smaller incremental gains relative to this harder benchmark, that should
not be interpreted as a failure of the framework. Rather, it indicates that factor-level
volatility scaling already approximates part of the selective de-risking logic that emerges
endogenously from a downside-focused quantile objective.

\subsection{Interpretation}

The benchmarks should therefore be read as disciplined reduced-form comparators. They
use lagged volatility to manage exposure, whereas the tail-managed portfolios solve a
different economic problem: they allocate across factors to improve a targeted part of the
conditional payoff distribution. The empirical comparison is informative precisely because all
portfolios are evaluated out of sample, under the same expanding-window protocol, and
under comparable investability and transaction-cost constraints.

% =========================
% Section 4 (Main text)
% =========================
\section{Simulated Data: A Multi-Period Choice with Volatility Regimes}
\label{app:simdynexample}

Before moving on to real-world data, we will consider a more realistic yet still tractable dynamic environment that incorporates regime-switching return dynamics and transaction costs. This does not have a closed-form solution, so we will use our algorithm to approximate the behaviour. The key behavioural shift in the earlier illustration was due to the fact that volatility transitions create nested conditional quantiles, breaking the all-or-nothing corner solutions and introducing interior solutions. To further illustrate this, we will introduce an additional random asset and volatility dynamics, extending the two-period example to a sequential cost-frictional setting. In this setting, today's action shifts the state distribution tomorrow, thereby affecting the future quantile of interest to the investor.

To achieve this, we use a three-regime vector autoregression to model the log-returns of risky assets, incorporating `bull, `neutral' and `bear' regimes, which follow a Markov transition, in a similar way as used in \cite{ang2002international,guidolin2007asset}, and account for portfolio rebalancing costs. More specifically, we consider two assets linked by VAR(1) dynamics across the three regimes, such that risky asset $r_{t,1}$ will have a higher premium than risky asset $r_{t,2}$, albeit with slightly higher variance, in a bull market (see \autoref{app:simdyn} for details). It will also have a slightly higher mean and substantially larger variance than asset two in a neutral market, which weakens its risk-adjusted advantage. Finally, it will have strongly negative returns with higher variance than asset two in a bear market. This makes asset two act as a hedge during market downturns (see the parameters of the regime-switching VAR in \autoref{tab:sim_var} in \autoref{app:simdyn}). The scenarios represented by the transition matrices determine how frequently the economy switches between regimes, thus controlling the long-run distribution of states and the exposure of portfolio policies to regime-dependent risks. In our analysis, we focus on these matrices. We define the Bull-Bear matrix as having a stationary distribution of 0.5, 0.1 and 0.4; the Neutral-Bear matrix as having a stationary distribution of 0.1, 0.4 and 0.5; and the Bull-Neutral matrix as having a stationary distribution of 0.5, 0.4 and 0.1 (see \autoref{tab:sim_Q} in \autoref{app:simdyn}).

The state collects the pre-trade portfolio from the previous period, the current returns and the active regime; the action is full allocation to the asset simplex (two risky portfolios and cash). Policies are learnt using a Dirichlet actor (with stochastic weights over the simplex) and a distributional critic that estimates the inverse CDF value across a grid of quantiles. See \autoref{app:simdyn} for formal definitions and training details.\footnote{Two practical elements make the experiment stable and interpretable: (i) a Dirichlet policy that samples weights on the simplex (including cash) to ensure interior exploration and well-behaved learning on the budget constraint, and (ii) transition-weighted losses that use the Markov matrix $Q$ to correct for enumerating next regimes during simulation (self-normalised importance weights). This means that the critic and actor ``see'' the true regime mixture implied by $Q$ (see \autoref{app:simdyn_training})}.

\begin{figure}[t!]
\includegraphics[width=0.31\textwidth]{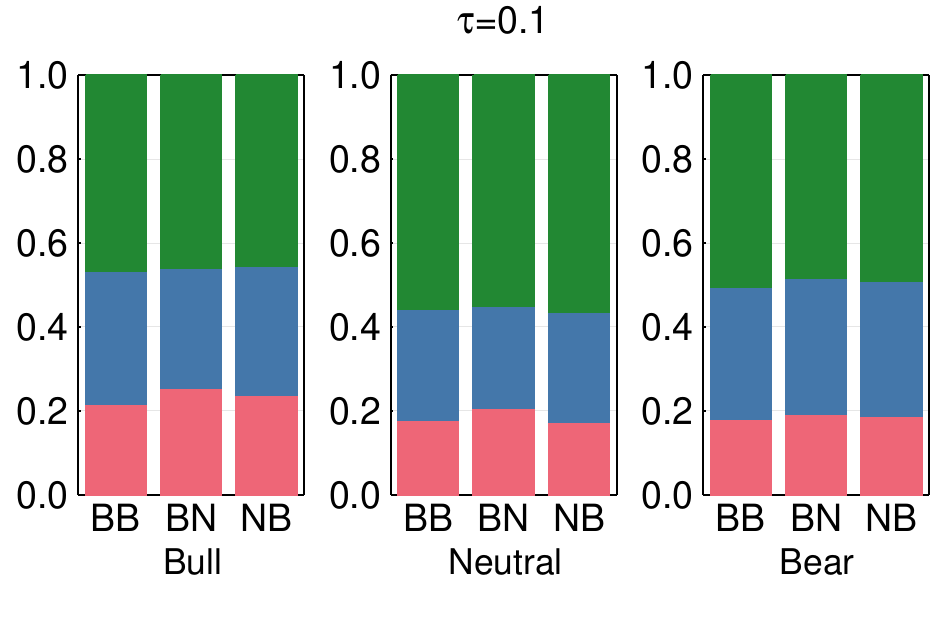}
\hspace{0.2cm}
\includegraphics[width=0.31\textwidth]{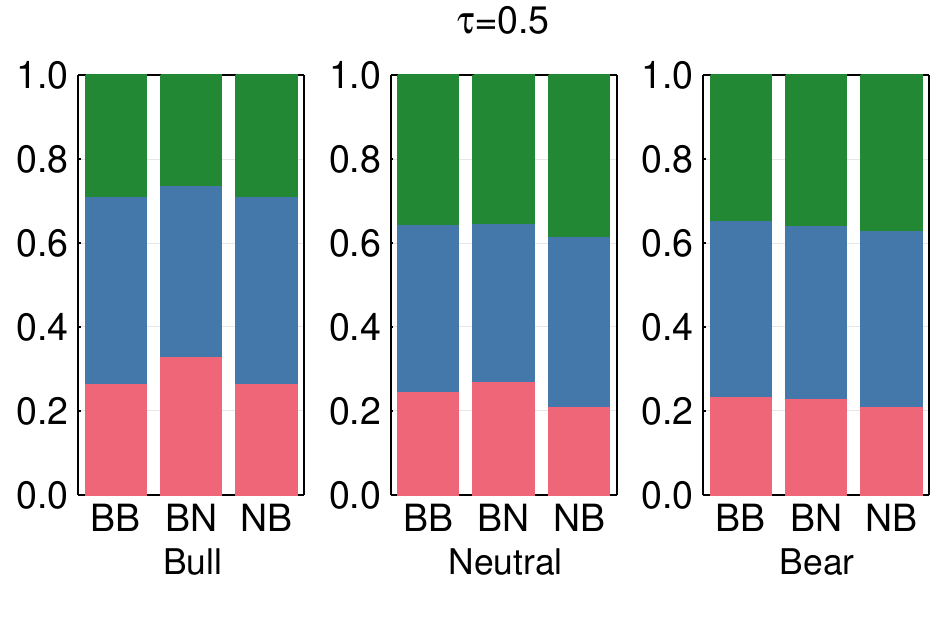}
\hspace{0.2cm}
\includegraphics[width=0.31\textwidth]{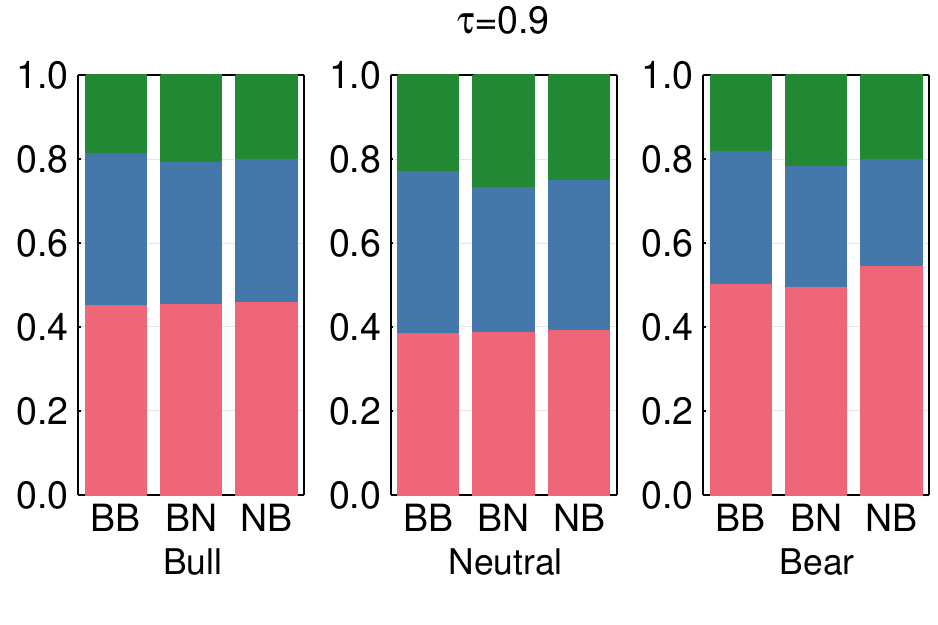}
\caption{\textbf{Regime-conditional portfolio weights for $\tau$-targeting policies.} Each stacked bar shows the allocation to the higher-dispersion portfolios $w_1$ (pink), the more defensive portfolio $w_2$ (blue), and cash (green) within a regime (Bull, Neutral, Bear) for three risk attitudes -- downside-focused, neutral, and upside-seeking -- captured by $\tau\in\{0.1,0.5,0.9\}$ and three transition-mix scenarios (BB = Bull--Bear, NB = Neutral--Bear, BN = Bull--Neutral) defined in \autoref{app:simdyn}. Bars sum to one. Together, the panels illustrate preference-consistent rebalancing: low-$\tau$ policies de-risk in high-variance environments, while high-$\tau$ policies concentrate in the higher-dispersion portfolio.}
\label{fig:sim_weights}
\end{figure}

From the resulting portfolios presented in \autoref{fig:sim_weights}, we can immediately see that the more risk-seeking the agent is, the less cash they are holding, with larger allocations to riskier assets depending on different states. As the parameter $\tau$ rises, the weight shifts towards the higher-dispersion factor (w1) and cash and low-dispersion factors decline; as $\tau$ falls, allocations move towards cash and the more defensive factor (w2). On average across regimes and transition scenarios, the mean weights change from approximately $(w1,w2,\text{cash})\!\approx\!(0.20,0.29,0.50)$ at $\tau{=}0.1$ to $(0.46,0.33,0.21)$ at $\tau{=}0.9$ (Appendix \autoref{tab:sim_avg_policy}).

The risk-neutral policy ($\tau = 0.5$) allocates more to asset 2, as it offers a more balanced risk-reward trade-off with lower volatility in each regime. In contrast, the risk-seeking policy ($\tau = 0.9$) focuses on the upper tail, where asset 1 offers a higher upside potential, despite its negative average returns in the bear regime. \autoref{fig:sim_weights} also shows how the optimal portfolio weights differ across regimes and scenarios. For example, in the bull-bear (BB) scenario, where there is a higher likelihood of entering or remaining in bear states, asset 2 acts as a hedge and the portfolio tilts more towards this asset. By contrast, the neutral-bear scenario exhibits the highest cash, as both the neutral and bear states are persistent, and the risk premium of asset 1 is weaker in this mix. Regardless of how persistent the regime is, the neutral regime always brings the highest cash to the portfolio across all values of $\tau$. The fine scenarios adjust the mix slightly, but do not change the ordering driven by $\tau$. This is due to the low drift and substantially higher variances, which eliminate diversification benefits. The neutral state is also highly persistent, which increases the expected duration of this high-volatility, low-drift scenario (see Appendix \autoref{tab:sim_by_scenario}).

\begin{figure}[H]\centering
\includegraphics[width=0.55\textwidth]{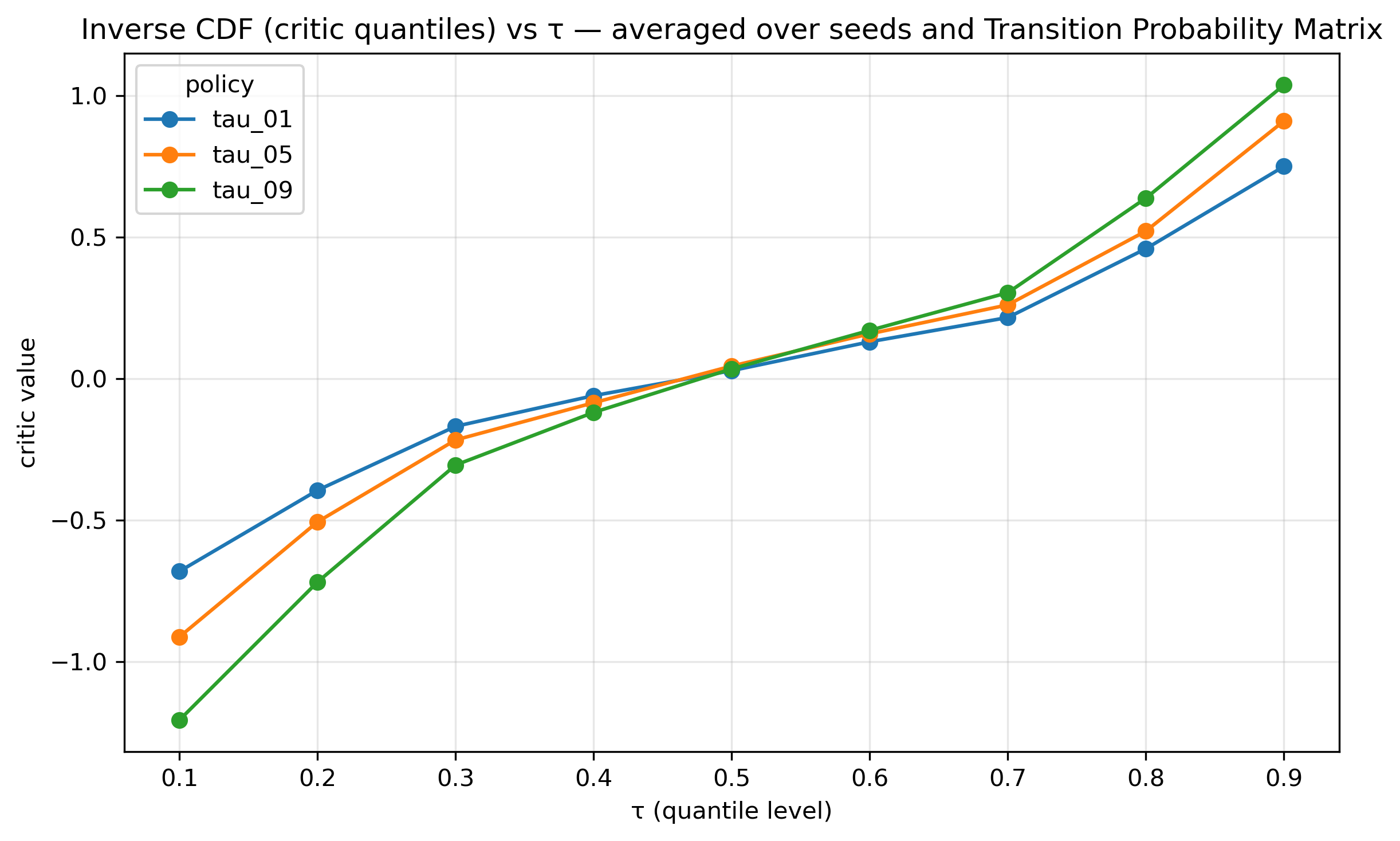}
\caption{Inverse-CDFs of learned value across policies (averaged across states and scenarios).}\label{fig:sim_icdf}
\end{figure}

Crucially, the inverse-CDFs intersect precisely where expected: the policy with a threshold of $\tau{=}0.1$ dominates the left tail, the policy with a threshold of  $\tau{=}0.9$ dominates the right tail, and the policy with a threshold of $\tau{=}0.5$ sits between them. This is evidence that the policy optimises quantiles rather than a mean-variance proxy (Appendix \autoref{tab:sim_icdf} and \autoref{fig:sim_icdf}).

In summary, whereas in the earlier illustration, volatility regimes created nested conditional quantiles and overturned corner solutions, even in a two-asset toy model, the same mechanism emerges in a richer dynamic state space here. Quantile investors adjust not only the scale, but also the composition, as regimes and costs evolve. Low-$\tau$ policies de-risk endogenously in adverse or high-variance states, while high-$\tau$ policies capitalise on the upside when the transition structure makes favourable regimes sufficiently likely. This provides a micro-founded account of volatility management within a dynamic, cost-aware environment.

\subsection{Multi-Period Choice with Realistic Dynamics - Full Details}
\label{app:simdyn}

\subsection{State dynamics, costs, and dynamic quantile objective}\label{app:simdyn_model}
Let $r_t\!\in\!\mathbb{R}^N$ be risky-asset log-returns and $k_t\!\in\!\{1,\dots,K\}$ the regime. Conditional on $k_t\!=\!k$,
\[
r_{t+1}=c_k+\Phi\, r_t+u^{(k)}_{t+1},\qquad u^{(k)}_{t+1}\sim \mathcal{N}(0,\Sigma_k),
\]
with $K{=}3$ regimes (Bull, Neutral, Bear). Regimes follow a first-order Markov chain with row-stochastic $Q$ so $\Pr(k_{t+1}{=}k'|k_t{=}k)=Q_{kk'}$. The state is 
\[
s_t=\big(w^{\text{prev}}_t,\; r_t,\; (k_t)\big), 
\]
and the action $w_t\in\Delta_{N+1}$ (risky factors $+$ cash) gives gross portfolio return
\[
R_p(w_t;R_{t+1})=w^\top_{\!t,\text{risky}}\, R_{t+1}+\big(1-\mathbf{1}^\top w_{t,\text{risky}}\big)R_f,
\]
with proportional $L^1$ turnover cost 
\(
T(w_t;w^{\text{prev}}_t,r_t)=\tfrac12\|w_t-w^{\text{pre}}_t\|_1
\)
on pre-trade weights $w^{\text{pre}}_t$ (drifted by returns). The \emph{dynamic} quantile value follows
\[
v_\tau^\pi(s_t)=Q_\tau\!\left[r(w_t,s_t)+\beta\,v_\tau^\pi(s_{t+1})\mid s_t\right],
\]
and the optimal policy solves $\max_{\pi} v_\tau^\pi(\cdot)$ (quantile of the discounted stream), not a mean of utilities. See \autoref{sec:theory} for how volatility transitions create nested conditional quantiles and interior solutions even in two-period examples.

\subsection{Policy/value approximation and training}\label{app:simdyn_training}
We use a \emph{Dirichlet} actor to parameterize $\pi_\phi(w|s)$ on $\Delta_{N+1}$ (risky factors plus cash), adding a small positive bias to the concentration vector to stabilize exploration. The critic $V_\omega$ predicts a vector of conditional quantiles $\{v_{\omega}^\tau(s)\}_{\tau\in\mathcal{T}}$ with pinball loss and a monotonicity penalty to avoid quantile crossing. Because we \emph{enumerate} next regimes during tuple generation, we weight per-sample losses by $q=Q_{kk'}$ and use self-normalized averages,
\[
\mathcal{L}=\frac{\sum_i q_i\,\ell_i}{\sum_i q_i},
\]
so training respects the true regime mixture implied by $Q$ (importance weighting). \autoref{alg:mbqac} summarizes the model-based QAC loop.

\begin{algorithm}[H]
\caption{Model-Based QAC with Dirichlet Policy (summary)}\label{alg:mbqac}
\begin{algorithmic}[1]
\State Input RS-VAR(1) params $(c_k,\Phi,\Sigma_k,Q)$; grids for $w^{\text{prev}}$ and $r_t$; discount $\beta$; transaction cost $c$.
\For{epochs}
 \State Build a transition buffer by looping $(w^{\text{prev}},r_t,k)$, sampling $w_t\!\sim\!\pi_\phi(\cdot|s_t)$, enumerating $k'$, simulating $r_{t+1}$, computing $u=U(R_p)-c\,T$, and storing $(s_t,w_t,u,s'_{t+1},q{=}Q_{kk'})$.
 \State Update critic by quantile TD with $q$-weighted pinball$+$order loss; update actor with $q$-weighted quantile advantage; soft-update target critic.
\EndFor
\end{algorithmic}
\end{algorithm}
\vspace{-0.75\baselineskip}

\subsection{Parameters and scenarios}\label{app:simdyn_params}
We fix hyperparameters across $\tau$ and average over multiple seeds (\autoref{tab:sim_hparams}). Regime-specific RS-VAR(1) parameters and three transition matrices (Bull-Bear, Neutral-Bear, Bull-Neutral) appear in \autoref{tab:sim_var}-\autoref{tab:sim_Q}; the stationary distributions match the economic narratives used later (tilts follow the regime mix).

\begin{table}[H]
\centering\small
\caption{Model and network hyperparameters}\label{tab:sim_hparams}
\begin{tabular}{l l}
\toprule
Episodes & 10 \quad (model-based training)\\
Discount $\beta$ & 0.96 \\
Learning quantiles & $\{0.1,0.5,0.9\}$ with 10 heads total \\
Actor/Critic layers & 2 hidden layers, 32 neurons each; L2=1e-4 \\
Actor output & Dirichlet (softplus), \\
Critic loss & Pinball $+$ order penalty (weight 5.0) \\
Transaction cost & $10^{-3}$ (proportional $L^1$ turnover) \\
Risk-free & $R_f=1.001$ \\
\bottomrule
\end{tabular}
\end{table}

\begin{table}[H]
\centering\small
\caption{Regime-specific RS-VAR(1) parameters ($K{=}3$, $N{=}2$)}
\label{tab:sim_var}
\begin{tabular}{l c c c}
\toprule
 & Bull & Neutral & Bear \\
\midrule
$\Phi$ & \multicolumn{3}{c}{$\begin{bmatrix}0.15 & 0.10\\ 0.10 & 0.15\end{bmatrix}$ (fixed across regimes)}\\[2pt]
$c_k$ & $[0.0040,\,0.0030]$ & $[0.0030,\,0.0028]$ & $[-0.0090,\,0.0030]$\\
$\Sigma_k$ & $\begin{bmatrix}0.0005&0.00010\\ 0.00010&0.00045\end{bmatrix}$ & 
$\begin{bmatrix}0.0018&0\\ 0&0.0014\end{bmatrix}$ &
$\begin{bmatrix}0.0050&-0.0030\\ -0.0030&0.0020\end{bmatrix}$\\
\bottomrule
\end{tabular}
\end{table}

\begin{table}[H]
\centering\small
\caption{Transition matrices $Q$ and stationary distributions $\pi$}
\label{tab:sim_Q}
\begin{tabular}{l c c}
\toprule
Scenario & $Q$ (rows: B, N, Br) & $\pi$ \\
\midrule
Bull-Bear & $\begin{bmatrix}0.74&0.02&0.24\\ 0.10&0.82&0.08\\ 0.30&0.02&0.68\end{bmatrix}$ & $[0.50,0.10,0.40]$ \\
Neutral-Bear & $\begin{bmatrix}0.82&0.08&0.10\\ 0.02&0.68&0.30\\ 0.02&0.24&0.74\end{bmatrix}$ & $[0.10,0.40,0.50]$ \\
Bull-Neutral & $\begin{bmatrix}0.74&0.24&0.02\\ 0.30&0.68&0.02\\ 0.10&0.08&0.82\end{bmatrix}$ & $[0.50,0.40,0.10]$ \\
\bottomrule
\end{tabular}
\end{table}

\subsection{Evaluation diagnostics and portfolio policies}\label{app:simdyn_results}
The critic's inverse-CDFs average over states, regimes, and scenarios and are ordered by the targeted quantile; they cross in the middle quantiles as expected (\autoref{tab:sim_icdf}). Portfolio weights exhibit (i) $\tau$-monotone shifts from cash $\to$ $w2$ $\to$ $w1$, and (ii) scenario-specific reallocations consistent with the stationary mix and covariances (\autoref{tab:sim_by_regime}-\autoref{tab:sim_avg_policy}).

\begin{table}[H]
\centering\small
\caption{Inverse-CDF values across quantiles (averaged over states, seeds, scenarios)}\label{tab:sim_icdf}
\begin{tabular}{lrrrrrrrrr}
\toprule
Policy & 0.1 & 0.2 & 0.3 & 0.4 & 0.5 & 0.6 & 0.7 & 0.8 & 0.9\\
\midrule
$\tau\!=\!0.1$ & -0.6807 & -0.3943 & -0.1687 & -0.0603 & \textbf{0.0280} & 0.1300 & 0.2157 & 0.4587 & 0.7493 \\
$\tau\!=\!0.5$ & -0.9120 & -0.5060 & -0.2167 & -0.0853 & \textbf{0.0443} & 0.1580 & 0.2603 & 0.5210 & 0.9097 \\
$\tau\!=\!0.9$ & -1.2057 & -0.7183 & -0.3053 & -0.1193 & \textbf{0.0343} & 0.1703 & 0.3033 & 0.6373 & 1.0367 \\
\bottomrule
\end{tabular}
\end{table}

\begin{table}[H]
\centering\small
\caption{Average portfolio weights by regime and $\tau$}\label{tab:sim_by_regime}
\begin{tabular}{lccc}
\toprule
Policy/Regime & $w1$ & $w2$ & Cash\\
\midrule
$\tau{=}0.1$ (Bear/Bull/Neutral) & 0.1872 / 0.2362 / 0.1862 & 0.3201 / 0.3042 / 0.2568 & 0.4927 / 0.4596 / 0.5570 \\
$\tau{=}0.5$ (Bear/Bull/Neutral) & 0.2257 / 0.2881 / 0.2437 & 0.4175 / 0.4329 / 0.3931 & 0.3568 / 0.2790 / 0.3632 \\
$\tau{=}0.9$ (Bear/Bull/Neutral) & 0.5168 / 0.4582 / 0.3917 & 0.2868 / 0.3473 / 0.3625 & 0.1965 / 0.1946 / 0.2459 \\
\bottomrule
\end{tabular}
\end{table}

\begin{table}[H]
\centering\small
\caption{Average portfolio weights by scenario (across $\tau$ and regimes)}\label{tab:sim_by_scenario}
\begin{tabular}{lccc}
\toprule
Scenario & $w1$ & $w2$ & Cash\\
\midrule
Bull-Bear & 0.2969 & \textbf{0.3584} & 0.3447\\
Bull-Neutral & \textbf{0.3151} & 0.3357 & 0.3492\\
Neutral-Bear & 0.2992 & 0.3463 & \textbf{0.3545}\\
\bottomrule
\end{tabular}
\end{table}

\begin{table}[H]
\centering\small
\caption{Average policy across risk aversion (across scenarios and regimes)}\label{tab:sim_avg_policy}
\begin{tabular}{lccc}
\toprule
Policy & $w1$ & $w2$ & Cash\\
\midrule
$\tau{=}0.1$ & 0.2032 & 0.2937 & 0.5031 \\
$\tau{=}0.5$ & 0.2525 & 0.4145 & 0.3330 \\
$\tau{=}0.9$ & 0.4555 & 0.3322 & 0.2123 \\
\bottomrule
\end{tabular}
\end{table}

\clearpage
\section{Additional OOS Empirical Results}
\label{app:OOSresults_additional}

% Auto-generated by Julia
\begin{table}[!htbp]
\centering
\caption{\textbf{Comparison against VM across states and crisis windows.}}
\label{tab:vm_horserace_crisis}
\small
\setlength{\tabcolsep}{4pt}
\begin{tabular}{llrrrrrrrrrrr}
\toprule
Strategy & Episode & Alpha & SE($\alpha$) & $t(\alpha)$ & Beta & $R^2$ & N & CumRet & CumActive & Mean & IR & $\Delta$CVaR5 \\
\midrule
$\tau$ = 0.1 & Dot-com bust & 9.09 & 3.25 & 2.80 & 0.26 & 0.504 & 638 & 16.94 & 49.52 & 16.80 & 1.25 & 1.56 \\
$\tau$ = 0.5 & Dot-com bust & 5.80 & 2.94 & 1.97 & 0.49 & 0.797 & 638 & 0.67 & 30.95 & 11.13 & 1.14 & 1.12 \\
$\tau$ = 0.9 & Dot-com bust & 3.05 & 3.90 & 0.78 & 0.60 & 0.754 & 638 & -9.36 & 18.86 & 7.24 & 0.80 & 0.80 \\
\addlinespace[0.3em]
$\tau$ = 0.1 & GFC & 1.02 & 3.28 & 0.31 & 0.60 & 0.926 & 356 & -17.86 & 14.95 & 10.31 & 1.09 & 1.23 \\
$\tau$ = 0.5 & GFC & 0.66 & 2.86 & 0.23 & 0.82 & 0.969 & 356 & -24.60 & 7.05 & 4.96 & 0.97 & 0.55 \\
$\tau$ = 0.9 & GFC & -0.66 & 2.88 & -0.23 & 1.00 & 0.980 & 356 & -31.07 & -0.92 & -0.61 & -0.20 & -0.06 \\
\addlinespace[0.3em]
$\tau$ = 0.1 & COVID crash & -13.59 & 12.32 & -1.10 & 0.59 & 0.986 & 24 & -13.39 & 7.14 & 75.81 & 2.90 & 3.34 \\
$\tau$ = 0.5 & COVID crash & -13.82 & 12.86 & -1.07 & 0.79 & 0.993 & 24 & -17.37 & 3.00 & 32.00 & 2.32 & 1.69 \\
$\tau$ = 0.9 & COVID crash & -20.29 & 12.77 & -1.59 & 0.85 & 0.995 & 24 & -19.05 & 1.15 & 12.50 & 1.23 & 1.27 \\
\addlinespace[0.3em]
$\tau$ = 0.1 & 2022 bear & 3.12 & 4.77 & 0.66 & 0.47 & 0.811 & 196 & -4.19 & 9.63 & 12.31 & 1.25 & 1.26 \\
$\tau$ = 0.5 & 2022 bear & 3.31 & 2.97 & 1.12 & 0.75 & 0.967 & 196 & -7.86 & 6.10 & 7.73 & 1.57 & 0.63 \\
$\tau$ = 0.9 & 2022 bear & 1.20 & 2.53 & 0.48 & 0.97 & 0.985 & 196 & -12.40 & 1.38 & 1.78 & 0.84 & 0.08 \\
\addlinespace[0.3em]
\bottomrule
\end{tabular}
\begin{tablenotes}
\footnotesize
\item \textit{Notes:} Each row reports the regression
\[
r_{p,t}=\alpha+\beta r_{VM,t}+\varepsilon_t
\]
estimated within the indicated state or crisis window. Reported alpha is annualized and expressed in percent. Standard errors are Newey--West. State definitions follow Table~\ref{tab:state_perf}. Crisis windows correspond to the dot-com bust, the global financial crisis, the COVID crash, and the 2022 bear market. Because these windows are short, alpha estimates are noisier than in the full sample and should be interpreted as supportive rather than definitive evidence.
\end{tablenotes}
\end{table}

\begin{figure}[ht]
    \begin{center}
        \includegraphics[width=0.6\textwidth]{tvp_alpha_ci_tau__0p1.pdf}
        \includegraphics[width=0.6\textwidth]{tvp_beta_ci_tau__0p1.pdf}
        \includegraphics[width=0.6\textwidth]{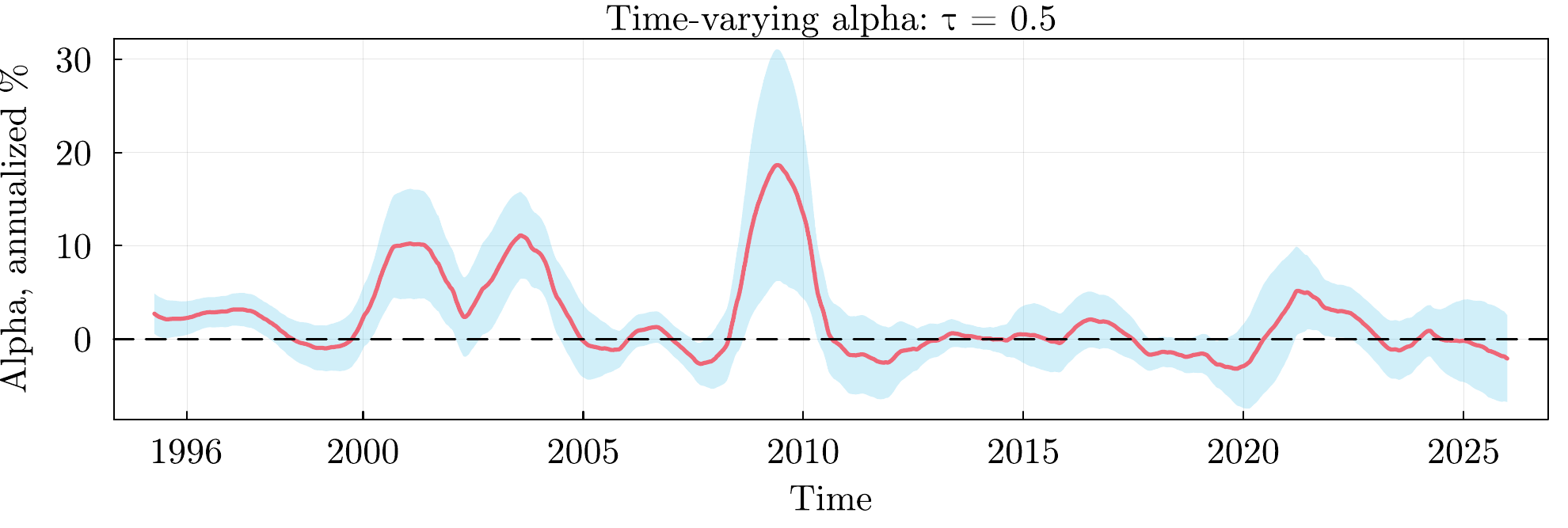}
        \includegraphics[width=0.6\textwidth]{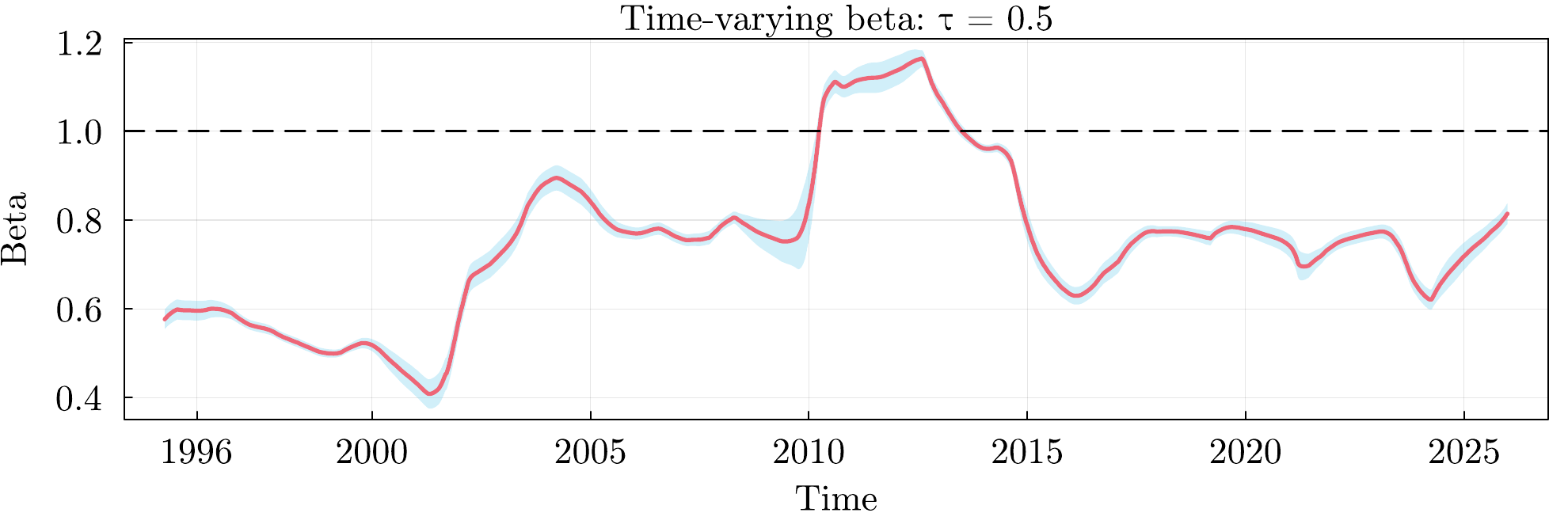}
        \includegraphics[width=0.6\textwidth]{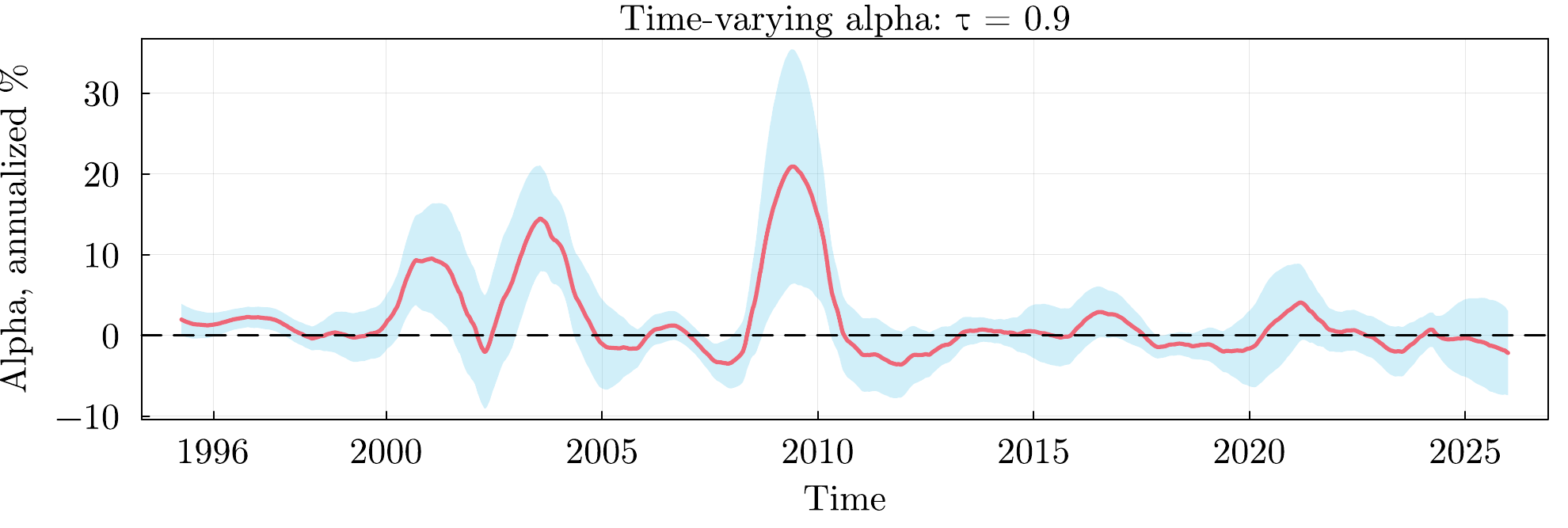}
        \includegraphics[width=0.6\textwidth]{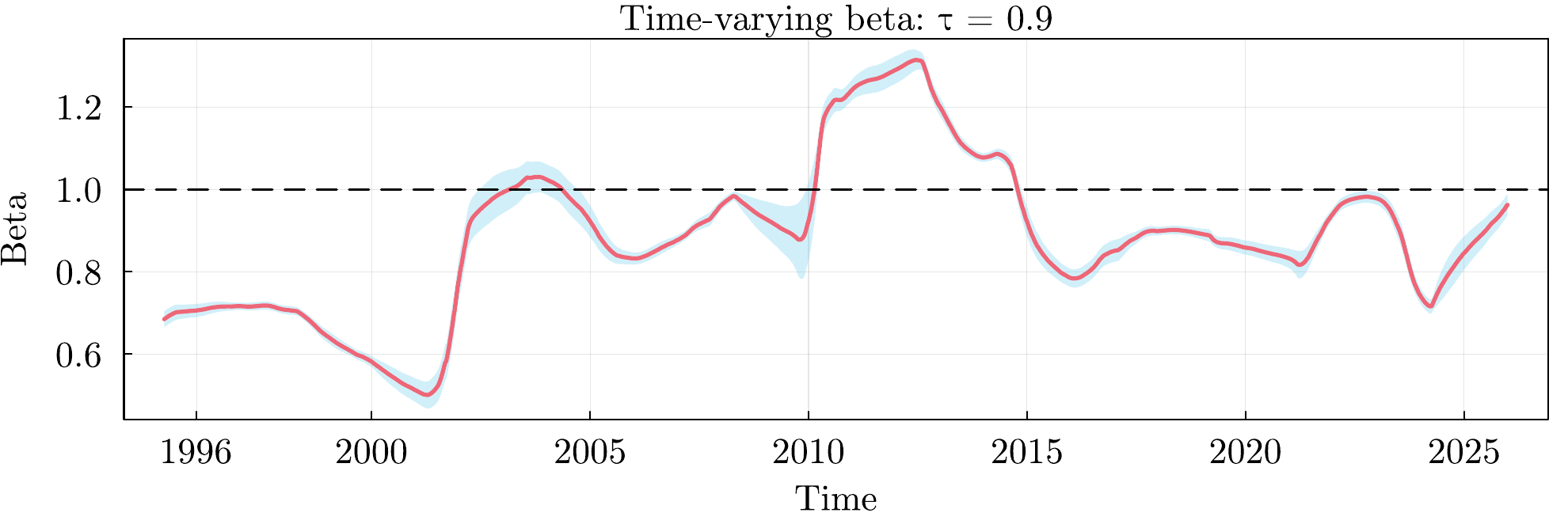}
    \end{center}
    \caption{\textbf{Conditional spanning by the volatility-managed benchmark.}
    This figure reports local time-varying estimates from the regression
    $r^{0.1}_{p,t}=\alpha_t+\beta_t r_{VM,t}+\varepsilon_t$ for the downside-focused
    tail-managed portfolio. The top panel plots the annualized alpha, expressed in
    percent, and the bottom panel plots the beta with respect to the market-volatility-managed
    benchmark. Shaded areas denote pointwise confidence bands based on local
    Newey--West standard errors. The dashed horizontal line in the top panel corresponds
    to zero alpha, and the dashed horizontal line in the bottom panel corresponds to unit
    beta. A beta persistently below one together with positive alpha in selected periods
    indicates that the low-$\tau$ portfolio is a defensive, preference-specific tilt of
    volatility management rather than a strategy fully spanned by the VM benchmark.}
    \label{fig:conditional_spanning_all}
\end{figure}

% Auto-generated by Julia
\begin{table}[!htbp]
\centering
\caption{\textbf{Comparison against factor-level volatility-scaled VM across states.}}
\label{tab:vm_horserace_states_assetwise}
\small
\setlength{\tabcolsep}{4pt}
\begin{tabular}{llrrrrrrrrrrr}
\toprule
Strategy & State & Alpha & SE($\alpha$) & $t(\alpha)$ & Beta & SE($\beta$) & $t(\beta)$ & $R^2$ & N & Mean & IR & $\Delta$CVaR5 \\
\midrule
\addlinespace[0.25em]
$\tau$ = 0.1 & Low volatility & -0.03 & 0.54 & -0.06 & 1.00 & 0.02 & 49.83 & 0.929 & 1542 & -0.03 & -0.02 & -0.03 \\
$\tau$ = 0.5 & Low volatility & 0.09 & 0.76 & 0.12 & 1.27 & 0.02 & 51.60 & 0.916 & 1542 & 0.72 & 0.32 & -0.27 \\
$\tau$ = 0.9 & Low volatility & 0.17 & 0.98 & 0.18 & 1.45 & 0.03 & 51.54 & 0.897 & 1542 & 1.22 & 0.38 & -0.43 \\
\addlinespace[0.3em]
$\tau$ = 0.1 & High volatility & -0.64 & 2.10 & -0.30 & 1.39 & 0.05 & 25.63 & 0.807 & 1542 & 1.47 & 0.24 & -0.69 \\
$\tau$ = 0.5 & High volatility & 0.01 & 3.09 & 0.00 & 1.84 & 0.07 & 26.06 & 0.781 & 1542 & 4.60 & 0.46 & -1.27 \\
$\tau$ = 0.9 & High volatility & -0.15 & 3.92 & -0.04 & 2.13 & 0.08 & 27.37 & 0.749 & 1542 & 6.03 & 0.46 & -1.67 \\
\addlinespace[0.3em]
$\tau$ = 0.1 & Downside market state & -0.05 & 1.91 & -0.03 & 1.40 & 0.05 & 26.55 & 0.828 & 1542 & 2.87 & 0.50 & -0.65 \\
$\tau$ = 0.5 & Downside market state & 0.38 & 2.80 & 0.14 & 1.87 & 0.07 & 27.63 & 0.807 & 1542 & 6.69 & 0.70 & -1.22 \\
$\tau$ = 0.9 & Downside market state & -1.07 & 3.55 & -0.30 & 2.18 & 0.07 & 30.49 & 0.779 & 1542 & 7.46 & 0.60 & -1.65 \\
\addlinespace[0.3em]
$\tau$ = 0.1 & Upside market state & 0.60 & 0.90 & 0.67 & 1.09 & 0.03 & 36.60 & 0.859 & 1542 & 1.01 & 0.43 & -0.17 \\
$\tau$ = 0.5 & Upside market state & -0.60 & 1.31 & -0.46 & 1.43 & 0.04 & 36.92 & 0.830 & 1542 & 1.27 & 0.31 & -0.49 \\
$\tau$ = 0.9 & Upside market state & -0.40 & 1.72 & -0.23 & 1.65 & 0.05 & 36.09 & 0.786 & 1542 & 2.46 & 0.44 & -0.73 \\
\addlinespace[0.3em]
\bottomrule
\end{tabular}
\begin{tablenotes}
\footnotesize
\item \textit{Notes:} Each row reports the regression
\[
r_{p,t}=\alpha+\beta r_{VM,t}+\varepsilon_t
\]
estimated within the indicated state or crisis window. Reported alpha is annualized and expressed in percent. Standard errors are Newey--West. State definitions follow Table~\ref{tab:state_perf}. Crisis windows correspond to the dot-com bust, the global financial crisis, the COVID crash, and the 2022 bear market. Because these windows are short, alpha estimates are noisier than in the full sample and should be interpreted as supportive rather than definitive evidence.
\end{tablenotes}
\end{table}

% Auto-generated by Julia
% Auto-generated by Julia
\begin{table}[!htbp]
\centering
\caption{Comparison against factor-level VM in crisis windows}
\label{tab:vm_horserace_crisis_assetwise}
\small
\setlength{\tabcolsep}{4pt}
\begin{tabular}{llrrrrrrrrrrr}
\toprule
Strategy & Episode & Alpha & SE($\alpha$) & $t(\alpha)$ & Beta & $R^2$ & N & CumRet & CumActive & Active Mean & IR & $\Delta$CVaR5 \\
\midrule
\multicolumn{7}{l}{\textit{VM asset-wise}} \\
\midrule
$\tau$ = 0.1 & Dot-com bust & 1.64 & 1.79 & 0.92 & 0.88 & 0.823 & 638 & 16.94 & 2.43 & 0.99 & 0.36 & 0.04 \\
$\tau$ = 0.5 & Dot-com bust & -5.77 & 3.23 & -1.79 & 1.20 & 0.689 & 638 & 0.67 & -11.51 & -4.69 & -0.87 & -0.41 \\
$\tau$ = 0.9 & Dot-com bust & -10.71 & 4.74 & -2.26 & 1.40 & 0.588 & 638 & -9.36 & -20.17 & -8.58 & -1.07 & -0.72 \\
\addlinespace[0.3em]
$\tau$ = 0.1 & GFC & -3.42 & 6.08 & -0.56 & 1.96 & 0.778 & 356 & -17.86 & -11.30 & -8.11 & -0.93 & -1.24 \\
$\tau$ = 0.5 & GFC & -5.93 & 8.27 & -0.72 & 2.54 & 0.744 & 356 & -24.60 & -18.32 & -13.46 & -1.03 & -1.92 \\
$\tau$ = 0.9 & GFC & -9.07 & 10.35 & -0.88 & 3.03 & 0.719 & 356 & -31.07 & -25.13 & -19.03 & -1.12 & -2.53 \\
\addlinespace[0.3em]
$\tau$ = 0.1 & COVID crash & 51.04 & 27.28 & 1.87 & 1.93 & 0.946 & 24 & -13.39 & -4.16 & -42.73 & -2.17 & -2.66 \\
$\tau$ = 0.5 & COVID crash & 71.52 & 32.35 & 2.21 & 2.57 & 0.947 & 24 & -17.37 & -8.36 & -86.55 & -2.71 & -4.30 \\
$\tau$ = 0.9 & COVID crash & 71.90 & 32.45 & 2.22 & 2.76 & 0.951 & 24 & -19.05 & -10.15 & -106.05 & -2.98 & -4.72 \\
\addlinespace[0.3em]
$\tau$ = 0.1 & 2022 bear & -3.27 & 2.49 & -1.31 & 1.12 & 0.944 & 196 & -4.19 & -2.68 & -3.46 & -1.51 & -0.22 \\
$\tau$ = 0.5 & 2022 bear & -7.17 & 6.10 & -1.18 & 1.53 & 0.853 & 196 & -7.86 & -6.22 & -8.04 & -1.24 & -0.84 \\
$\tau$ = 0.9 & 2022 bear & -12.53 & 9.56 & -1.31 & 1.89 & 0.784 & 196 & -12.40 & -10.69 & -13.99 & -1.35 & -1.40 \\

\addlinespace[0.3em]
\bottomrule
\end{tabular}
\vspace{0.4em}
\begin{minipage}{0.94\linewidth}
\footnotesize Notes: Each row reports the regression $r_{p,t} = \alpha + \beta r_{VM,t} + \varepsilon_t$ estimated within the indicated crisis window. Standard errors are Newey--West and alpha is annualized in percent. Because crisis windows are short, the alpha estimates should be interpreted more cautiously than the full-sample or state-based estimates.
\end{minipage}
\end{table}

\begin{figure}[ht]
    \begin{center}
        \includegraphics[width=0.82\textwidth]{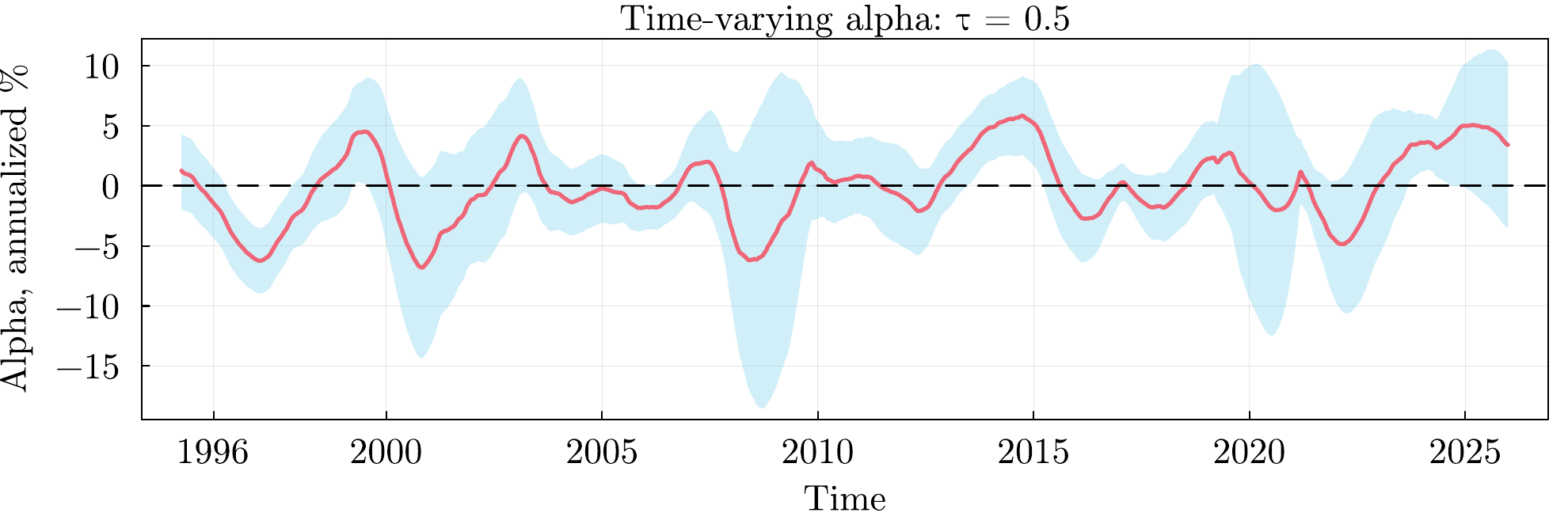}
        \includegraphics[width=0.82\textwidth]{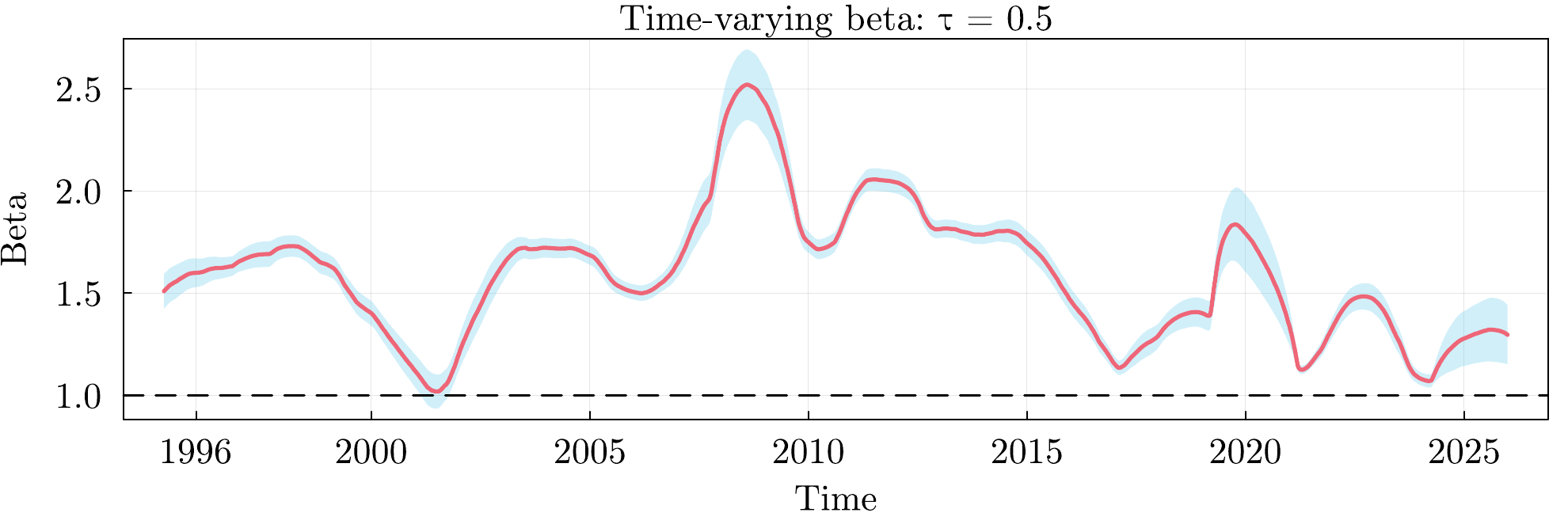}
        \includegraphics[width=0.82\textwidth]{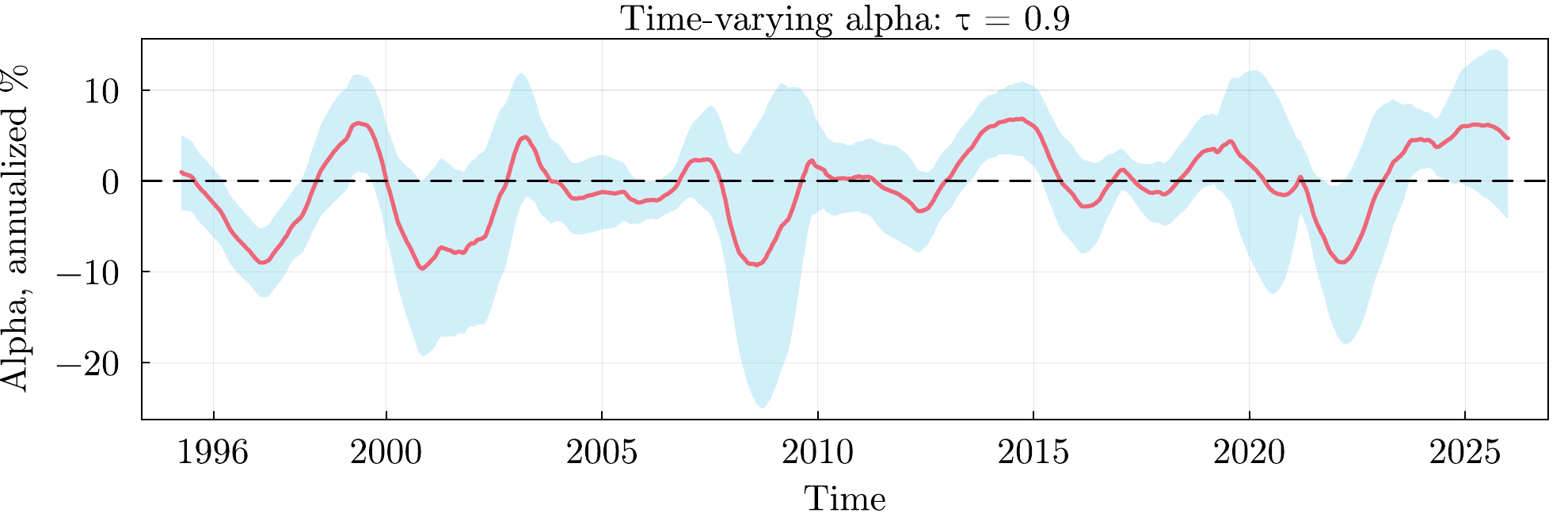}
        \includegraphics[width=0.82\textwidth]{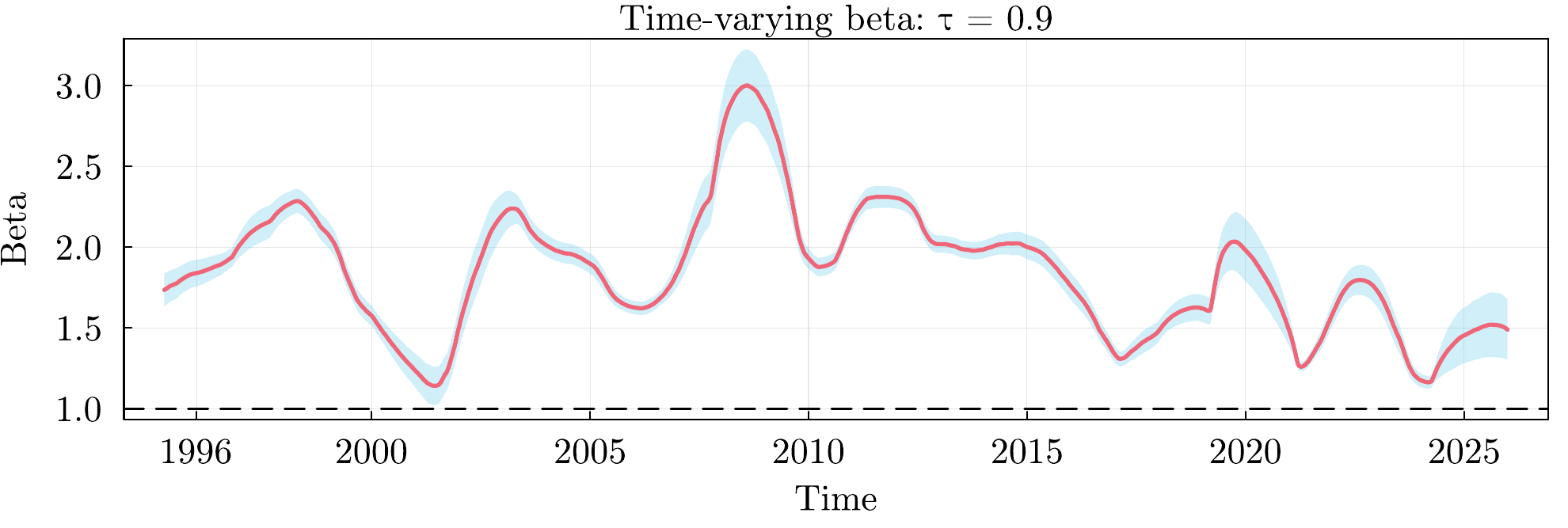}
    \end{center}
    \caption{\textbf{Conditional spanning by the factor-level volatility-scaled benchmark:
    additional quantile levels.}
    This figure reports local time-varying estimates from
    $r^{\tau}_{p,t}=\alpha_t+\beta_t r^{FV}_{t}+\varepsilon_t$ for
    $\tau=0.5$ and $\tau=0.9$. The top panel within each pair plots the annualized alpha,
    expressed in percent, and the bottom panel plots the beta with respect to the
    factor-level volatility-scaled benchmark. Shaded areas denote pointwise confidence
    bands based on local Newey--West standard errors.}
    \label{fig:conditional_spanning_factor_vm_appendix}
\end{figure}

\clearpage
\section{Additional OOS Empirical Results: Alternative Universe}
\label{ref:add_11alt_oos}

This sextion contains additional results using alternative universe of six-factors mixed with market neutral portfolios such as betting against beta, or quality as well as tail-shape such as PUT, or PPUT, specifically BAB, BXM, CMA, HML, HML Devil, MOM, MKT, PPUT, QMJ RMW, SMB.

\begin{table}[htbp]
\centering
\caption{\textbf{Conditional out-of-sample performance by volatility and market-tail states.}}
\label{tab:state_perf_alt}
\footnotesize
\setlength{\tabcolsep}{6pt}

\begin{tabular}{lccccc}
\toprule
Metric & UC & VM & $\tau=0.1$ & $\tau=0.5$ & $\tau=0.9$ \\
\midrule

\multicolumn{6}{l}{\textit{Panel A: Volatility states}} \\
\addlinespace[0.25em]
\multicolumn{6}{l}{\quad \textit{Low volatility state}} \\
Ann. Mean (\%) & -0.76 & -0.59 & -0.41 & -0.24 & \textbf{-0.07} \\ 
Sharpe (ann.) & -0.24 & -0.20 & -0.15 & -0.08 & \textbf{-0.02} \\ 
Sortino (ann.) & -0.33 & -0.26 & -0.20 & -0.11 & \textbf{-0.03} \\ 
CVaR 5\% (\%) & -0.47 & -0.46 & \textbf{-0.43} & -0.44 & -0.46 \\ 
Tail-Adj Sharpe (CVaR5) & -1.59 & -1.28 & -0.96 & -0.54 & \textbf{-0.16} \\ 
\addlinespace[0.4em]
\multicolumn{6}{l}{\quad \textit{High volatility state}} \\
Ann. Mean (\%) & 5.44 & \textbf{5.67} & 3.36 & 3.62 & 3.91 \\ 
Sharpe (ann.) & 0.70 & \textbf{0.90} & 0.45 & 0.48 & 0.49 \\ 
Sortino (ann.) & 0.91 & \textbf{1.24} & 0.59 & 0.62 & 0.65 \\ 
CVaR 5\% (\%) & -1.23 & \textbf{-0.93} & -1.15 & -1.18 & -1.21 \\ 
Tail-Adj Sharpe (CVaR5) & 4.42 & \textbf{6.10} & 2.92 & 3.08 & 3.22 \\ 
\addlinespace[0.4em]
\midrule
\multicolumn{6}{l}{\textit{Panel B: Tail states}} \\
\addlinespace[0.25em]
\multicolumn{6}{l}{\quad \textit{Low-tail state}} \\
Ann. Mean (\%) & 1.18 & \textbf{1.19} & -0.25 & 0.26 & 0.99 \\ 
Sharpe (ann.) & 0.17 & \textbf{0.20} & -0.04 & 0.04 & 0.13 \\ 
Sortino (ann.) & 0.22 & \textbf{0.27} & -0.04 & 0.04 & 0.16 \\ 
CVaR 5\% (\%) & -1.11 & \textbf{-0.90} & -1.16 & -1.19 & -1.22 \\ 
Tail-Adj Sharpe (CVaR5) & 1.06 & \textbf{1.33} & -0.22 & 0.22 & 0.81 \\ 
\addlinespace[0.4em]
\multicolumn{6}{l}{\quad \textit{High-tail state}} \\
Ann. Mean (\%) & 2.34 & 2.34 & \textbf{5.01} & 4.92 & 4.94 \\ 
Sharpe (ann.) & 0.44 & 0.55 & \textbf{1.23} & 1.19 & 1.14 \\ 
Sortino (ann.) & 0.60 & 0.80 & \textbf{1.92} & 1.84 & 1.73 \\ 
CVaR 5\% (\%) & -0.78 & -0.61 & \textbf{-0.55} & -0.56 & -0.58 \\ 
Tail-Adj Sharpe (CVaR5) & 2.99 & 3.85 & \textbf{9.18} & 8.85 & 8.46 \\ 
\bottomrule
\end{tabular}

\vspace{0.35em}

\noindent
\begin{minipage}{\textwidth}
\footnotesize
\textit{Notes:} This table reports conditional out-of-sample performance for UC, VM, and the tail-managed portfolios with $\tau \in \{0.1,0.5,0.9\}$. In Panel A, volatility states are defined using lagged 21-day realized market volatility; low- and high-volatility states correspond to the bottom and top quintiles of the volatility distribution. In Panel B, market-tail states are defined using lagged 21-day compounded market returns; downside and upside states correspond to the bottom and top quintiles of the return distribution. Annualized mean returns, Sharpe ratios, and Tail-Adj Sharpe (CVaR5) are reported within each state. CVaR 5\% is the average return in the worst 5\% of observations within the corresponding state. The table shows where the gains from tail targeting arise: VM is strongest in calm states, whereas tail-managed portfolios become more competitive when volatility is high or when market states make tail considerations more relevant. Bold entries indicate the best value in each row.
\end{minipage}

\end{table}

% Auto-generated by Julia
\begin{table}[!htbp]
\centering
\caption{Horse-race against VM market by volatility and tail states}
\label{tab:vm_horserace_states_alt}
\small
\setlength{\tabcolsep}{4pt}
\begin{tabular}{llrrrrrrrrrrr}
\toprule
Strategy & State & Alpha & SE($\alpha$) & $t(\alpha)$ & Beta & SE($\beta$) & $t(\beta)$ & $R^2$ & N & Active Mean & IR & $\Delta$CVaR5 \\
\midrule
$\tau$  = 0.1 & Low volatility & -0.79 & 1.27 & -0.62 & 0.30 & 0.04 & 7.83 & 0.223 & 927 & -1.67 & -0.42 & 0.28 \\
$\tau$  = 0.5 & Low volatility & -0.64 & 1.29 & -0.50 & 0.32 & 0.04 & 7.84 & 0.239 & 927 & -1.50 & -0.38 & 0.27 \\
$\tau$  = 0.9 & Low volatility & -0.50 & 1.33 & -0.38 & 0.34 & 0.05 & 7.53 & 0.249 & 927 & -1.33 & -0.34 & 0.24 \\
\addlinespace[0.3em]
$\tau$  = 0.1 & High volatility & 1.22 & 3.23 & 0.38 & 0.39 & 0.05 & 8.66 & 0.289 & 927 & -2.12 & -0.24 & 0.30 \\
$\tau$  = 0.5 & High volatility & 1.35 & 3.24 & 0.42 & 0.41 & 0.04 & 9.25 & 0.307 & 927 & -1.86 & -0.21 & 0.28 \\
$\tau$  = 0.9 & High volatility & 1.57 & 3.30 & 0.48 & 0.43 & 0.04 & 9.56 & 0.300 & 927 & -1.57 & -0.18 & 0.24 \\
\addlinespace[0.3em]
$\tau$  = 0.1 & Low tail & -1.01 & 2.82 & -0.36 & 0.42 & 0.04 & 10.75 & 0.388 & 927 & -2.06 & -0.25 & 0.37 \\
$\tau$  = 0.5 & Low tail & -0.54 & 2.85 & -0.19 & 0.44 & 0.04 & 11.68 & 0.411 & 927 & -1.55 & -0.19 & 0.34 \\
$\tau$  = 0.9 & Low tail & 0.15 & 2.92 & 0.05 & 0.46 & 0.04 & 12.17 & 0.410 & 927 & -0.83 & -0.10 & 0.31 \\
\addlinespace[0.3em]
$\tau$  = 0.1 & High tail & 4.68 & 2.16 & 2.17 & 0.07 & 0.03 & 2.14 & 0.012 & 927 & 0.47 & 0.07 & 0.37 \\
$\tau$  = 0.5 & High tail & 4.57 & 2.18 & 2.09 & 0.08 & 0.04 & 2.17 & 0.013 & 927 & 0.38 & 0.05 & 0.36 \\
$\tau$  = 0.9 & High tail & 4.64 & 2.26 & 2.06 & 0.07 & 0.04 & 1.67 & 0.009 & 927 & 0.40 & 0.06 & 0.33 \\
\addlinespace[0.3em]
\bottomrule
\end{tabular}
\vspace{0.4em}
\begin{minipage}{0.94\linewidth}
\footnotesize Notes: Each row reports the regression $r_{p,t} = \alpha + \beta r_{VM,t} + \varepsilon_t$ estimated within the indicated conditional state. Volatility states are defined using lagged 21-day realized market volatility; tail states are defined using lagged 21-day compounded market returns. Standard errors are Newey--West. Reported alpha is annualized in percent.
\end{minipage}
\end{table}

% Auto-generated by Julia
\begin{table}[!htbp]
\centering
\caption{Horse-race against VM asset-wise by volatility and tail states}
\label{tab:vm_horserace_states_alt}
\small
\setlength{\tabcolsep}{4pt}
\begin{tabular}{llrrrrrrrrrrr}
\toprule
Strategy & State & Alpha & SE($\alpha$) & $t(\alpha)$ & Beta & SE($\beta$) & $t(\beta)$ & $R^2$ & N & Active Mean & IR & $\Delta$CVaR5 \\
\midrule
$\tau$  = 0.1 & Low volatility & 0.01 & 0.91 & 0.01 & 0.71 & 0.08 & 9.42 & 0.571 & 927 & 0.18 & 0.09 & 0.03 \\
$\tau$  = 0.5 & Low volatility & 0.19 & 0.92 & 0.21 & 0.73 & 0.08 & 9.26 & 0.571 & 927 & 0.35 & 0.17 & 0.02 \\
$\tau$  = 0.9 & Low volatility & 0.37 & 0.96 & 0.39 & 0.76 & 0.09 & 8.86 & 0.563 & 927 & 0.52 & 0.24 & -0.00 \\
\addlinespace[0.3em]
$\tau$  = 0.1 & High volatility & 2.25 & 3.73 & 0.60 & 0.20 & 0.11 & 1.73 & 0.028 & 927 & -2.31 & -0.26 & -0.22 \\
$\tau$  = 0.5 & High volatility & 2.54 & 3.80 & 0.67 & 0.19 & 0.12 & 1.64 & 0.025 & 927 & -2.06 & -0.23 & -0.25 \\
$\tau$  = 0.9 & High volatility & 2.83 & 3.86 & 0.73 & 0.19 & 0.12 & 1.61 & 0.023 & 927 & -1.77 & -0.19 & -0.28 \\
\addlinespace[0.3em]
$\tau$  = 0.1 & Low tail & -0.49 & 3.39 & -0.15 & 0.20 & 0.13 & 1.59 & 0.029 & 927 & -1.44 & -0.17 & -0.26 \\
$\tau$  = 0.5 & Low tail & 0.03 & 3.49 & 0.01 & 0.20 & 0.13 & 1.50 & 0.025 & 927 & -0.92 & -0.11 & -0.29 \\
$\tau$  = 0.9 & Low tail & 0.77 & 3.59 & 0.21 & 0.19 & 0.13 & 1.40 & 0.021 & 927 & -0.20 & -0.02 & -0.32 \\
\addlinespace[0.3em]
$\tau$  = 0.1 & High tail & 3.92 & 1.83 & 2.13 & 0.47 & 0.07 & 6.36 & 0.243 & 927 & 2.67 & 0.63 & 0.06 \\
$\tau$  = 0.5 & High tail & 3.82 & 1.86 & 2.05 & 0.47 & 0.07 & 6.33 & 0.236 & 927 & 2.59 & 0.61 & 0.05 \\
$\tau$  = 0.9 & High tail & 3.82 & 1.93 & 1.98 & 0.48 & 0.08 & 6.22 & 0.229 & 927 & 2.61 & 0.59 & 0.02 \\
\addlinespace[0.3em]
\bottomrule
\end{tabular}
\vspace{0.4em}
\begin{minipage}{0.94\linewidth}
\footnotesize Notes: Each row reports the regression $r_{p,t} = \alpha + \beta r_{VM,t} + \varepsilon_t$ estimated within the indicated conditional state. Volatility states are defined using lagged 21-day realized market volatility; tail states are defined using lagged 21-day compounded market returns. Standard errors are Newey--West. Reported alpha is annualized in percent.
\end{minipage}
\end{table}

% ======================================================================
% APPENDIX TABLES
% Place in the Internet Appendix or Appendix section.
% ======================================================================
\clearpage
\section{Appendix Tables for Mandate-Flow Evidence}

\begin{table}[ht]
\centering
\caption{Full regression: income and growth mandate interactions}
\label{tab:appendix_income_growth_full}
\begin{tabular}{lcc}
\toprule
Variable & Estimate & $t$-statistic \\
\midrule
q10\_perf & 0.061*** & 5.387 \\ 
q50\_perf & 0.080*** & 7.590 \\ 
q90\_perf & 0.044*** & 3.853 \\ 
income\_q10 & 0.097*** & 6.205 \\ 
income\_q50 & 0.012 & 0.538 \\ 
income\_q90 & 0.047*** & 2.853 \\ 
growth\_q10 & -0.039*** & -2.671 \\ 
growth\_q50 & 0.003 & 0.167 \\ 
growth\_q90 & -0.019 & -1.238 \\ 
log\_assets & 0.004*** & 10.400 \\ 
lag\_flow & 0.173*** & 58.787 \\ 
lag\_ret & 0.011*** & 3.077 \\ 
vol\_perf & 0.099*** & 4.240 \\ 
\bottomrule
\end{tabular}
\begin{minipage}{0.92\textwidth}
\footnotesize
\emph{Notes:} This table reports the full coefficient vector for Equation~\eqref{eq:mandate_quantile_flows}. The dependent variable is next-month net flow. The regression includes fund and month fixed effects. Standard errors are heteroskedasticity robust. $^{***}$, $^{**}$, and $^{*}$ denote significance at the 1\%, 5\%, and 10\% levels.
\end{minipage}
\end{table}

\begin{table}[ht]
\centering
\caption{Full regression: downside-protection demand and risk salience}
\label{tab:appendix_downside_salience_full}
\begin{tabular}{lcc}
\toprule
Variable & Estimate & $t$-statistic \\
\midrule
mean\_perf & 0.241*** & 14.896 \\ 
worst\_perf & -0.047*** & -5.127 \\ 
best\_perf & -0.001 & -0.156 \\ 
vol\_perf & 0.006 & 0.205 \\ 
downside\_worst & 0.297*** & 4.972 \\ 
downside\_best & -0.266*** & -4.862 \\ 
downside\_vol & 1.105*** & 5.535 \\ 
growth\_worst & 0.013 & 0.655 \\ 
growth\_best & -0.034* & -1.947 \\ 
growth\_vol & 0.118** & 1.972 \\ 
log\_assets & 0.004*** & 10.236 \\ 
lag\_flow & 0.173*** & 58.700 \\ 
lag\_ret & 0.010*** & 2.828 \\ 

\bottomrule
\end{tabular}
\begin{minipage}{0.92\textwidth}
\footnotesize
\emph{Notes:} This table reports the full coefficient vector for Equation~\eqref{eq:downside_salience}. The dependent variable is next-month net flow. The regression includes fund and month fixed effects. Standard errors are heteroskedasticity robust. $^{***}$, $^{**}$, and $^{*}$ denote significance at the 1\%, 5\%, and 10\% levels.
\end{minipage}
\end{table}

\begin{table}[ht]
\centering
\caption{Robustness: salience dummy specification}
\label{tab:appendix_salience_dummies}
\begin{tabular}{lcc}
\toprule
Variable & Estimate & $t$-statistic \\
\midrule
mean\_perf & 0.179*** & 14.354 \\ 
high\_vol\_state & -0.001*** & -3.847 \\ 
bad\_downside\_state & -0.002*** & -5.867 \\ 
downside\_high\_vol & -0.006* & -1.931 \\ 
downside\_bad\_state & -0.001 & -0.348 \\ 
growth\_high\_vol & -0.001 & -0.772 \\ 
growth\_bad\_state & 0.002*** & 3.495 \\ 
log\_assets & 0.003*** & 9.589 \\ 
lag\_flow & 0.173*** & 58.838 \\ 
lag\_ret & 0.014*** & 3.778 \\ 

\bottomrule
\end{tabular}
\begin{minipage}{0.92\textwidth}
\footnotesize
\emph{Notes:} This table reports a dummy-based robustness specification. High-volatility months are defined using the top quintile of recent fund return volatility, and bad-downside months are defined using the bottom quintile of recent worst-month performance. The dependent variable is next-month net flow. The regression includes fund and month fixed effects. Standard errors are heteroskedasticity robust. $^{***}$, $^{**}$, and $^{*}$ denote significance at the 1\%, 5\%, and 10\% levels.
\end{minipage}
\end{table}

\end{document}